\documentclass[usenatbib]{emulateapj}
\usepackage{graphicx}
\usepackage[flushleft]{threeparttable}
\usepackage[usenames,dvipsnames,svgnames,table]{xcolor}
\usepackage{hyperref}
\definecolor{darkblue}{rgb}{0.0,0.0,0.9}
\hypersetup{colorlinks,breaklinks,
            linkcolor=darkblue,urlcolor=darkblue,
            anchorcolor=darkblue,citecolor=darkblue}
\usepackage{amsmath,amssymb}
\usepackage{amsmath}

\usepackage{natbib}
\usepackage{bm}
\usepackage{lineno}
\begin{document}

\def\etal{et al.\ \rm}
\def\ba{\begin{eqnarray}}
\def\ea{\end{eqnarray}}
\def\etal{et al.\ \rm}
\def\Fdw{F_{\rm dw}}
\def\Tex{T_{\rm ex}}
\def\Fdis{F_{\rm dw,dis}}
\def\Fnu{F_\nu}
\def\WD{\rm WD}

\newcommand{\ra}{r_\mathrm{a}}
\newcommand{\rb}{r_\mathrm{b}}
\newcommand{\rc}{r_\mathrm{c}}
\newcommand{\md}{\mathrm{d}}
\newcommand{\me}{\mathrm{e}}
\newcommand{\mi}{\mathrm{i}}
\newcommand{\rp}{r_\mathrm{p}}
\newcommand{\fmer}{f_\mathrm{m}}
\newcommand{\rt}{\mathrm{t}}

\newcommand{\fnow}{f_\mathrm{m}(12\,\mathrm{Gyr})}
\def\p{\partial}
\newcommand{\epsGR}{\epsilon_\mathrm{GR}}
\newcommand{\nn}{\nonumber}
\newcommand{\jmin}{j_\mathrm{min}}
\newcommand{\Rg}{\mathbf{R}_\mathrm{g}}
\newcommand{\Rb}{\mathbf{R}_\mathrm{b}}

\newcommand{\epsstr}{\epsilon_\mathrm{strong}}
\newcommand{\epsweak}{\epsilon_\mathrm{weak}}
\newcommand{\epspibytwo}{\epsilon_{\pi/2}}

\newcommand{\tmin}{t_\mathrm{min}}
\newcommand{\imin}{i_\mathrm{min}}
\newcommand{\jmax}{j_\mathrm{max}}
\newcommand{\pmin}{p_\mathrm{min}}
\newcommand{\emax}{e_\mathrm{max}}
\newcommand{\elim}{e_\mathrm{lim}}
\newcommand{\tsec}{t_\mathrm{sec}}
\newcommand{\bw}{\mathbf{w}}

\newcommand\cmtrr[1]{{\color{red}[RR: #1]}}
\newcommand\cmtch[1]{{\color{red}[CH: #1]}}
\newcommand\newrr[1]{{\color{magenta} #1}}
\newcommand\newch[1]{{\color{red} #1}}
\newcommand\red[1]{{\color{red} #1}}

%%%%%%%%%%%%%%%%%%%%%%%%%%%%%%%%%%%
%%%%%%%%%%%%%%%%%%%%%%%%%%%%%%%%%%%
%%%%%%%%%%%%%%%%%%%%%%%%%%%%%%%%%%%
%%%%%%%%%%%%%%%%%%%%%%%%%%%%%%%%%%%

\title{Relativistic phase space diffusion of compact object binaries\\
in stellar clusters and hierarchical triples}

\author{Chris Hamilton\altaffilmark{1,$\dagger$} \& Roman R. Rafikov\altaffilmark{1,2}}
\altaffiltext{1}{Institute for Advanced Study, Einstein Drive, Princeton, NJ 08540}
\altaffiltext{2}{Department of Applied Mathematics and Theoretical Physics, University of Cambridge, Wilberforce Road, Cambridge CB3 0WA, UK}
\altaffiltext{$\dagger$}{chamilton@ias.edu}

%%%%%%%%%%%%%%%%%%%%%%%%%%%%%%%%%%%
%%%%%%%%%%%%%%%%%%%%%%%%%%%%%%%%%%%
%%%%%%%%%%%%%%%%%%%%%%%%%%%%%%%%%%%
%%%%%%%%%%%%%%%%%%%%%%%%%%%%%%%%%%%

\begin{abstract}
The LIGO/Virgo detections of compact object mergers have posed a challenge for theories of binary evolution and coalescence. One promising avenue for producing mergers dynamically is through secular eccentricity oscillations driven by an external perturber, be it a tertiary companion (as in the Lidov-Kozai (LK) mechanism) or the tidal field of the stellar cluster in which the binary orbits. The simplest theoretical models of these oscillations use a `doubly-averaged' (DA) approximation, averaging both over the binary's internal Keplerian orbit and its `outer' barycentric orbit relative to the perturber. However, DA theories do not account for fluctuations of the perturbing torque on the outer orbital timescale, which are known to increase a binary's eccentricity beyond the maximum DA value, potentially accelerating mergers. Here we reconsider the impact of these short-timescale fluctuations in the test-particle quadrupolar limit for binaries perturbed by arbitrary spherical cluster potentials (including LK as a special case), {in particular including 1pN} general relativistic (GR) apsidal precession of the internal orbit. Focusing on the behavior of the binary orbital elements around peak eccentricity, we discover a new effect, \textit{relativistic phase space diffusion} (RPSD), in which a binary can jump to a completely new dynamical trajectory on an outer orbital timescale, violating the approximate conservation of DA integrals of motion. RPSD arises from an interplay between secular behavior at extremely high eccentricity, short-timescale fluctuations, and rapid GR precession, and can change the subsequent secular evolution dramatically. This {effect occurs even in hierarchical triples},
but has not been uncovered until now.
\end{abstract}
%%%%%%%%%%%%%%%%%%%%%%%%%%%%%%%%%%%%%%%%%%%%%%%%%%

%%%%%%%%%%%%%%%%%%%%%%%%%%%%%%%%%%%%%%%%%%%%%%%%%%

%%%%%%%%%%%%%%%%% BODY OF PAPER %%%%%%%%%%%%%%%%%%

%%%%%%%%%%%%%%%%%%%%%%%%%%%%%%%%%%%%%%%%%%%%%%%%%%
%%%%%%%%%%%%%%%%%%%%%%%%%%%%%%%%%%%%%%%%%%%%%%%%%%

    \section{Introduction}
    
    The compact object 
%    (black hole and/or neutron star)
    --- black hole (BH) and/or neutron star (NS) ---
    binary mergers discovered by the LIGO/Virgo collaboration in recent years \citep{abbott2021population}
    have reinvigorated the detailed study of secular evolution of binaries in external tidal fields.
    The most famous scenario of this kind 
   is a hierarchical triple, in which the compact object binary in question is orbited by a bound tertiary companion. In this case the tertiary companion can drive secular oscillations of the binary orbital elements --- known as Lidov-Kozai (LK) oscillations \citep{Lidov1962,Kozai1962} --- on timescales much longer than either orbital period.
    Alternatively, very similar secular evolution arises if one considers a binary perturbed not by a tertiary point mass but 
   by the global tidal field 
    of the stellar cluster in which it resides \citep{Hamilton2019a,Hamilton2019b,Hamilton2019c,Bub2020}.
    In either scenario, the key idea is that 
    {secular tidal forcing can increase a binary's eccentricity dramatically.
    Provided this forcing is strong enough to overcome 1pN general relativistic (GR) apsidal precession --- which acts to decrease the maximum eccentricity \citep{Wen2003,Hamilton2021} --- 
    the binary can thereby achieve a small pericenter distance,
    allowing it to radiate gravitational waves (GWs) efficiently and thus rapidly merge.}
    Similar mechanisms have been invoked to explain
    %ascribed to 
    the {origin} of other 
    exotic 
    %astrophysical systems
    objects, such as Type 1a supernovae \citep{Katz2012},
    blue stragglers \citep{Leigh2018}, hot Jupiters \citep{Fabrycky2007},
    and so on.
    
Regardless of the particular
    system under consideration, the same two key questions always arise: 
    (i) under
    what circumstances can a binary reach extremely high eccentricity on an
    astrophysically relevant timescale? and (ii) how does the binary behave when
    it reaches such extreme eccentricities?
   
    Question (i) is straightforward to answer using secular theories, the
    simplest of which involve truncating the perturbing potential at quadrupole
    order, taking the `test particle' approximation and then `double-averaging'
    (DA) --- that is, averaging the dynamics over both the binary's `inner' Keplerian orbit 
    {and over its 'outer' barycentric motion relative to the perturbing potential}.
    In \citet{Hamilton2019a,Hamilton2019b,Hamilton2021} --- hereafter Papers I, II and III respectively --- 
    we
    developed the most comprehensive such DA theory to date, capable of describing
    the secular evolution of any binary perturbed by any {fixed} axisymmetric 
    potential {(in the test particle, quadrupole limit),
    accounting for 1pN GR precession of the inner orbit.
    In this theory, {which includes the LK scenario as a special case,}
the binary's
    maximum eccentricity $e_\mathrm{max}$ can be calculated (semi-)analytically
    as a function of the initial conditions. As a result one can easily determine
    the region of parameter space that leads to extremely high $e_\mathrm{max}$.
    {In the case of spherical cluster potentials (including the Keplerian LK case) which we will focus on exclusively throughout this paper,} one always finds that $e_\mathrm{max}$ is limited by the initial
    relative inclination $i_0$ between the binary's inner and outer orbital planes:
    %\footnote{Strictly, for a generic axisymmetric potential the outer orbit is not confined to a plane, but instead
    %fills a torus whose symmetry axis coincides with that of the potential; in this case, inclination is measured relative to the plane perpendicular to this axis.}
    %%%%%%%%%%%%%%%%%%%%%%%%%%%%%%%%%%%%
    \begin{align}
    e_\mathrm{max} \leq e_\mathrm{lim} \equiv (1-\Theta)^{1/2},
    \end{align}
        %%%%%%%%%%%%%%%%%%%%%%%%%%%%%%%%%%%%
    where
        %%%%%%%%%%%%%%%%%%%%%%%%%%%%%%%%%%%%
    \begin{align}
  \Theta \equiv (1-e_0^2)\cos^2 i_0,
    \end{align}
    %%%%%%%%%%%%%%%%%%%%%%%%%%%%%%%%%%%%
    and $e_0$ is the initial eccentricity. Hence, a necessary (but not always sufficient) part of the answer to
    question (i) is that $\Theta \ll 1$.
    
    However, DA theories often do not provide an accurate answer to question
    (ii). That is because DA theory ignores a component of the torque that
    fluctuates on the timescale of the outer orbit, and normally washes out to zero upon
    averaging {over that timescale}.  This becomes problematic at extremely high eccentricity, when
    the relative changes in the binary's (very small) angular momentum due to
    this fluctuating torque can become $\mathcal{O}{(1)}$. {As} a result, the
    DA theory {can fail} to capture the dynamics in detail.  A more accurate (if more cumbersome)
    description is provided by the singly-averaged (SA) theory, in which one only 
 averages over the binary's inner Keplerian orbit, and hence
    captures fully the fluctuations in the orbital elements on the outer orbital
    timescale. In particular, these short-timescale fluctuations\footnote{{Throughout this paper, we
    use the term `short-timescale fluctuations' to mean those fluctuations that arise in SA theory when
    compared with DA theory.  We do not consider other fluctuations that might
    occur on short timescales, e.g. flyby encounters with passing stars.}} (sometimes
    called `SA fluctuations') can increase a binary's maximum eccentricity
    beyond $e_\mathrm{max}$ \citep{Ivanov2005}.  Because of this they can be of great significance
    when predicting LK-driven merger rates of black hole (BH) or neutron star
    (NS) binaries, blue straggler formation rates, white dwarf collision rates,
    and so on (e.g.
\citealt{Katz2012,Antonini2012,Bode2014,Antonini2014,Antognini2014,Luo2016,Grishin2018,Lei2018,Lei2019,Mangipudi2022}).
    
    Faced with this assessment, one might decide simply to abandon DA theory
    altogether and only work with the SA equations of motion {(e.g. \citealt{Bub2020})}.  Alternatively one
    might choose to forego all averaging, and instead to integrate the `N-body'
    equations of motion directly {(see e.g. the 3-body integration results of \citealt{Antonini2014})}.    
    There are three main objections
    to these approaches.  First, numerical integration of the SA or {direct} N-body
    equations is prohibitively expensive if one wants to evolve millions
    of binary initial conditions, as 
    done in e.g.
    %we did in 
    \cite{Hamilton2019c}. Second, SA and N-body approaches
    necessarily demand more initial data, inflating the parameter space.  Third,
    whatever one gains though brute-force computation, one also often sacrifices
    in terms of analytical and physical insight. Instead, our approach will be
    to understand the SA,
    at high eccentricity in an approximate
    analytical fashion, guided by the DA theory and by numerical integrations
    where appropriate. 
    {We will restrict ourselves to studying the test particle quadrupole limit, and we will include GR precession of the inner orbit but we will ignore GW emission (though see \citealt{hamilton2022anatomy}).}
    %As a corollary we will be able to
    %calibrate a merger timescale equation  (Hamitlon \& Rafikov, in prep.).
    
%In fact, by dissecting
In particular, by examining in detail the numerical solutions to the SA equations of motion,
we have encountered an important phenomenon which we call 
\textit{relativistic phase space diffusion} (RPSD).
This phenomenon 
must have been present in many authors' direct numerical integrations of the LK problem,
but {seems not to have been discussed explicitly in the past}.
%acknowledged 
{The only exception we know of is the recent study by \cite{Rasskazov2023}, who confirmed our results numerically, and who extended them by showing that RPSD is present and important even when GW emission is included.}

Since RPSD is the key result of this paper, let us now motivate our study with an example of it.

%%%%%%%%%%%%%%%%%%%%%%%%%%%%%%%%%%%%%%%%%%%%%%%%%%%%%%%
%%%%%%%%%%%%%%%%%%%%%%%%%%%%%%%%%%%%%%%%%%%%%%%%%%%%%%%

\subsection{Example of relativistic phase space diffusion}
\label{sec:phase_dependence}

%    However, they have not yet been explored in the more general case of
   % binaries perturbed by arbitrary axisymmetric potentials, e.g. that of a host
    %globular cluster, nuclear cluster or galaxy in which the binary orbits. %The first 
  %  aim of this paper is to extend the well-known LK results to these
   % more general systems, and to expand upon them. 
    It is well known that when a binary is driven secularly to very high eccentricity, then
    even in the DA approximation, its orbital elements (i.e. its eccentricity, argument of pericenter, and longitude of ascending node)  exhibit $\mathcal{O}(1)$ fractional changes on the 
    timescale (see Paper III):
    \begin{equation}
        t_\mathrm{min} \sim \jmin t_\mathrm{sec},
        \label{eqn:tmin_jmin}
        \end{equation}
     where $\jmin \equiv (1-\emax^2)^{1/2}$ is the minimum dimensionless angular momentum, $\emax$ is the maximum eccentricity, and $\tsec$ is the secular timescale.  Clearly, for very large eccentricities, $\tmin \ll \tsec$.
     The central result of this paper is 
     that in the SA approximation,
    if $\tmin$ is so short as to be comparable to or smaller than the outer orbital period $T_\phi$,
    and GR apsidal precession is included,
    %More precisely, one often finds $T_\phi \, \md \ln j/\md t
    %\sim 1$ when $j \ll 1$ small enough, while $T_\phi \, \md \ln \omega/\md t
    %\sim 1$ and $T_\phi \, \md \ln \Omega/\md t \sim 1$ because of the scaling
    %$\md{\omega}/\md t, \,\md{\Omega}/\md t \propto j^{-1}$ in this limit. 
    then the binary can very quickly `jump'
    to a new phase
    space orbit, such that the DA approximation would fail completely to track
    the next secular cycle.
    Said differently, RPSD drives abrupt shifts of the binary's 
    approximate DA integrals of motion, just like GW emission or stellar encounters could, but in their absence. 
    %This effect is not specific to cluster potentials --- it occurs even for hierarchichal triples (the LK limit)
    %but has not been shown
    %before.
    
    %Next, all LK studies mentioned above have assumed that, for the purposes of
   % calculating short-timescale fluctuations, one may freeze the
    %secularly-evolving quantities --- i.e. the binary's dimensionless angular
    %momentum $j = (1-e^2)^{1/2}$, apsidal angle $\omega$ and nodal angle
    %$\Omega$ --- for the duration of an outer orbit $T_\phi$. 

%As mentioned several times in this paper, a very striking example of the
%difference between the GR and non-GR scenarios comes from examining the phase
%dependence of binary dynamical evolution.  Let us now illustrate this further.

To illustrate the RPSD phenomenon, we present Figures \ref{fig:Example_Hernquist_Divergence} and \ref{fig:Example_Hernquist_GR_Divergence}. To create these Figures we considered a NS-NS binary with semimajor axis $a=50$ AU, orbiting in (and tidally perturbed by) a spherical Hernquist potential of mass $\mathcal{M} = 10^7 M_\odot$ and radial scale $b=1$ pc, which constitutes a simple model of a nuclear stellar cluster. The outer orbit's azimuthal period is $T_\phi = 0.064$ Myr.
(The full set of initial conditions used here will be described in \S\ref{sec:Example_Hernquist_A} after we have introduced our notation more fully).  In each Figure, we integrate the (test-particle, quadrupole) SA equations of motion for roughly one secular period $t_\mathrm{sec}$.  We ran this integration seven times, each time with identical initial conditions except that we give the binary's outer orbit a {different initial radial phase}.  We also integrated the DA equations of motion for the same initial conditions (in the DA case the outer orbital phase information is irrelevant, since we have averaged over the outer motion). The only difference between Figures \ref{fig:Example_Hernquist_Divergence} and \ref{fig:Example_Hernquist_GR_Divergence} is that in Figure \ref{fig:Example_Hernquist_GR_Divergence}, we switched on the effect of 1pN GR apsidal precession.
This is a weak effect which only becomes important at very high eccentricity (the dimensionless strength of GR here is $\epsGR \sim 10^{-3}$ --- see \S\ref{sec:GR_Notation}).
{We repeat that we do not include GW emission in any of our calculations in this paper.}

%In Figure \ref{fig:Example_Hernquist_Divergence} we
%rerun the first secular period of SA evolution from Figure
%\ref{fig:Example_Hernquist} (i.e. with GR switched off), except
%using many different values of the initial radial phase

%\footnote{In practice we
%achieve this simply by performing the outer orbit integration first from $R_0 =
%\ra$, and then shifting the initial time of the SA integration by the appropriate
%amount in
%$(0, T_R )$.},

In both Figures we plot the SA eccentricity, $\log_{10}(1 - e)$,
for each run (different colored lines) at three stages of the evolution: (a)
the very earliest stages near $t = 0$, (b) around the eccentricity peak, and (c)
the latest stages near $t = t_\mathrm{sec}$. The thick dashed blue line in each
panel is the DA solution. In Figure \ref{fig:Example_Hernquist_Divergence}, we see that 
although the different SA realizations track each other very
closely (at least in log space) in panel (a), around the
eccentricity peak in panel (b) they differ quite considerably. Some
runs do not even reach the DA maximum eccentricity $\log_{10} (1 -
e_\mathrm{max} ) \approx -4.75$, while {some runs
achieve an extremely high eccentricity, $\log_{10}(1 - e_\mathrm{max}) \approx -5.65$.}
These SA fluctuations at high-$e$ can have a major effect on merger timescales, as is well known \citep{Grishin2018,Mangipudi2022}.
Nevertheless, upon emerging from the eccentricity peak, the different runs converge together
%again agree nicely 
(see panel (c)), closely tracking the `underlying' DA solution.

\begin{figure*}\centering
            \includegraphics[width=0.99\linewidth]{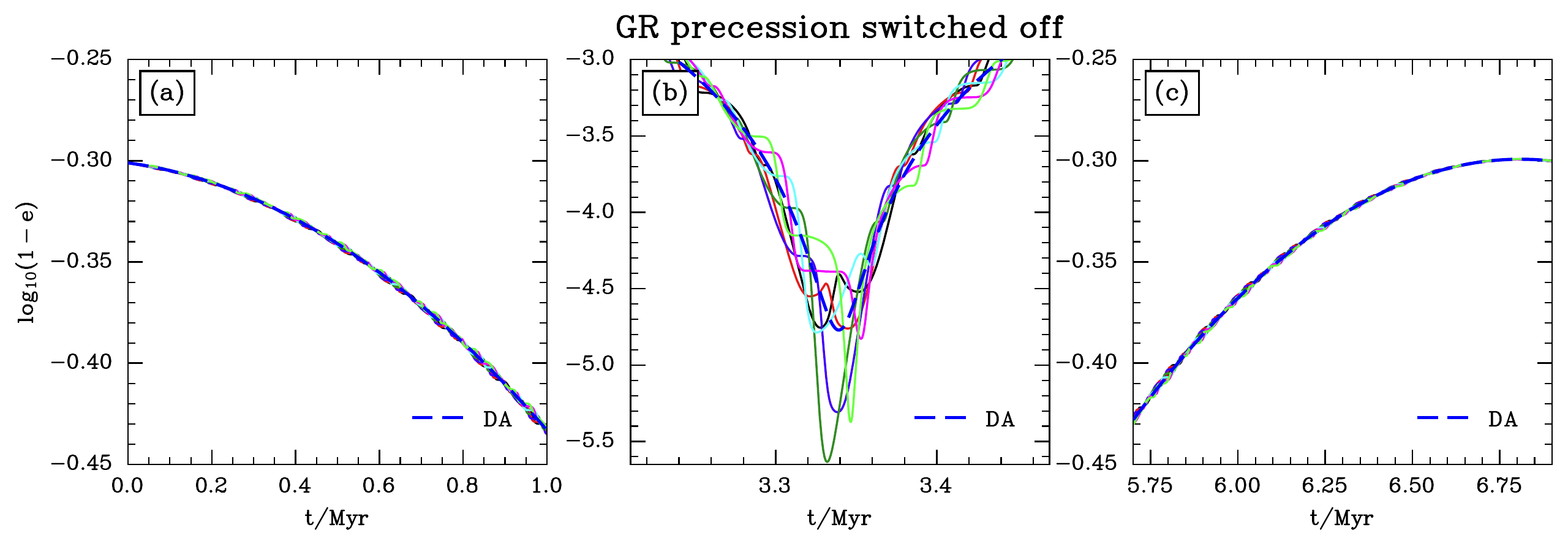}
      \caption{Example of {eccentricity evolution} for a NS-NS binary orbiting a Hernquist potential with outer azimuthal period $T_\phi = 0.064$ Myr, for {seven} different
   values of the outer orbit's initial radial phase.  Panels (a) and (c) show the beginning ($t\approx 0$)
   and end ($t\approx t_\mathrm{sec}$) of the calculation, respectively, while panel (b) focuses on the
   eccentricity peak around $t\approx t_\mathrm{sec}/2$.
   The DA solution (independent of radial phase) is shown
   with a blue dashed line.}
      \label{fig:Example_Hernquist_Divergence}
   \end{figure*}%
   %\hfill <-- it is superfluous 
   \begin{figure*}\centering
                          \includegraphics[width=0.99\linewidth]{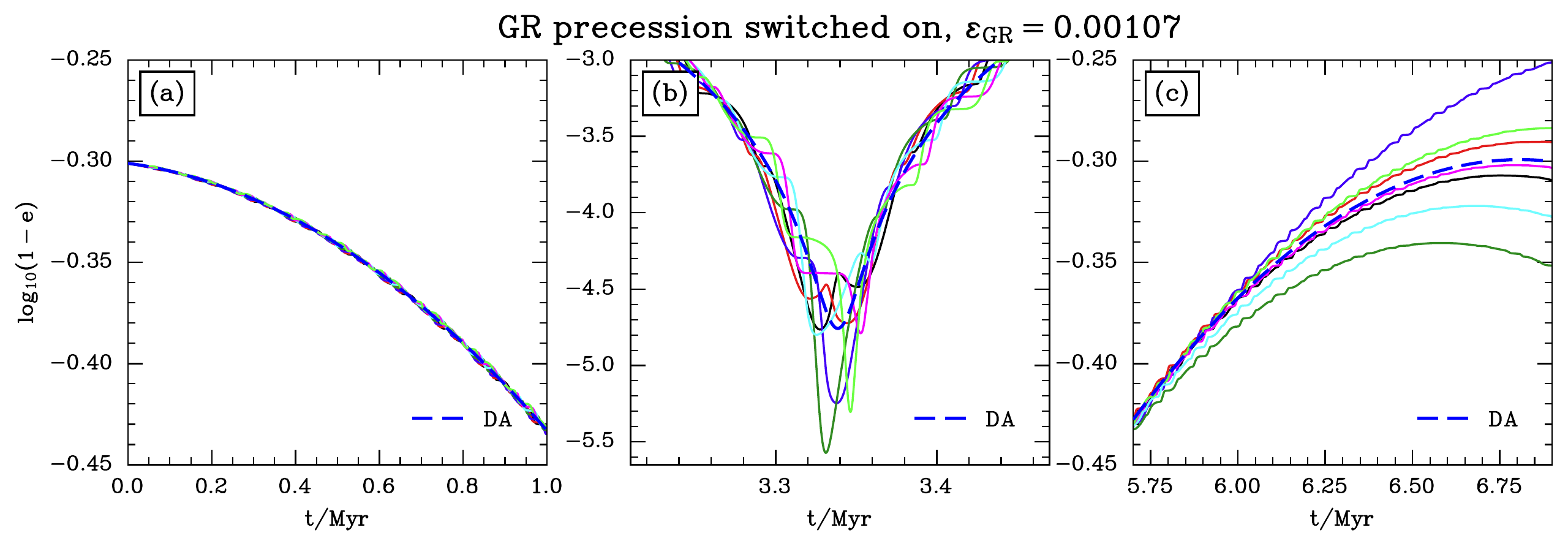}
      \caption{As in Figure \ref{fig:Example_Hernquist_Divergence}, except with 1pN GR apsidal precession switched
   on.  The fact that the trajectories have diverged by $t\sim t_\mathrm{sec}$ is a manifestation of 
   \textit{relativistic phase space diffusion} (RPSD), which is the central result of this paper.}
      \label{fig:Example_Hernquist_GR_Divergence}
\end{figure*}
Now we compare this with Figure \ref{fig:Example_Hernquist_GR_Divergence}.
%in which we perform exactly the same calculation except with GR switched on ---
%in other words, we are rerunning panel (n) of Figure
%\ref{fig:Example_Hernquist_GR} for many different initial radial
%phases. The thick dashed blue line in each panel is the DA solution from Figure
%\ref{fig:Example_Hernquist_GR}n. 
The early 
%stage 
evolution (compare Figures \ref{fig:Example_Hernquist_Divergence}a and \ref{fig:Example_Hernquist_GR_Divergence}a) is essentially identical with and without GR, as expected away from high-$e$.  Next, comparing Figures
\ref{fig:Example_Hernquist_Divergence}b and
\ref{fig:Example_Hernquist_GR_Divergence}b, we see that the
maximum eccentricity reached by a given binary is {very} slightly diminished by the
inclusion of GR {(the green line now reaches $1-e_\mathrm{max} \approx -5.6$ rather than $-5.65$)}, as we would expect given that the binary resides in the weak GR regime (Paper III), but otherwise the evolution is {almost indistinguishable from} the case without GR.
However, by the time the DA
eccentricity has returned
%is returning 
to its minimum around $6.8$ Myr (panel (c)) {most of the} 
SA trajectories have diverged significantly both from the DA solution and from each other.  They will subsequently follow
%undergo 
entirely different secular evolutionary tracks, {many} of which will be badly approximated by the original DA solution.  

As we will see in
\S\ref{sec:Effect_of_GR}, over the course of many secular cycles
this trajectory divergence results in
%corresponds to 
a `diffusion' of the value of the DA Hamiltonian, which is an approximate
integral of motion around which each SA solution would normally fluctuate (as in Figure \ref{fig:Example_Hernquist_Divergence}), {and will be defined properly in \S\ref{sec:dynamical_framework}}. This {diffusion}
behavior breaks down at sufficiently  high eccentricity provided GR is switched on, so we call it RPSD.
{However, by using the term `diffusion'
we do not mean to suggest that the system's integrals of motion follow any simple
Brownian walk or that their evolution can be described by a diffusion equation.
In fact, the `diffusion' we have found does not seem to obey any simple statistical 
behavior, as we will see in \S\ref{sec:RPSD_Statistical}.}

{We note here that \cite{Luo2016} also found that short-timescale fluctuations can cause the mean trajectory of a binary to drift gradually away from the DA prediction, even in the test-particle quadrupole LK problem, and 
derived a `corrected DA Hamiltonian' to account for this deviation\footnote{{This effect, which seems to have been first investigated at arbitrary eccentricities and inclinations
by \cite{brown1936stellar}, 
was recently put on a solid mathematical footing by \cite{tremaine2023hamiltonian}.}}.
However, this effect is distinct from RPSD, for several reasons:
(i) it has nothing to do with GR precession; 
(ii) it does not require extremely high eccentricity behavior, and (iii) it occurs due to 
an accumulation of nonlinear quadrupolar perturbations, which are normally averaged out in the standard LK theory, over many secular periods (see \citealt{tremaine2023hamiltonian}; these nonlinear effects are too small to be discernible in Figure \ref{fig:Example_Hernquist_Divergence}).
On the contrary, RPSD only occurs when GR is included, and
happens on a much shorter timescale, requiring just one (very high) eccentricity peak.  We discuss this comparison further in \S\ref{sec:literature}.}

% but that their mechanism%

%Averaging breaks down for a short time, but this minor violation
%has a major impact on the subsequent evolution
%\
%Moreover, adding in the extra
%degree of freedom that we have not altered here --- the azimuthal phase ---
%would compound the phase dependence still further.

    \subsection{Plan for the rest of this paper}
    
   The rest of this paper is structured as follows. In
   \S\ref{sec:dynamical_framework} we briefly recap some
   key results from Papers I-III and establish our notation. In
   \S\ref{sec:qualitative} we provide several numerical examples
   that illustrate the phenomenology of short-timescale fluctuations when GR is
   \textit{not} included, particularly with regard to high eccentricity
   behavior. We then proceed to explain the observed behavior quantitatively,
   and derive an approximate expression for the magnitude of angular momentum
   fluctuations at high $e$.  In \S\ref{sec:Effect_of_GR} we
   switch on GR precession and give several numerical examples of systems
   exhibiting RPSD. We then analyse this phenomenon more
   quantitatively and offer a physical explanation for it. In
   \S\ref{sec:discussion} we consider the astrophysical
   importance of RPSD, and discuss our results in
   the context of the existing LK literature.  We summarise in
   \S\ref{sec:summary}.
    
    %%%%%%%%%%%%%%%%%%%%%%%%%%%%%%%%%%%%%%%%%%%%%%%%%%%%%%%%%%%
    %%%%%%%%%%%%%%%%%%%%%%%%%%%%%%%%%%%%%%%%%%%%%%%%%%%%%%%%%%%
    %%%%%%%%%%%%%%%%%%%%%%%%%%%%%%%%%%%%%%%%%%%%%%%%%%%%%%%%%%%

    \section{Dynamical framework}
    \label{sec:dynamical_framework}

In this section we recap the basic formalism for describing a binary perturbed by quadrupole-order tides, including 1pN GR precession. For more details see Papers I-III.

\subsection{Inner and outer orbits}

    Consider a binary with component masses $m_1$ and $m_2$, inside some {spherically symmetric} host system (the `cluster') whose potential is $\Phi(\mathbf{R})$.  The binary's barycentre $\mathbf{R}_\mathrm{g}(t)$ is assumed to move as a test particle in this potential: {
    we call this the `outer orbit'.}
    %{, 
    %and assume it is confined to the $(X, Y)$ plane of a fixed inertial Cartesian coordinate system $(X, Y, Z)$.}  
    Meanwhile, the binary's 
internal Keplerian orbital motion (the `inner' orbit) is 
described by orbital elements: semi-major axis $a$, eccentricity $e$, inclination $i$, longitude of the ascending node $\Omega$, argument of pericenter $\omega$, and mean anomaly $M$, {
defined as in Paper I.}
%Here $i$ is measured relative to the {outer orbital} $(X,Y)$ plane, and $\Omega$ is measured relative to the fixed $X$ axis. 
%In the case of an axisymmetric potential we take the symmetry axis along $Z$; 
%For a spherical potential (which includes the LK case of a Keplerian potential), we take $(X, Y)$ to be the plane of the outer orbit.

An alternative description of this inner orbit is provided by 
the Delaunay actions $\mathbf{I} = (L, J, J_z)$, with $L=\sqrt{G(m_1+m_2) a}$, $J=L\sqrt{1-e^2}$, and $J_z = J\cos i$, and their conjugate angles $\bm{\psi} = (M, \omega, \Omega)$. Sometimes we will find it useful to refer to the dimensionless versions of these variables:
%%%%%%%%%%%%%%%%%%%
\begin{align}  
\label{eqn:def_j}
j &\equiv J/L = (1-e^2)^{1/2}, \\
\label{eq:def_jz}
j_z &\equiv J_z/L = (1-e^2)^{1/2}\cos i, 
\\
\Theta &\equiv j_z^2 = (1-e^2)\cos^2 i.
\end{align} 
%%%%%%%%%%%%%%%%%%%
%Obviously $j$ is just the dimensionless angular momentum. Moreover, $e$ and $j$ must obey
%\begin{align}    
%0\leq e\leq e_\mathrm{lim} \equiv \sqrt{1-\Theta},~~~~~~~~~~~\Theta^{1/2}\leq j\leq 1,
%\label{eq:AMconstr}
%\end{align}
%to be physically meaningful for a given $\Theta$.

\subsection{Dynamical equations}

{Let us ignore GR precession for now, so the only forces that the binary feels are the internal 
two-body Keplerian attraction and the tidal perturbation from the cluster.}
Let these forces be encoded in the Hamiltonian $H(\bm{\psi}, \mathbf{I}, t)$.  Then the Delaunay variables evolve according to 
Hamilton's equations of motion: 
\begin{equation}
    \frac{\md \bm{\psi}}{\md t} = \frac{\partial H}{\partial \mathbf{I}}, \,\,\,\,\,\,\,\,\,\,\,\,\,\,\,    \frac{\md \mathbf{I}}{\md t} = -\frac{\partial H}{\partial \bm{\psi}}.
    \label{eqn:Hamiltons_Equations}
\end{equation}

After averaging over the {binary's inner orbit},  $H=H_0+H_\mathrm{1}$ where $H_0(\mathbf{I}) =  -\mu^2/(2L^2)$ (with $\mu \equiv G(m_1+m_2)$) is just the Keplerian energy. The `singly-averaged' (SA), {test-particle quadrupole tidal} Hamiltonian is then\footnote{In Papers I-II we referred to
$H_\mathrm{1,SA}$ as $\langle H_1 \rangle_M$. We will stick to the
$H_\mathrm{1,SA}$ notation in what follows.  An analogous statement holds for
the upcoming $H_\mathrm{1,DA}$ and $H_\mathrm{GR}$.} (Paper I):
    %%%%%%%%%%%%%%%%%%%%%%%%
    \begin{align} H_\mathrm{1,SA}(\bm{\psi}, \mathbf{I},t) = \frac{1}{2}\sum_{\alpha \beta}\Phi_{\alpha \beta} \langle r_\alpha r_\beta \rangle_M .
    \label{eqn:H1SA}
    \end{align}
    The averages $\langle r_\alpha r_\beta \rangle_M$ are given explicitly in
    terms of orbital elements, expressible through $(\bm{\psi}, \mathbf{I})$, in Appendix A of Paper I.      
    The singly-averaged Hamiltonian $ H_\mathrm{1,SA}(\bm{\psi}, \mathbf{I},t)$ ends up being a function of the
    variables $J, J_z, \omega, \Omega$ and the time $t$ (through the dependence of $\Phi_{\alpha \beta}$ on $\Rg(t)$). 

    The SA equations of
    motion follow by differentiating \eqref{eqn:H1SA} according to \eqref{eqn:Hamiltons_Equations} --- these are 
 given explicitly in Appendix
    \ref{sec:SA_equations} (equations
    \eqref{eqn:domegadt_SA}-\eqref{eqn:dJzdt_SA}).  {In particular $L=\sqrt{\mu a}$
   is conserved under SA dynamics, so the binary's semimajor axis is constant.}
    
If we further average the Hamiltonian \eqref{eqn:H1SA} over
    the outer orbital motion $\Rg(t)$ (i.e. over the orbital ellipse, annulus or torus --- see Paper I),
    the resulting doubly-averaged (DA) perturbing {tidal}
    Hamiltonian is 
    %%%%%%%%%%%%%%%%%%%%%%%%%%%%%%%
    \begin{align}
H_\mathrm{1,DA} =& \,
    \frac{1}{2}\sum_{\alpha\beta}\overline{\Phi}_{\alpha \beta} \langle r_\alpha
    r_\beta \rangle_M \nn 
    %= \frac{\overline{\Phi}_{xx} \langle x^2 + y^2 \rangle_M + \overline{\Phi}_{zz} \langle z^2 \rangle_M}{2}
 \nn \\ =& \, \frac{A}{8 \mu^2} \times  J^{-2}\Big[ (J^2 - 3\Gamma J_z^2)( 5L^2-3J^2) \nn \\ &- 15\Gamma(J^2-J_z^2)(L^2-J^2) \cos 2\omega \Big].
 \label{eqn:H1_Doubly_Averaged}
\end{align}  
    %%%%%%%%%%%%%%%%%%%%%%%%%%%%%%%
Here $A$ and $\Gamma$ are constants (see \S6 of Paper I) that depend on the potential and outer orbit; $A$ measures the strength of the tidal potential and has units of (frequency)$^2$, whereas $\Gamma$ is dimensionless. In the {Keplerian (LK)} limit {we find} $\Gamma=1$ and $A=G\mathcal{M}/[2
a_\mathrm{g}^3(1-e_\mathrm{g}^2)^{3/2}]$, where $a_\mathrm{g}$ and $e_\mathrm{g}$ are respectively the semimajor axis and eccentricity of the outer orbit and $\mathcal{M}$ is the perturber mass.    {The value of $\Gamma$ affects the phase space morphology significantly, and as such it was central to the analysis we performed in Papers II-III.
However, for reasons we outline in Appendix \ref{sec:Note_Morphology}, the value of $\Gamma$ is of little consequence for analyzing short-timescale fluctuations, and so we will mostly not mention it explicitly from now on.}

    %Notice that 
    Being time-independent, $H_\mathrm{1,DA}$ is a constant of motion in the DA approximation. Moreover, the DA Hamiltonian \eqref{eqn:H1_Doubly_Averaged} does not depend on the longitude of ascending node $\Omega$, meaning that in the DA approximation, $J_z$ is also a constant of motion. The nontrivial DA equations of evolution of $\omega$, $J$ and $\Omega$ arising from \eqref{eqn:H1_Doubly_Averaged} are
    given in equations (12)-(14) of Paper III.
%The remaining nontrivial DA equations of arising from \eqref{eqn:H1_Doubly_Averaged} --- governing the evolution of $\omega$, $J$ and $\Omega$ ---  are given in equations (12)-(14) of Paper III.

%%%%%%%%%%%%%%%%%%%%%%%%%%%%%%%%%%%%%%%%%%%%%%%%%%%%%%%%%%%%%%%%%%%%%%%%%%%%%%%%%%%%%%%%%%%%%
%%%%%%%%%%%%%%%%%%%%%%%%%%%%%%%%%%%%%%%%%%%%%%%%%%%%%%%%%%%%%%%%%%%%%%%%%%%%%%%%%%%%%%%%%%%%%

\subsubsection{1pN GR precession}
\label{sec:GR_Notation}

    {Next we wish to we include the effects of 1pN GR apsidal precession on the binary's inner orbit.}
 Whether we use SA or DA theory, {we can achieve this by adding} to our Hamiltonian a term 
    %%%%%%%%%%%%%%%%%%%%%%%%%%%%%%%
    \begin{align}
      \label{eqn:HGR}
{       H_{\mathrm{GR}} = -\frac{AL^5}{8\mu^2} \times \frac{\epsGR}{J} = -\frac{Aa^2}{8} \times \frac{\epsGR}{j}, }
    \end{align}
    %%%%%%%%%%%%%%%%%%%%%%%%%%%%%%%
    where the strength of the precession is measured by the dimensionless parameter (see Papers II \& III)
\begin{align} 
\epsilon_\mathrm{GR} &\equiv \frac{24G^2(m_1+m_2)^2}{c^2Aa^4} 
\label{eq:epsGRformula} \\
&=  0.258 \times \left( \frac{A^*}{0.5}\right)^{-1}\left( \frac{\mathcal{M}}{10^5M_\odot}\right)^{-1}\left( \frac{b}{\mathrm{pc}}\right)^{3} \nn \\ & \times \left( \frac{m_1+m_2}{M_\odot}\right)^{2}  \left( \frac{a}{20 \, \mathrm{AU}}\right)^{-4}. 
\label{eqn:epsGRnumerical}
\end{align}
In the numerical estimate
\eqref{eqn:epsGRnumerical} we have assumed a spherical cluster
of mass $\mathcal{M}$ and scale radius $b$, and $A^* \equiv A/(G\mathcal{M}/b^3)$ --- see Paper I.
  The inclusion of GR precession obviously affects the equation of motion for $\md \omega/\md t$ (whether we work in the SA or DA approximation) by adding
  an extra term {$\partial H_\mathrm{GR}/\partial J = AL^5\epsGR/(8\mu^2 J^2)$},
% The physical effect of GR precession is typically to quench the cluster tide-driven eccentricity oscillations, 
as we explored in detail in Paper III (see also \citealt{Miller2002,Fabrycky2007,Bode2014}). As we showed there, there are typically no large $e$ oscillations in the `strong GR' regime, {defined by} $\epsGR \gtrsim \epsstr \equiv 3(1+5\Gamma)$.

\subsubsection{Integrals of motion}

    We know from Paper III that in the
    DA approximation, i.e. under the dynamics prescribed by the total DA
    Hamiltonian $H_\mathrm{DA} \equiv H_\mathrm{1,DA} + H_\mathrm{GR}$, there
    are two independent integrals of motion. We could take these to be the value of the Hamiltonian $H_\mathrm{DA}$ and the $z$-component of angular momentum $J_z$, but for the purposes of this paper
    it
    will be most useful to {take} them to be $j_z \equiv J_z/L$ and $D$, defined as (see
    equation (19) of Paper III):
    %%%%%%%%%%%%%%%%%%%%%%%%%%%%%%%
    \begin{align} 
      \label{eqn:def_D}
       D &\equiv  e^2\left(1+\frac{10\Gamma}{1-5\Gamma}\sin^2i\sin^2\omega\right) - \frac{\epsGR}{3(1-5\Gamma)\sqrt{1-e^2}}.
      \end{align} 
      %%%%%%%%%%%%%%%%%%%%%%%%%%%%%%%
      
In the SA approximation, i.e. under the dynamics prescribed by the total SA
    Hamiltonian $H_\mathrm{SA} \equiv H_\mathrm{1,SA} + H_\mathrm{GR}$, the
    quantities $j_z(t)$, $D(t)$ are not precise integrals of motion\footnote{To be clear, in the SA approximation 
    the value of $D$ is 
    found by evaluating the right hand
    side of \eqref{eqn:def_D} using the SA orbital elements.}.
    {Nevertheless, 
    they can be usually be regarded as adiabatic invariants, i.e.
    quantities which fluctuate on the timescale of the outer orbital period
    but are approximately conserved upon averaging over this period, provided it is sufficiently short.}
    In other words, 
    under normal circumstances a binary's $j_z$ and $D$ values simply
    fluctuate around the underlying DA solution corresponding to one of the
    level curves in the characteristic $(\omega,e)$ phase space --- see Paper II.  This 
    allows one to consider
    short-timescale fluctuations as a perturbation on top of a dominant secular
    effect \citep{Ivanov2005,Luo2016}. 
    On the other hand, in
    \S\ref{sec:Effect_of_GR} we will see that for non-zero
    $\epsGR$ and very high eccentricities, this perturbative assumption can
    break down --- {for instance, the time-averaged value of $D$ can change dramatically
    and abruptly, reflective of a violation of the adiabatic invariance condition, and this is what gives rise to RPSD.}

%    Finally, To avoid confusion, we note here that $D$ is always defined in terms of the
   % DA Hamiltonian $H_\mathrm{DA}$.  When we refer to $D_\mathrm{DA}$ or
   % $D_\mathrm{SA}$, we simply mean the value of $D$ that arises from evaluating
    %the formula \eqref{eqn:def_D} using the DA or SA solution to 
    %the equations of motion, respectively.
    
    %%%%%%%%%%%%%%%%%%%%%%%%%%%%%%%%%%%%%%%%%%%%%%%%%%%%%%%%%%%%%%%%%%%%%%%%%%%%%%%%%%%%%%%%%%%%%%%%%%%%%%%%%%%%%%%%%%%%%%%%%%%%%%%%%%%%%%%%%%%%%%%%%%%%%%%%%%%%%%%%%%%%%%%%%%%%%%%%%%%%%%%%%%%%%%%%%%%%%%%%%%%%%%%%%%%%%%%%%%%%%%%%%%%%%%%%%%%%%%%%%%%%%%%%%%%%%%%%%%%%%%%%%%%%%%%%%%%%%%%%%%%%%%%%%%%%%%%%%%%%%%%%%%%%%%%%%%%%%%%%%%%%%%%%%%%%%%%%%%%%%%%%%%%%%%%%%%%%%%%%%%%%%%%%%%%%%%%%%%%%%%%%%%%%%%%%%%%%%%%%%%%%%%%%%%%%%%%%%%%%
    
    \section{Short-timescale fluctuations and high eccentricity behavior} 
    \label{sec:qualitative}
    
    {In this section we first provide a qualitative discussion of 
    some numerical examples which demonstrate the
    phenomenology of short-timescale fluctuations 
    (\S\ref{sec:Example_Hernquist})}. Crucially, for
    simplicity and in order to cleanly separate certain physical effects,
    \textit{we do \textbf{not} include GR precession in any of these examples}
    (GR precession will be added in \S\ref{sec:Effect_of_GR}). In
    \S\ref{sec:quantitative} we provide a quantitative analysis
    of the behavior we have uncovered. Finally in
    \S\ref{sec:characteristic_delta_j} we derive an approximate
    expression for the magnitude of angular momentum fluctuations at highest DA
    eccentricity, which gives us the maximum achievable $e$.

    %We note that in all numerical examples presented in this paper,
    %we checked that the numerical method for integrating the SA equations had
    %converged. {In particular we uses a timestep
    %$\Delta t$ which was small enough that further shortening of the %did not affect the SA results. This often requires very small timesteps near peak eccentricity,
    %$\Delta t / T_\phi \lesssim 10^{-3}$.}

    \subsection{Numerical examples}
    \label{sec:Example_Hernquist}

        \subsubsection{Fiducial example in the Hernquist potential, $i_0 = 90.3^\circ$}
    \label{sec:Example_Hernquist_A}

    \begin{figure*}\centering
                  \includegraphics[width=0.95\linewidth]{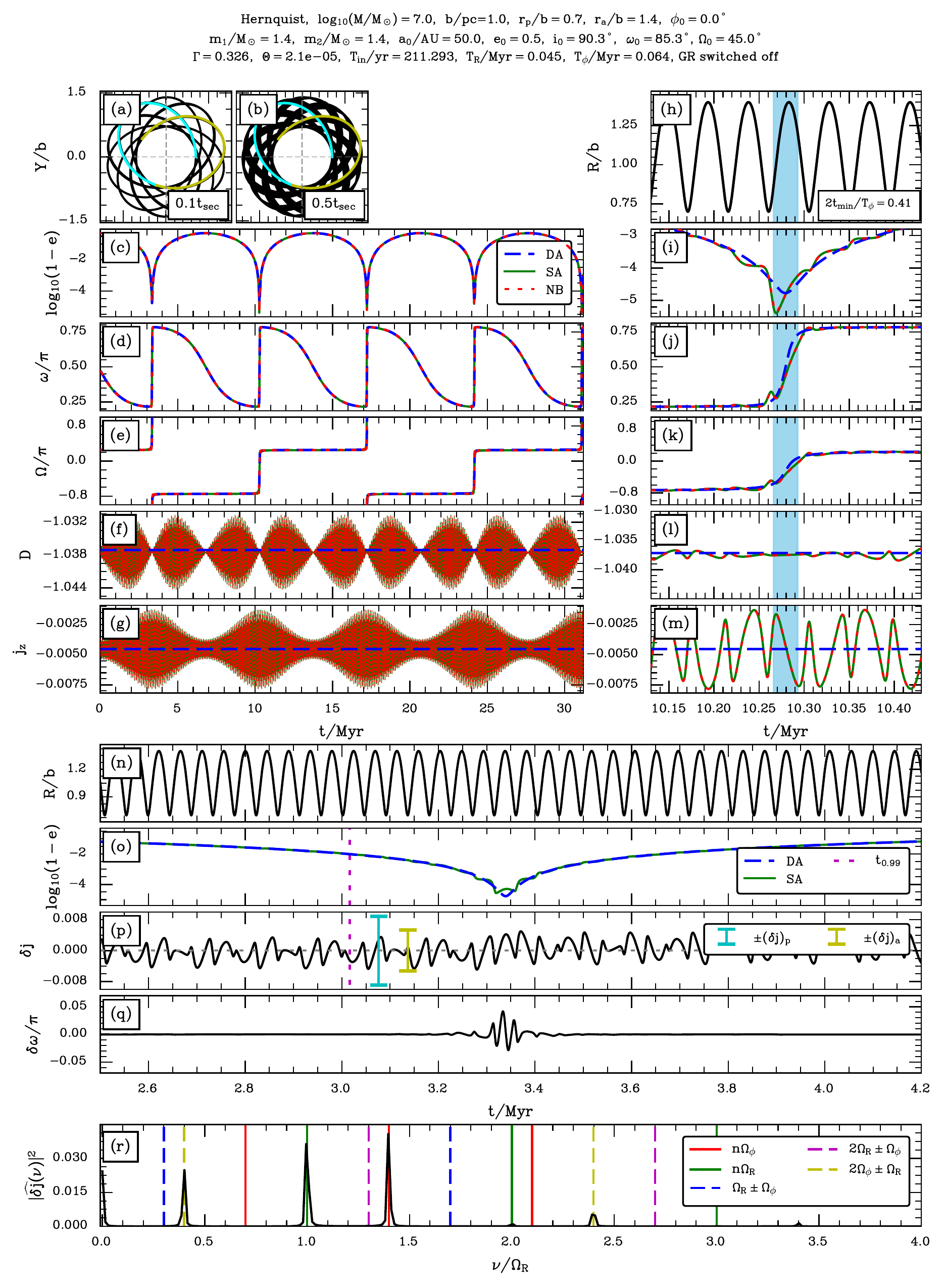}
         \caption{Example of a binary that undergoes significant short-timescale fluctuations at very high eccentricity (the same one we used in Figure \ref{fig:Example_Hernquist_Divergence}). The details of the plot are explained fully in \S\ref{sec:Example_Hernquist}.  Note the initial inclination $i_0 = 90.3^\circ$.
         {The DA secular period is $t_\mathrm{sec} \approx 6.9 $ Myr.}}
         \label{fig:Example_Hernquist}
      \end{figure*}%
      %\hfill <-- it is superfluous 

    In Figure \ref{fig:Example_Hernquist} we give an example of a
    binary that undergoes significant short-timescale fluctuations at high
    eccentricity. This figure is very rich in information and exhibits several
    interesting features that we wish to explore throughout the paper.  We will
    also see several other figures with this or similar structure.  It is
    therefore worth describing the structure of Figure
    \ref{fig:Example_Hernquist} in detail. 
    
    At the very top of the figure (top line of text) we provide the values of 6
    input parameters ($ \Phi, \, \mathcal{M}, \, b, \, \rp, \, \ra, \, \phi_0 $) that
    define the perturbing potential as well as the outer orbit's initial
    conditions. In this case (which is the same setup as in Figure \ref{fig:Example_Hernquist_Divergence}) we are considering a binary in a Hernquist
    potential $\Phi(r) = -G\mathcal{M}/(b+r)$, with total mass $\mathcal{M}=10^7
    M_\odot$ and scale radius $b=1\mathrm{pc}$.  {Since this potential is spherical, the shape of the outer orbit is determined by two numbers: its
    pericenter distance $\rp = 0.7b$, its apocenter $\ra = 1.4b$.  Unless otherwise specified, 
    in this paper we always
    initiate the outer orbit at $t=0$ from $(R,\phi)=(\rp,\phi_0)$ with $\dot{\phi} > 0$,
    where $\phi$ is the outer orbit's
    azimuthal angle relative to the $X$ axis (see Paper I).}
    
    In the second line of text we list 7 input parameters $(m_1, \, m_2, \, a_0,
    \, e_0, \, i_0, \, \omega_0, \, \Omega_0)$ that concern the binary's inner
    orbit; the subscript `$0$' denotes initial values.  In this example we are
    considering a NS-NS binary ($m_1=m_2=1.4M_\odot$) with initial semimajor
    axis $a_0 = 50\mathrm{AU}$. Note also that the initial inclination $i_0$ is
    chosen close to $90^\circ$, which is necessary to achieve very large
    eccentricities {starting from $e_0 \ll 1$ (since this requires $\Theta \ll 1$)}.
    
    In the third and final line of text at the top of the figure, we list 5
    important quantities that follow from the choices of 13 input parameters
    above: $\Gamma, \Theta$, the inner orbital period $T_\mathrm{in} =
    2\pi\sqrt{\mu/a^3}$, the outer orbit's radial period $T_R$, and its azimuthal
    period $T_\phi$. In this
    instance we have a $\Gamma$ value of $0.326$ %(which is $> 1/5$,
    and $\Theta = 2.1\times 10^{-5}$, which allows $e_\mathrm{max}$ to become extremely
    high. Lastly, we emphasise that we have artificially switched off GR precession in this example. Thus we set $\epsGR=0$ in the equations of motion and in
    evaluating $D$, {which is equivalent to taking the speed of light $c\to\infty$}. This choice is also indicated in the third line of text.  
    GR precession will be incorporated in
    \S\ref{sec:Effect_of_GR},
    allowing for direct comparison with the results of this section.
    
    Now we move on to the figure proper. In panels (a) and (b) we display the
    trajectory of the outer orbit through the $(X,Y)$ plane, integrated using
    \texttt{GALPY} \citep{Bovy2015}. In both panels we show the trajectory
    from $t=0$ to $t=T_R$ (cyan line), and from $t=T_R$ to $t=2T_R$ (yellow
    line). In black we show the entire trajectory traced up to time $t=0.1
    t_\mathrm{sec}$ (panel (a)) and $t=0.5 t_\mathrm{sec}$ (panel (b)), where
    $t_\mathrm{sec}$ is the period of secular oscillations, computed using equation
    (33) of Paper II.
    
    In total we integrated the outer orbit $\Rg(t)$ until $t=4.5
    t_\mathrm{sec}$.  We then fed the resulting
    $\Phi_\mathrm{\alpha\beta}(\Rg(t))$ time series into the SA {quadrupolar} equations of
    motion
    \eqref{eqn:domegadt_SA}-\eqref{eqn:dJzdt_SA}
    and integrated them numerically.  In panels (c), (d) and (e) we compare the
    numerical integrations of the SA equations of motion for $e$, $\omega$,
    $\Omega$ (green curves) against the prediction of DA theory (blue dashed curves).
    We also show the results of direct {orbit integration} (red dotted
    curves), {where we evolved the exact two-body equations of motion of the binary in the presence of the smooth time-dependent external field
    $\Phi(\Rg(t))$ without any tidal approximation, using the N-body code \texttt{REBOUND} \citep{Rein2012}}.\footnote{{In these direct integrations, the orbital elements are deduced as a post-processing step after the simulation is complete;  at a given time they are defined according to the two-body system's instantaneous relative position and velocity, ignoring the external potential.}} In panel (c) we see that the binary reaches extremely high
    eccentricity, with the DA result $1-e_\mathrm{DA}$ reaching {a minimum at
    $\approx 10^{-4.8}$}.  In this panel we already see that the maximum eccentricity
    reached in the SA approximation can be rather different from the DA value,
    and changes from one eccentricity peak to the next --- near the second peak, around $10$ Myr, $1-e_\mathrm{SA}$ plunges to {$\sim 10^{-5.4}$}. 
    {On the other hand, the agreement between the SA and direct orbit integration integrations is excellent, which gives us confidence that the numerics is accurate and that the quadrupolar approximation is a good one (see also \S\S\ref{sec:SA_breakdown}-\ref{sec:literature}).}
    {Another striking feature of}
    panels (d) and (e) is the step-like jumps in $\omega$ and $\Omega$ that
    occur near maximum eccentricity in both SA and DA integrations. These are just what we expect
    from our investigation in Appendix C of Paper III.
    
    In panels (f) and (g) we show the evolution of the quantities $D$ (equation
    \eqref{eqn:def_D}) and $j_z$ (equation
    \eqref{eq:def_jz}).  %We emphasise again that although the definition
    %\eqref{eqn:def_D} of $D$ involves $\epsGR$  (and we report a
    %finite $\epsGR$ value in the third line at the top of the figure), we do not
    %include GR at all in this example, i.e. we set $\epsGR \equiv 0$  for the purposes
    %of calculating $D$.
    In the DA approximation, $D$ and $j_z$ are integrals of motion --- hence the
    blue dashed DA result is simply a straight horizontal line.  We see that the SA
    result oscillates around the constant DA value in both cases,
    with an envelope that has {period $t_\mathrm{sec}$ for $j_z$ and $\tsec/2$
    for $D$}. In panel (f) there is {in fact} a small offset between $D_\mathrm{DA}$ and
    the mean value of $D_\mathrm{SA}$, which is due to an initial phase offset
    of the outer orbit (\citealt{Luo2016,Grishin2018}). We also notice a
    characteristic behavior which is that fluctuations in $D$ are minimised
    around the eccentricity peak, while fluctuations in $j_z$ are maximised
    there.
    
    In the right hand column, in panels (i)-(m) we simply reproduce panels
    (c)-(g), except we zoom in on the sharp eccentricity peak at around $10.28
    \,\mathrm{Myr}$.  At the top of this column we have panel (h), which shows
    the outer orbital radius $R(t)$ during this high-eccentricity episode.  
    In addition, in each panel (h)-(m) we shade in light blue the region {
       %%%%%%%%%%%%%%%%%%%%%%%%%%%%%%%%%%%%
    \begin{align}
         t(j_\mathrm{min}) -  t_\mathrm{min} < t <  t(j_\mathrm{min}) +  t_\mathrm{min}.
        \label{eqn:t_blue_bands}
    \end{align} 
    %%%%%%%%%%%%%%%%%%%%%%%%%%%%%%%%%%%%}
    }
    Here $t(j_\mathrm{min})$ is the time corresponding to the minimum of
    $j_\mathrm{DA}$ (i.e. peak DA eccentricity), namely when $j_\mathrm{DA} =
    j_\mathrm{min}$,  and $t_\mathrm{min}$ is the time taken for $j_\mathrm{DA}$
    to change from $j_\mathrm{min}$ to $\sqrt{2} j_\mathrm{min}$ --- see
    \S4.2 of Paper III. In panel (h) we also
    indicate the value of the ratio $2t_\mathrm{min}/T_\phi$, which will turn
    out to be very important when we switch on GR in
    \S\ref{sec:Effect_of_GR}.
    In this particular case we see that $2t_\mathrm{min}/T_\phi = 0.41$, so that
    most of the interesting (very highest eccentricity) behavior happens on a
    timescale shorter than an outer azimuthal period.
    
    From panels (i)-(m) we see that the {SA integration (green lines) agrees very well with the direct orbit integration (red dashed lines)}
    even at very high eccentricities, giving us confidence that the SA
    approximation is a good one here\footnote{The SA approximation can itself break down at
    extremely high eccentricity --- see \S\ref{sec:SA_breakdown}
    for an example and a rough criterion.}.  
    However, the SA prediction differs markedly from the DA
    prediction at very high eccentricity.  In particular, from panel (i) we see
    that $1-e_\mathrm{DA}$ becomes $\approx 10^{-4.6}$ at its minimum, while
    $1-e_\mathrm{SA}$ reaches a significantly smaller value still, $\approx
    10^{-5.4}$. Panels (j) and (k) reveal that the large jumps in $\omega$ and
    $\Omega$ both happen on a timescale $\sim 2t_\mathrm{min}$. In panels (l)
    and (m) we show how the integrals of motion $j_z$ and $D$ fluctuate around
    the maximum eccentricity.
    
    In panels (n)-(q) we again show the time series of $R$ and $1-e$ around the
    second eccentricity peak (although over a wider timespan), as well as the
    differences 
    %%%%%%%%%%%%%%%%%%%%%%%%%%%%%%%%%%%%%%%%%%
    \begin{align} 
    \delta j (t) \equiv j_\mathrm{SA}(t) -j_\mathrm{DA}(t), \,\,\,\,\,\,\,\, \delta \omega(t) \equiv \omega_\mathrm{SA}(t)-\omega_\mathrm{DA}(t),
    \label{eqn:deltaj_jSA_minus_jDA}
    \end{align}
    %%%%%%%%%%%%%%%%%%%%%%%%%%%%%%%%%%%%%%%%%%
    between the results of SA and DA integration. The vertical dotted magenta
    line in panels (o) and (p) corresponds to $t=t_{0.99}$, which is the time
    when $e_\mathrm{DA}$ first reaches $0.99$.  From panel (p) we see that
    $\delta j$ fluctuates in a complex but near-periodic manner, with period
    $\sim 5T_R$. The fluctuations themselves are not perfectly centered around
    zero; before the eccentricity peak, the mean value of $\delta j$ 
  is slightly
    negative, whereas afterwards it is slightly positive. The blue and cyan bars
    in panel (p) correspond to simple approximations to the amplitude of $\delta
    j$ at peak eccentricity --- see
    \S\ref{sec:characteristic_delta_j}. Meanwhile, from panel (q)
    we see that the fluctuation $\delta \omega$ is negligible until the very
    highest eccentricities are reached, where there is a sharp pulse before it
    decays to zero again.  This pulse is approximately antisymmetric in time
    around $t=t(j_\mathrm{min})$. The pulse episode lasts for $\sim 2T_R$.

   {Finally, in panel (r) we show the power spectrum of $\delta j$ 
   fluctuations, which is the square of the Fourier transform $\widehat{\delta j}(\nu) \equiv \int \md t \exp(i\nu t) \delta j(t)$.  We calculated this Fourier transform numerically using the $\delta j(t)$ data from panel (p).  We see that the signal is concentrated at 
   frequencies $\nu = n_1\Omega_R + n_2\Omega_\phi$ for certain pairs of integers $n_1$, $n_2$, where $\Omega_i \equiv 2\pi/T_i$ is the outer orbital frequency. 
   We discuss these power spectra briefly 
   in \S\ref{sec:characteristic_delta_j} and \S\ref{sec:RPSD_Statistical}.}

 {
    \subsubsection{Further examples}

    In Appendix \ref{sec:App_Further_Examples}
    we give two more numerical examples, which display behavior mostly similar to that just described.
    They differ from Figure \ref{fig:Example_Hernquist} in that in the first one (Figure \ref{fig:Example_Hernquist_4}), the initial inclination is changed to $i_0 = 93.3^\circ$,
    and in the second one (Figure \ref{fig:Example_Plummer}), the Hernquist potential is replaced with a Plummer potential. They will also give rise to very different behavior when they are run with GR precession switched on, as we will see in \S\ref{sec:Effect_of_GR}.
    We will refer to these Figures throughout the rest of the paper.
}

    \subsection{Analysis of fluctuating behavior} 
\label{sec:quantitative}

%%%%%%%%%%%%%%%%%%%%%%%%%%%%%%
We now wish to explain more quantitatively the behavior that we observed in
{ \S\ref{sec:Example_Hernquist}}. Precisely, we aim to understand the
 characteristic behaviors of $\delta j$ and $\delta \omega$ around the
 eccentricity peak and to understand the envelopes of $D$ and $j_z$
 fluctuations over secular timescales. {We will address the separate problem
of determining the \textit{amplitude} of fluctuations $\delta j$ around the peak
eccentricity {in}
\S\ref{sec:characteristic_delta_j}.} {All this is necessary for understanding the nature of RPSD in Section \ref{sec:GR_physical_quantitative}.}

%%%%%%%%%%%%%%%%%%%%%%%%%%%%%%%%%%%
\subsubsection{Notation}
\label{sec:fluctuation_notation}

%%%%%%%%%%%%%%%%%%%%%%%%%
To achieve these aims we must first introduce a clean, precise notation to
describe fluctuations. Let us define the vector $\mathbf{w} \equiv
[\omega,J,\Omega,J_{z}]$. Then the `SA solution'
%%%%%%%%%%%%%%%%%%%%%%%%%%%%%%%%%%%
\begin{align} 
   \mathbf{w}_\mathrm{SA}(t) \equiv [\omega_\mathrm{SA}(t),J_\mathrm{SA}(t),\Omega_\mathrm{SA}(t),J_{z,\mathrm{SA}}(t)],
\end{align}
%%%%%%%%%%%%%%%%%%%%%%%%%%%%%%%%%%%
is found by self-consistently integrating the SA equations
\eqref{eqn:domegadt_SA}-\eqref{eqn:dJzdt_SA},
which are the Hamilton equations resulting from
$H_\mathrm{SA}(\omega_\mathrm{SA},J_\mathrm{SA},\Omega_\mathrm{SA},J_{z,\mathrm{SA}},t)
\equiv H_\mathrm{SA}(\mathbf{w}_\mathrm{SA},t)$. Meanwhile the `DA solution' 
%%%%%%%%%%%%%%%%%%%%%%%%%%%%%%%%%%%
\begin{align} 
   \mathbf{w}_\mathrm{DA}(t) \equiv [\omega_\mathrm{DA}(t),J_\mathrm{DA}(t),\Omega_\mathrm{DA}(t),J_{z,\mathrm{DA}}],
\end{align}
%%%%%%%%%%%%%%%%%%%%%%%%%%%%%%%%%%%
is found by self-consistently integrating the DA equations of motion,
which are the
Hamilton equations for
$H_\mathrm{DA}(\omega_\mathrm{DA},J_\mathrm{DA},J_{z,\mathrm{DA}}) \equiv
H_\mathrm{DA}(\mathbf{w}_\mathrm{DA})$. Consistent with equation
\eqref{eqn:deltaj_jSA_minus_jDA} we formally define: 
%%%%%%%%%%%%%%%%%%%%%%%%%%%%%%%%%%%
\begin{align} \delta \bw(t) \equiv \bw_\mathrm{SA}(t) - \bw_\mathrm{DA}(t).
   \label{eqn:deltaw_SA_minus_DA}
\end{align}
%%%%%%%%%%%%%%%%%%%%%%%%%%%%%%%%%%%
Next, we will also find it useful to define a `fluctuating Hamiltonian':
%%%%%%%%%%%%%%%%%%%%%%%%%%%%%%%%%%%
\begin{align}
    \Delta H(\mathbf{w},t) \equiv H_\mathrm{SA}(\bw,t)-H_\mathrm{DA}(\bw),
    \label{eqn:DeltaH_HSA_minus_HDA}
\end{align}
%%%%%%%%%%%%%%%%%%%%%%%%%%%%%%%%%%%
which is written out explicitly in Appendix
\ref{sec:fluctuating_Hamiltonian}. Using $\Delta H$, we
can \textit{define} four new quantities 
%%%%%%%%%%%%%%%%%%%%%%%%%%%%%%%%%%%
\begin{align}
    \Delta \bw (\bw,t) \equiv [\Delta \omega(\bw,t), \Delta J(\bw,t), \Delta \Omega(\bw,t), \Delta J_z(\bw,t)],
\end{align}
%%%%%%%%%%%%%%%%%%%%%%%%%%%%%%%%%%%
as the solution to the equations of motion
%%%%%%%%%%%%%%%%%%%%%%%%%%%%%%%%%%%
\begin{align}
\frac{\md \Delta \bw(\bw, t)}{ \md t} &\equiv 
\left[ \frac{\partial}{ \partial J},  
-\frac{\partial }{ \partial \omega},
\frac{\partial} { \partial J_z},
-\frac{\partial}{ \partial \Omega}\right]\Delta H(\bw, t).
\label{eqn:diag}
\end{align}
%%%%%%%%%%%%%%%%%%%%%%%%%%%%%%%%%%%
As an example, the partial derivative $\partial \Delta H/\partial \omega $ is given explicitly
for spherical potentials in equation
\eqref{eqn:minusddeltaHdomega}.
\begin{figure*}
   \centering
   \includegraphics[width=0.99\linewidth]{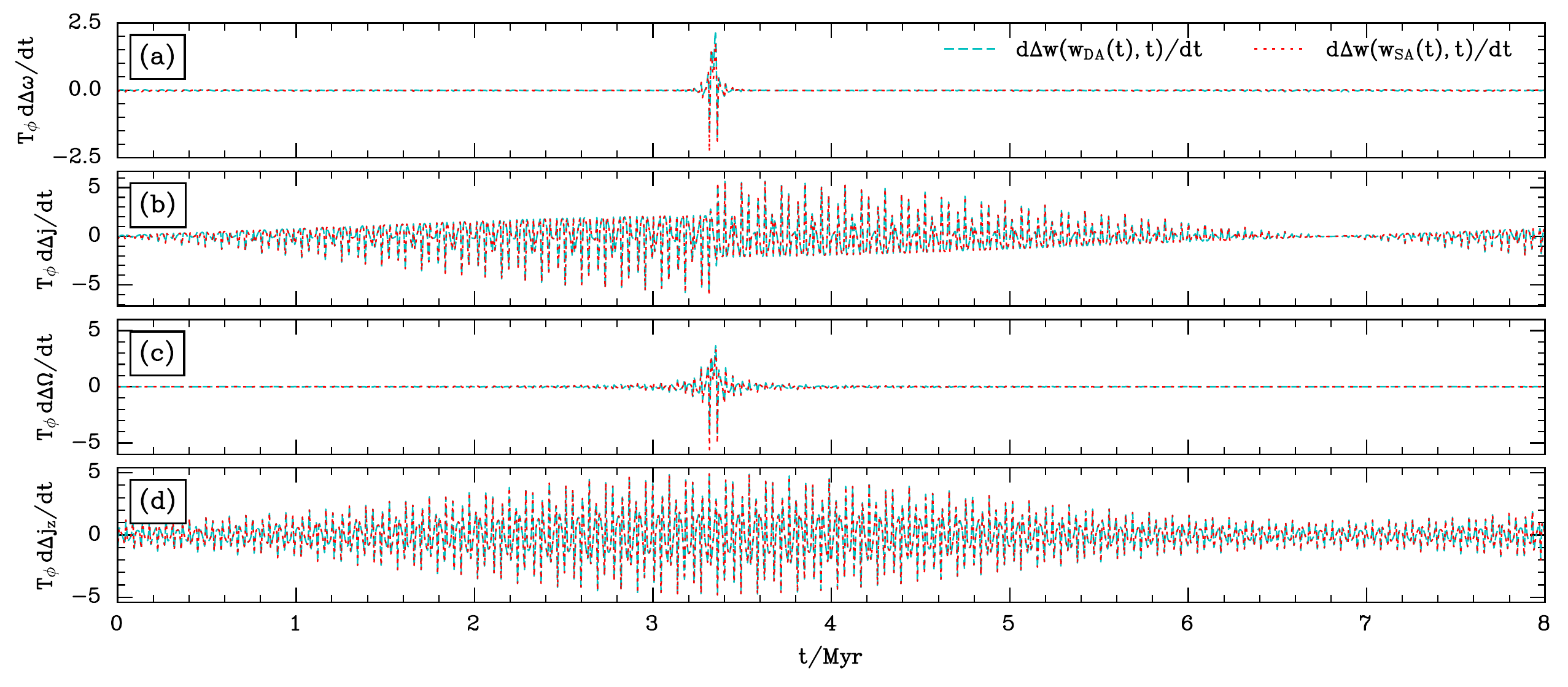}
   \caption{{Time series of $\md \Delta \bw/\md t$ following the definition \eqref{eqn:diag}, for the first $8$ Myr of evolution from Figure \ref{fig:Example_Hernquist} (the first peak in DA eccentricity occurs at $t\approx 3.34$Myr; the subsequent eccentricity minimum is at $t\approx 6.8$Myr).
   Cyan dashed lines and red dotted lines show $\md \Delta \mathbf{w}(\bw_\mathrm{SA},t)/\md t$ and $\md \Delta \mathbf{w}(\bw_\mathrm{SA},t)\md t$ respectively.}}
   \label{fig:dDeltawdtHernquist}
   \end{figure*}
   %%%%%%%%%%%%%%%%%%%%%%%%%
{Note that equation \eqref{eqn:diag}
is defined for arbitrary arguments $\bw$.
In Figure \ref{fig:dDeltawdtHernquist} we plot the time series of $\md \Delta \bw(\bw, t)/\md t$
for $\bw = \bw_\mathrm{DA}$ and $\bw = \bw_\mathrm{SA}$ using the 
data from the first $8$ Myr of evolution from Figure \ref{fig:Example_Hernquist}.
Repeating the same exercise for other examples gives plots that look qualitatively the same as 
Figure \ref{fig:dDeltawdtHernquist}.}

{Now}  we must bear in mind that in general, 
%%%%%%%%%%%%%%%%%%%%%%%%%%%%%%%%%%%
\begin{align}
\delta \bw(t) \neq \Delta \bw (\bw_\mathrm{SA}(t), t) \neq \Delta \bw (\bw_\mathrm{DA}(t), t).
\end{align}
%%%%%%%%%%%%%%%%%%%%%%%%%%%%%%%%%%%
%This is because both terms on the right hand side of \eqref{eqn:DeltaH_HSA_minus_HDA} take the same argument $\mathbf{w}$.  In reality $\delta \bw$ cannot be reproduced exactly with a single set of Hamilton equations. 
Nevertheless, for our purposes it is normally sufficient to approximate
%%%%%%%%%%%%%%%%%%%%%%%%%%%%%%
\begin{align} \delta \mathbf{w}(t) \approx \Delta \mathbf{w}(\bw_\mathrm{SA}(t),t)
\approx \Delta \mathbf{w}(\bw_\mathrm{DA}(t),t).
\label{eqn:deltaw_DeltawSA_DeltawDA}
\end{align}
%%%%%%%%%%%%%%%%%%%%%%%%%%%%%%
{In Appendix \ref{sec:Deltaw_Explanation} we explain why this is the case, and justify it with a numerical example.}

%%%%%%%%%%%%%%%%%%%%%%%%%%%%%%%%%%%%%%%%%%%%%%%%%%%%%%%%%%%%%%%%%%%%%%%%%%%%%%%%%%%%%%%%%%%%%%%%%%%%%%%%%%%%%%%%%%%%%%%%%%

\subsubsection{Scaling of fluctuations at high eccentricity in spherical cluster potentials}
\label{sec:scaling_highe}

Having established the approximation
\eqref{eqn:deltaw_DeltawSA_DeltawDA}, we can use the equations of
motion \eqref{eqn:diag} to gain a better understanding of the
behavior of fluctuating quantities $\delta \bw$ in Figures
\ref{fig:Example_Hernquist} and \ref{fig:Example_Plummer}.
To do this we take derivatives of {$\Delta H$ as given in} \eqref{eqn:DeltaH_spherical}
(which is valid for spherical potentials only) and then take the high
eccentricity limit\footnote{{Note we are \textit{not} assuming $J^2 \to J_z^2$, i.e. we are assuming nothing about the ratio $J_z/J \equiv \cos i$ other than that it is $\in (-1,1)$.
This means that the results \eqref{eqn:scalings} hold regardless of the 
complicated behavior of the inclination near high-$e$ as seen in e.g. Figure \ref{fig:Inclination_Plummer}}.
} {$L^2 \gg J^2 \gtrsim J_z^2$}. As a result we find the following
scalings:
%%%%%%%%%%%%%%%%%%%%%%%%
\begin{align}
   \frac{\md }{\md t} \delta j \propto J^0,
   \,\,\,\,\,\,\,\,
\frac{\md }{\md t} \delta j_z \propto J^0, \nn
\\
\frac{\md }{\md t} \delta \omega \propto J^{-1}
, \,\,\,\,\,\,\,\,
\frac{\md }{\md t} \delta \Omega \propto J^{-1}.
\label{eqn:scalings}
\end{align}
Thus, we expect the fluctuations $\delta j$, $\delta j_z$ to be independent
of $j$ as $e\to 1$, i.e. as {$j^2, \, j_z^2 \to 0$}. In other words, as the binary approaches maximum
eccentricity we do not
expect any sharp peak in $\delta j(t)$ or $\delta j_z(t)$, but we do expect a spike in $\delta \omega(t)$ and
$\delta \Omega(t)$. Such behavior is
precisely what we found in panels (m), (p)
and (q) of Figures
\ref{fig:Example_Hernquist} and \ref{fig:Example_Plummer}, and is also exhibited clearly in Figure \ref{fig:Difference_Equations_Hernquist}.

\subsubsection{Envelope of fluctuations in $D$ and $j_z$}
\label{sec:jz_D_envelopes}

In the numerical examples above,
{fluctuations in $j_{z,\mathrm{SA}}$  consisted of 
rapid oscillations on the timescales $\sim T_\phi, T_R$ (reflective of the torque fluctuating on the outer orbital timescale), modulated by an envelope with period $\tsec$.
Similarly, $D_\mathrm{SA}$ oscillated on the timescale $\sim T_\phi$, 
modulated by an envelope with period $\tsec/2$.
 We now explain
 each of these envelope behaviors in turn.}

 First, we consider the envelope of $j_{z,\mathrm{SA}}$ fluctuations. It is clear
 from the numerical examples that the amplitude of this envelope is largest 
around the eccentricity peak (minimum $j$),
and smallest around the eccentricity minimum (maximum $j$).
This is easily explained by 
evaluating the torque formula $\md j_{z,\mathrm{SA}} / \md t$ 
 using equations \eqref{eqn:diag} and 
 \eqref{eqn:minusddeltaHdOmega}, and examining the scaling with $j$ ({which is well illustrated by the example in  Figure \ref{fig:dDeltawdtHernquist}d)}. 
The  envelope of $j_z$ fluctuations simply reflects 
 the amplitude of the fluctuating torque}.
% The rapid `swing' between extreme
% values of $\omega$ that occurs around maximum eccentricity is unimportant for determining this envelope, since it lasts such a short time.

\begin{figure*}
   \centering
   \includegraphics[width=0.99\linewidth]{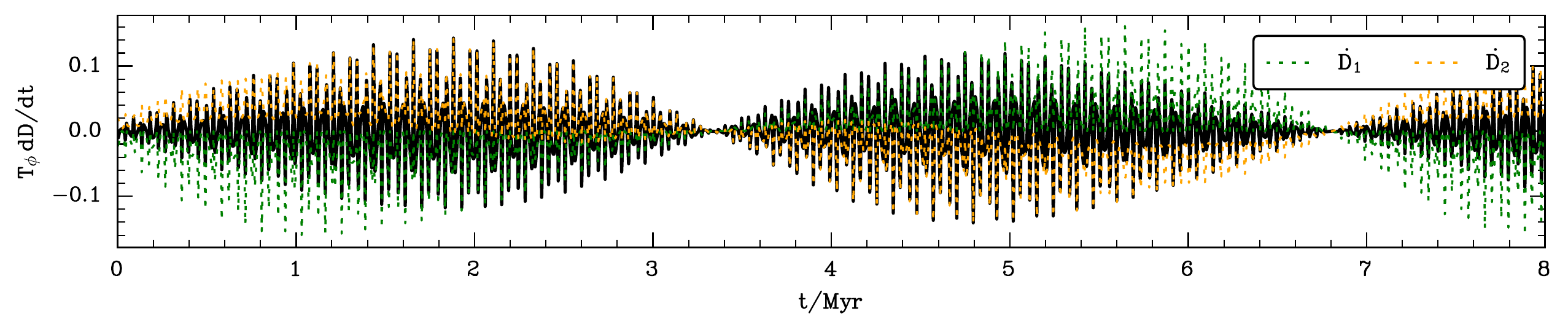}
   \caption{{Time series of $\md D/\md t$ (black solid line) for the first $8$ Myr of evolution from Figure \ref{fig:Example_Hernquist}, following equation \eqref{eqn:dDdt_SA}.  The green and orange {dotted lines show contributions from $\dot{D}_1$ (equation \eqref{eqn:Ddot_1}) and 
   $\dot{D}_2$ (equation \eqref{eqn:Ddot_2}) respectively.} The amplitude of both terms is smallest at the extrema of $e_\mathrm{DA}$.}}
   \label{fig:dDdtHernquist}
   \end{figure*}
{Second, we address the fluctuations in $D_\mathrm{SA}$.  In this case
 the amplitude of the envelope exhibits minima at times corresponding to both
 $j_\mathrm{DA}=j_\mathrm{max}$ and $j_\mathrm{DA}=j_\mathrm{min}$, and maxima
 in-between}.  To see why, we differentiate \eqref{eqn:def_D} {with $\epsGR$ set to zero}:
%%%%%%%%%%%%%%%%%%%%%%%%%%%%%%
{\begin{align}
\frac{\md D}{\md t} = \dot{D}_1 + \dot{D}_2
   \label{eqn:dDdt_SA}
\end{align}
where
\begin{eqnarray}
\dot{D}_1 &=& \frac{2 \, D(\epsGR=0)}{e}\frac{\md e}{\md t} =  -\frac{2j \, D(\epsGR=0)}{1-j^2}\frac{\md j}{\md t},
   \label{eqn:Ddot_1}
\\
\dot{D}_2 &=& \frac{10\Gamma e^2}{1-5\Gamma}
   \frac{\md }{\md t}
   \left( \sin^2 i \sin^2 \omega\right).
   \label{eqn:Ddot_2}
\end{eqnarray}
%%%%%%%%%%%%%%%%%%%%%%%%%%%%%%
{In Figure \ref{fig:dDdtHernquist} we plot $\md D/\md t$ following equation \eqref{eqn:dDdt_SA}, again 
over the first $8$ Myr of evolution from Figure \ref{fig:Example_Hernquist}.
With green and orange dotted lines we overplot the contributions coming from {$\dot{D}_1$ and $\dot{D}_2$ respectively}.}
Like with $j_{z,\mathrm{SA}}$, the envelope of fluctuations in 
$D_\mathrm{SA}$ reflects the envelope of $\md D/\md t$.
In particular, the amplitudes of both $\dot{D}_1$ and $\dot{D}_2$} are minimized at the eccentricity extrema, and maximized in-between.
A much more detailed discussion of $\md D/\md t$ at high eccentricity is postponed to \S\ref{sec:RPSD_Mechanism}.

\subsection{Characteristic amplitude of $\delta j$ fluctuations}
\label{sec:characteristic_delta_j}

%Under SA dynamics, the maximum eccentricity reached by the binary is typically modified from the DA value  $e_\mathrm{max} = (1-j^2_\mathrm{min})^{1/2}$ to 
%%%%%%%%%%%%%%%%%%%%%%%%%%%%%%%%%%%%%%%%
%\begin{align} \widetilde{e}_\mathrm{max} = (1-\left[j_\mathrm{min}-(\delta j)_\mathrm{max} \right]^2)^{1/2}, \label{SAFluctuations_eqn_etilde} \end{align} 
%%%%%%%%%%%%%%%%%%%%%%%%%%%%%%%%%%%%%%%%
 Perhaps the most important consequence of short-timescale fluctuations is that
 they enhance the value of $\emax$ when $e$ gets very large, which can lead to
 e.g. more rapid compact object binary mergers \citep{Grishin2018}. With
 this in mind, we wish to estimate $(\delta j)_\mathrm{max}$, which we define to
 be the absolute value of the maximum fluctuation $\delta j$ in the vicinity of
 maximum eccentricity. Unfortunately, a given binary does not have a single value of
 $(\delta j)_\mathrm{max}$, because the precise details of the fluctuating behavior differ 
 from one
 secular eccentricity peak to the next. (We already saw something closely related to this in Figure \ref{fig:Example_Hernquist_Divergence}b, where binaries which approached the same secular eccentricity peak with different outer orbital phases ended up exhibiting very different behavior around $e_\mathrm{max}$).
 %This is because the specific torque at the time of the passage through $e_\mathrm{max}$ depends in a complex way on the relative phases of $\Omega(t)$ and the outer orbit $\phi(t), R(t)$.  This phase dependence is demonstrated explicitly by comparing Figures \ref{fig:Example_Plummer} and \ref{fig:Example_Plummer_2}.
{In Appendix \ref{sec:App_deltaj}, we show how to estimate the characteristic
size of such fluctuations and then demonstrate numerically that our estimate is
a reasonable one, culminating in equations 
\eqref{SAFluctuations_eqn_deltaj} and
\eqref{SAFluctuations_eqn_Frpra}.}

The problem with equations \eqref{SAFluctuations_eqn_deltaj} and
\eqref{SAFluctuations_eqn_Frpra} as they stand is that we do not know precisely
which quarter-period in $\phi$ will provide the dominant fluctuation, because
this would require knowledge of $R(t)$ and $\phi(t)$.  Thus, we cannot evaluate
\eqref{SAFluctuations_eqn_Frpra} directly for an arbitrary outer orbit --- and
even if we could, we would not expect the resulting $(\delta j)_\mathrm{max}$ to
be exactly correct because of the non-stationarity of $\omega,j,\Omega$. 

However, we can make a very rough estimate of the importance of these fluctuations 
if we note that $F$ is normally of the same order of magnitude as the azimuthal frequency of the outer orbit,
$2\pi/T_\phi$.  Then using $\cos \imin \sim 1$ and the weak GR result $\jmin^2  \sim  \Theta \sim \cos^2 i_0$,
we get
\begin{equation}
    \frac{(\delta j)_\mathrm{max}}{\jmin} \sim \frac{1}{\vert \cos i_0 \vert }\frac{T_\mathrm{in}}{T_\phi} . 
    \label{eqn:SA_strength_estimate}
\end{equation}
This equation is useful for estimating whether the effect of short-timescale fluctuations can drastically change $j$ at very high eccentricity.  In particular, it tells us that SA effects are crucial for binaries with initial inclinations in the range 
\begin{equation}
    \vert \cos i_0 \vert \lesssim  \frac{T_\mathrm{in}}{T_\phi}.
    \end{equation}
%Naively one might assert that, at least in cuspy clusters, we should choose to integrate over the quarter-period in azimuth that encompasses the point $R=\rp$ since the instantaneous tidal torque should be largest there.  However we must keep in mind that a strongly radial outer orbit spends much less time near $\rp$ than near, say, $\ra$.  In general the dominant fluctuation could come from any part of the outer orbit.  Additionally, suppose that the dominant fluctuation really did arise around $R=\rp$.  Then the \textit{maximum possible} fluctuation would arise if the function $\sin[2(\phi(t) - \Omega)]$ was peaked at exactly at the time when $R(t)= \rp$. However such a coincidence will not be realised in general: instead, for a large ensemble of binaries we expect $R=\rp$ to coincide with a uniform spread of values $\phi - \Omega \in (0,\pi/2)$.  
We can also relate the estimate \eqref{eqn:SA_strength_estimate} to the ratio $\tmin/T_\phi$, which will be a very important parameter in our discussion of RPSD (see \S\ref{sec:Effect_of_GR}). 
We know from Paper II that $\tsec \sim T_\phi^2 / T_\mathrm{in}$.  Then the time spent near highest eccentricity (equation \eqref{eqn:tmin_jmin}) is on the order of $\tmin \sim \jmin t_\mathrm{sec} \sim \jmin T_\phi^2 / T_\mathrm{in}$.  Using this to eliminate $T_\mathrm{in}$ from the right hand side of \eqref{eqn:SA_strength_estimate} gives
\begin{equation}
    \frac{(\delta j)_\mathrm{max}}{\jmin} \sim \frac{T_\phi}{t_\mathrm{min}} . 
    \label{eqn:SA_strength_estimate_2}
\end{equation}
Thus short-timescale fluctuations are important in precisely those regions of parameter space where $\tmin$ is comparable to or smaller than $T_\phi$.\footnote{{Unfortunately, this is also the regime in which the calculation we have performed in this section is not really valid, since this relied on our freezing the DA quantities while we calculated the fluctuations --- see the discussion below \eqref{SAFluctuations_dJdtfinal}.  See \S\ref{sec:literature} for more.}}

\section{The effect of GR precession}
\label{sec:Effect_of_GR}

In the previous section we gained {insight into} how short-timescale fluctuations in the
 tidal torque affect high eccentricity behavior, but if {our results are to be relevant 
 for studying dynamical compact object merger channels, then it is
 vital that we also account for 1pN GR precession.} As we will see in this section,
 including GR precession can change the picture significantly. To begin with, in
 \S\ref{sec:GR_Hernquist}
 we rerun the numerical calculations from
 \S\ref{sec:Example_Hernquist}
 except this time with GR precession switched on, and simply describe the
 altered phenomenology. In \S\ref{sec:Example_Kepler} we compare
 the GR and non-GR calculations in the special case of the LK problem. The main
 new result that arises in each of these subsections {is RPSD, which stems from 
non-conservation of the}
 approximate integral of motion $D$ in the SA approximation.  In
 \S\ref{sec:GR_physical_quantitative} we offer a physical
 explanation for this new phenomenon, {explain the criteria for its existence,
 and attempt a rough statistical analysis.}
 
%%%%%%%%%%%%%%%%%%%%%%%%%%%%%%%%%%%%
%%%%%%%%%%%%%%%%%%%%%%%%%%%%%%%%%%%%
%%%%%%%%%%%%%%%%%%%%%%%%%%%%%%%%%%%%

\subsection{Numerical examples with GR precession}
\label{sec:GR_Hernquist}

%%%%%%%%%%%%%%%%%%%%%%%%%%%%%%%%%%%%
\subsubsection{Fiducial Hernquist example}

\begin{figure*}\centering
            \includegraphics[width=0.95\linewidth]{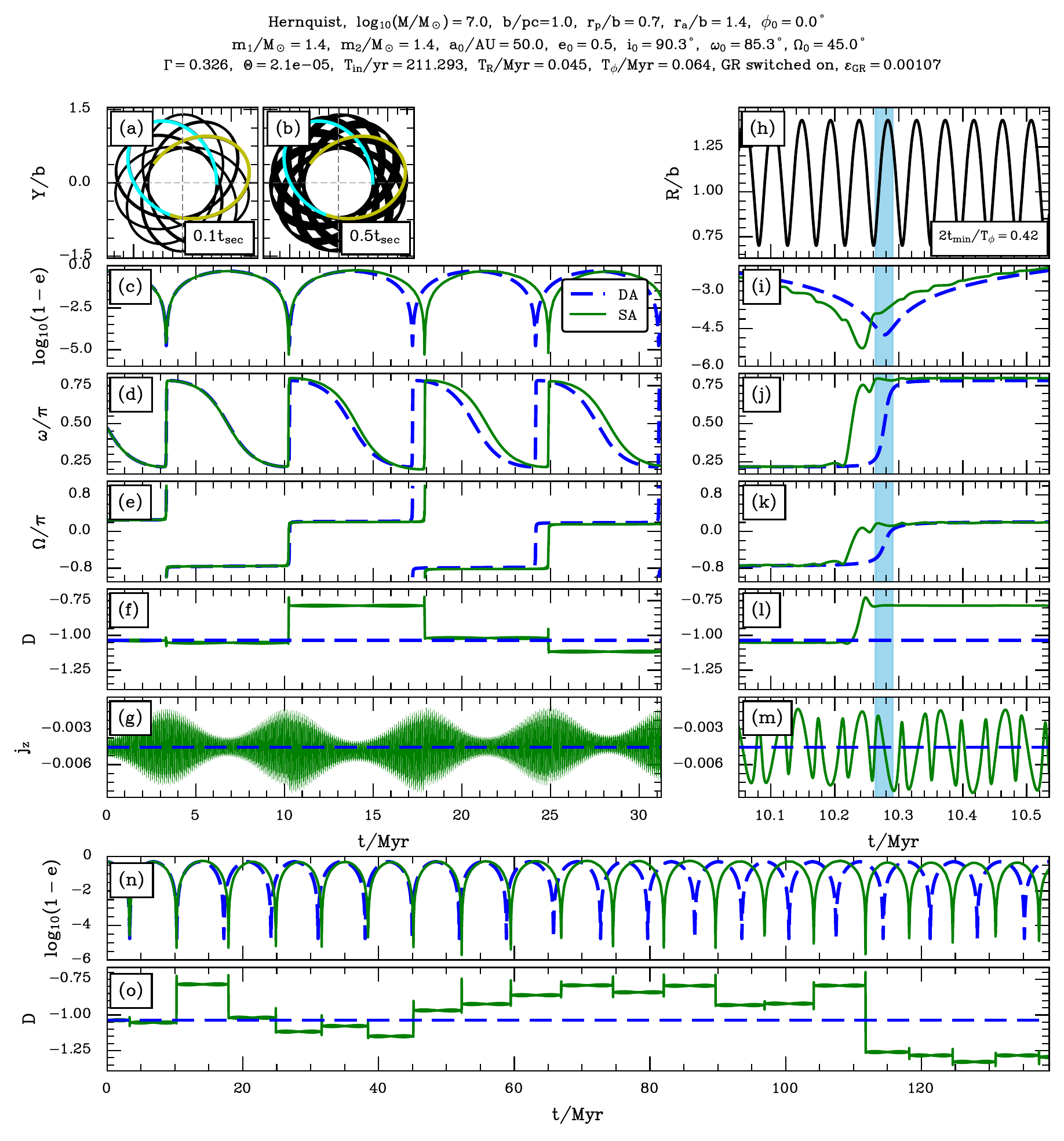}
      \caption{As in Figure \ref{fig:Example_Hernquist}, except
   we switch on GR precession.  This causes the time-averaged value of
   $D_\mathrm{SA}$ to diffuse away from $D_\mathrm{DA}$, with discrete jumps
   occurring during episodes of extremely high eccentricity. We call this
   relativistic phase space diffusion (`RPSD').}
      \label{fig:Example_Hernquist_GR}
   \end{figure*}%
   %\hfill <-- it is superfluous 

In Figure \ref{fig:Example_Hernquist_GR} we rerun the calculation
from Figure \ref{fig:Example_Hernquist}, except we switch on the
GR precession term (with strength $\epsGR = 0.00107$) in the SA and DA equations
of motion. We now discuss Figure \ref{fig:Example_Hernquist_GR}
in some detail.

The structure of panels (a)-(m) is identical to those of Figure
\ref{fig:Example_Hernquist}, except that we have dispensed with
{direct orbit integration} since it is prohibitively computationally expensive {(to capture accurately the 
fast GR precession during eccentricity peaks tends to require an extremely tiny timestep).}
%We have also chosen in 
%panels (h)-(m) we have chosen to zoom in on the first secular eccentricity
%peak rather than the second. 
Comparing panels (a)-(m) with those of Figure
\ref{fig:Example_Hernquist} we immediately notice several
qualitative differences. Whereas in Figure
\ref{fig:Example_Hernquist} the DA and SA predictions for
$\log_{10}(1-e)$ agreed almost perfectly except at extremely high eccentricity,
now in Figure \ref{fig:Example_Hernquist_GR} (with GR precession switched on)
they disagree manifestly after the second eccentricity maximum. Moreover, while
the period of secular oscillations is fixed in the DA case, the SA secular
period changes from one eccentricity peak to the next. By the time of the third
eccentricity peak the DA and SA curves in panels (c)-(e) are completely out of
sync, as we intimated would happen back in \S\ref{sec:phase_dependence}, when we were discussing Figure \ref{fig:Example_Hernquist_GR_Divergence}.

\begin{figure}
   \centering
   \includegraphics[width=0.98\linewidth]{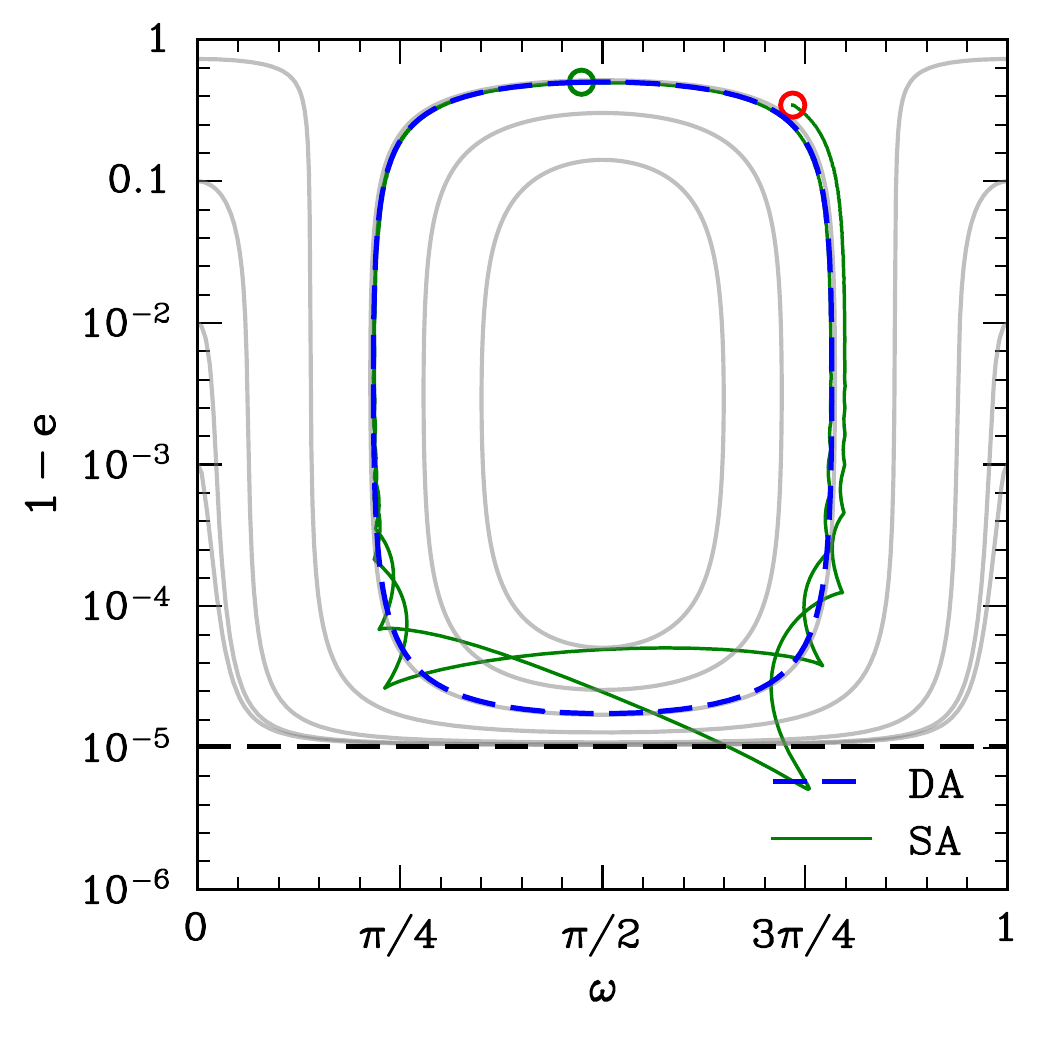}
   \caption{{Illustration of RPSD {in phase space}.} We plot the trajectory of the binary from Figure \ref{fig:Example_Hernquist_GR} through the $(\omega, e)$ phase space, over the first $\sim 2\tsec$.
   {We show the DA trajectory in blue and the SA trajectory in green.}
   The grey lines are contours of constant DA Hamiltonian (equations \eqref{eqn:H1_Doubly_Averaged} and \eqref{eqn:HGR}), and the black dashed line gives the maximum achievable DA eccentricity $e_\mathrm{lim} \equiv \sqrt{1-\Theta}$, 
   all calculated at $t=0$.  The binary's initial phase space location is shown with a green circle,
   {and this of course coincides with the blue dashed contour on which the DA solution lives indefinitely.} 
   The binary's final SA location is shown with a red circle. {We see that the SA trajectory jumps to a new `parent' DA contour {(i.e. off the blue dashed DA contour)} during {the second} high eccentricity episode.}}
   \label{fig:Phase_Space_Hernquist_GR}
   \end{figure}

{Crucially}, from panel (f) we see that $D_\mathrm{SA}$ no longer fluctuates
around $D_\mathrm{DA}$ indefinitely like it did in Figure
\ref{fig:Example_Hernquist}f, but rather exhibits discrete jumps
during very high eccentricity episodes.  In panel (l) we zoom in on the
$D_\mathrm{SA}$ behavior around the second eccentricity peak. We see that
$D_\mathrm{SA}$ jumps from $\approx -1.05$ to $\approx -0.8$ and that this jump
happens on the timescale $\sim 2 t_\mathrm{min}$ {(the width of the blue shaded band)}.  
{This jump in the approximate integral of motion $D$ means that the binary has 
jumped to a new constant-Hamiltonian
contour in the DA $(\omega,e)$ phase space (Papers II-III), as we confirm in Figure \ref{fig:Phase_Space_Hernquist_GR}.
Since this behavior
depends crucially on the presence of finite GR precession we choose to call it
`relativistic phase space diffusion' (RPSD).}
{Furthermore, each phase space contour has its own secular period, i.e. the secular period $\tsec$ depends on $D$ (see \S2.6 of Paper II, and especially Figure 3 of that paper).}
Hence it is unsurprising that a jump in $D_\mathrm{SA}$ leads to
a modified SA secular period compared to the fixed DA period\footnote{For instance, in Figure
\ref{fig:Example_Hernquist_GR}, making $D$ less negative while
keeping $j_z$ --- and therefore $\Theta$ --- fixed moves the binary closer to
the separatrix in the $(D,\Theta)$ phase space (see Figure 3c of Paper II), which accounts for the increase in the period of the
subsequent secular oscillation.
\\
(Strictly speaking, $t_\mathrm{sec}$ is not the exact DA secular period
when we include GR precession, as it is calculated assuming no GR (equation (33) of
Paper II), but for $\epsGR \ll 3(1+5\Gamma)$  --- the `weak-to-moderate GR regime', see Paper III --- it is a very
good approximation.)}.

{Meanwhile, we observe in panel (g) that
the envelope of $j_{z,\mathrm{SA}}$ fluctuations
 does undergo abrupt, although very modest, changes that coincide with the 
 second and third eccentricity peaks.
Thus, the adiabatic invariance of $j_{z,\mathrm{SA}}$ is also slightly broken, 
but its time average is much better preserved than is the time average of $D_\mathrm{SA}$.}

To investigate the RPSD behavior further, we next ran the integration for a much longer
time, $20 t_\mathrm{sec}$.
In panels (n) and (o) of Figure \ref{fig:Example_Hernquist_GR} we
plot $1-e$ and $D$ respectively as functions of time over this entire duration.
The same picture holds in that $D_\mathrm{SA}$ is roughly static between
eccentricity peaks, but often makes a discernible jump during a peak. These
jumps seem to have no preferred sign. 
%%%%%%%%%%%%%%%%%%%%%%%%%%%%%%%%%%%%%%%%%%%%%%%%%

%Finally, in Figure \ref{fig:Phase_Space_Hernquist_GR}
%we plot 
%In particular, this figure makes it clear that large jumps in phase space occur within a single outer orbital period.

%\subsubsection{Circular outer orbit}

%In Figure \ref{fig:Example_Hernquist_2_GR} we rerun the
%calculation from Figure \ref{fig:Example_Hernquist_2} with GR
%switched on, and again we choose to zoom in on the first secular eccentricity
%peak in the right hand column. We again find abundant RPSD, and see from panel
%(l) that the $D_\mathrm{SA}$ jump happens essentially within the blue stripe
%\eqref{eqn:t_blue_bands}.  

%%%%%%%%%%%%%%%%%%%%%%%%%%%%%%%%%%%%%%%%%%%%%%%%%

\subsubsection{Changing the initial inclination to $i_0=93.3^\circ$}
      \begin{figure*}\centering
            \includegraphics[width=0.95\linewidth]{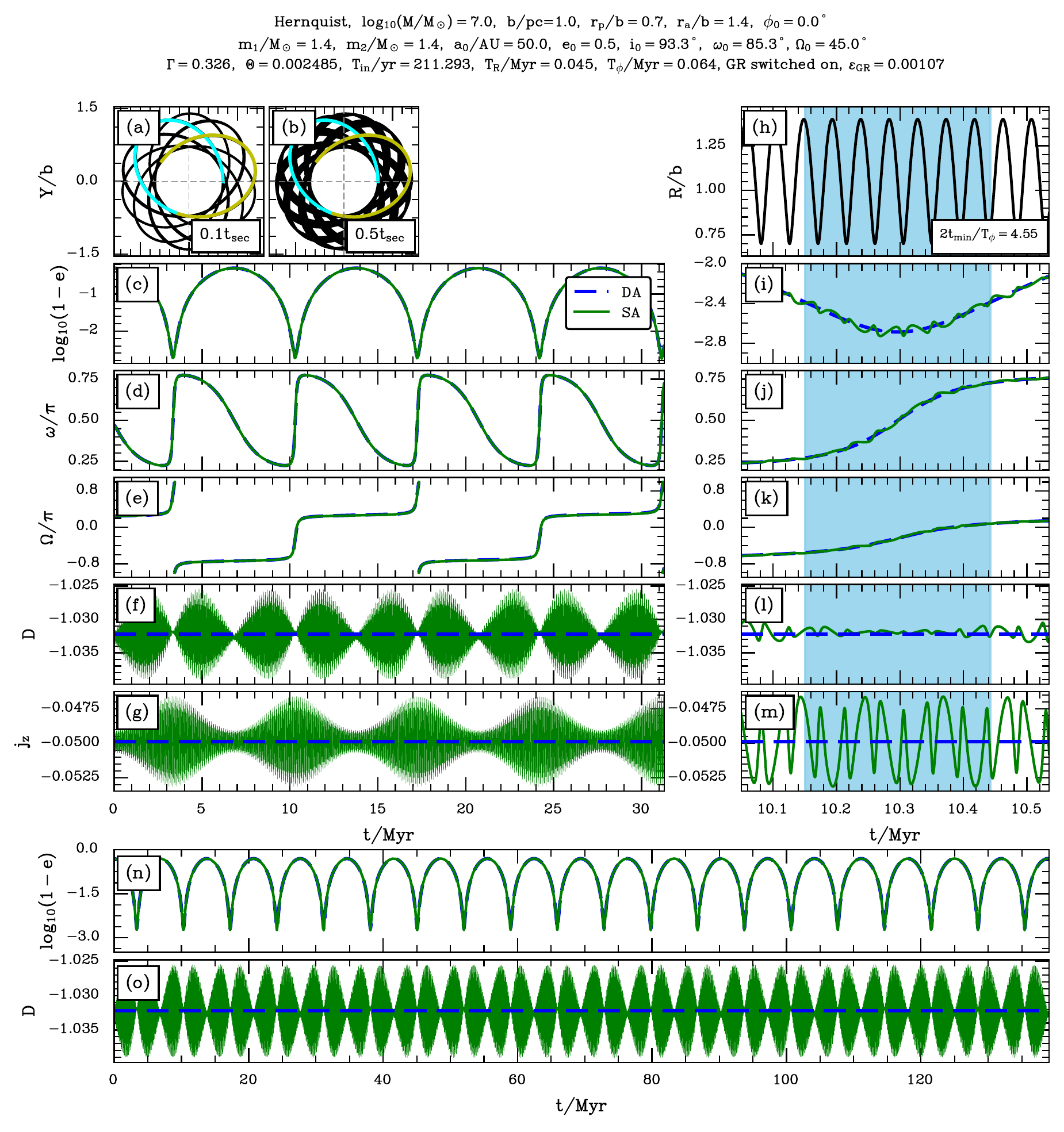}
      \caption{As in Figure
   \ref{fig:Example_Hernquist_GR} except we take $i_0=93.3^\circ$
   --- in other words, the same as Figure
   \ref{fig:Example_Hernquist_4} except we switch on GR
   precession.  There is no RPSD in this case.}
      \label{fig:Example_Hernquist_4_GR}
   \end{figure*}
{RPSD does} \textit{not} {occur}
in Figure \ref{fig:Example_Hernquist_4_GR}, in which we keep the initial
conditions exactly the same as Figure
\ref{fig:Example_Hernquist_GR} except that we change $i_0$ from
$90.3^\circ$ to $93.3^\circ$. (In other words, we rerun the same calculation as
in Figure \ref{fig:Example_Hernquist_4} except with GR switched
on). Note that this case has $2t_\mathrm{min}/T_\phi =4.55$, {as opposed to Figure \ref{fig:Example_Hernquist_GR} which had $2t_\mathrm{min}/T_\phi =0.42$.}

%%%%%%%%%%%%%%%%%%%%%%%%%%%%%%%%%%%%%%%%%%%%%%%%%
%%%%%%%%%%%%%%%%%%%%%%%%%%%%%%%%%%%%%%%%%%%%%%%%%
    %%%%%%%%%%%%%%%%%%%%%%%%%%%%%%%%%%%%%%%%%%%%%%%%%
%%%%%%%%%%%%%%%%%%%%%%%%%%%%%%%%%%%%%%%%%%%%%%%%%
%%%%%%%%%%%%%%%%%%%%%%%%%%%%%%%%%%%%%%%%%%%%%%%%%
\subsubsection{An example in the Plummer potential}
\label{sec:GR_Plummer}

%\subsubsection{(A) Fiducial Plummer example including GR precession}

\begin{figure*}\centering
\includegraphics[width=0.95\linewidth]{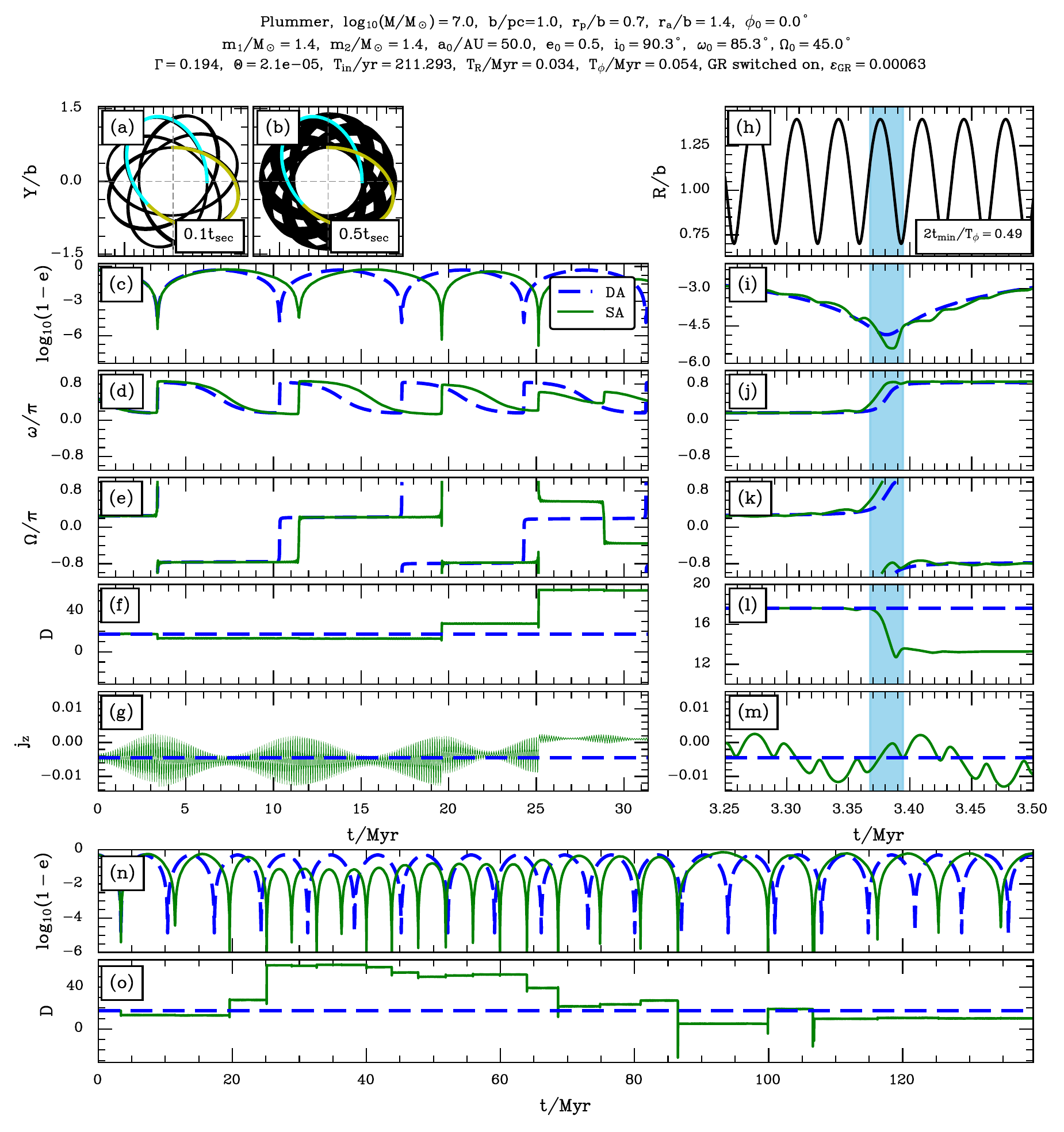}
      \caption{Fiducial Plummer example including GR precession. As in Figure \ref{fig:Example_Hernquist_GR}
 except we use the Plummer potential --- in other words, the same as Figure
 \ref{fig:Example_Plummer} except we switch on GR precession. {In this case, after a few secular periods the DA prediction does not even qualitatively describe the SA dynamics.}}
      \label{fig:Example_Plummer_GR}
   \end{figure*}%
   
In Figure \ref{fig:Example_Plummer_GR} we give another example of
RPSD, this time in the Plummer potential.  This example is identical to that in
Figure \ref{fig:Example_Plummer} except that we switched on GR
precession, and zoomed in on the first eccentricity peak in the right column
rather than the second.
{Panel (l) shows that this first eccentricity peak coincides with a large jump in $D$.
As a result, the subsequent SA evolution is entirely different from the 
original DA prediction.}

%\subsubsection{
\subsubsection{Dependence on phase angles}
\label{sec:Phase_Angles_GR}
 Analogous to the discussion in \S\ref{sec:Phase_Angles_no_GR}, we have also run several more numerical calculations identical to those shown in this section but varying the outer orbit's initial radial phase angle and the initial value of $\phi - \Omega$. {As expected from} Figure \ref{fig:Example_Hernquist_GR_Divergence}, we find that RPSD is {highly phase-dependent,
 meaning that SA simulations run with slightly different 
 initial conditions can have dramatically different outcomes.
This suggests we should attempt a statistical analysis --- see \S\ref{sec:RPSD_Statistical}.}

%%%%%%%%%%%%%%%%%%%%%%%%%%%%%%%%%%%%%%%%%%%%%%%%%
%%%%%%%%%%%%%%%%%%%%%%%%%%%%%%%%%%%%%%%%%%%%%%%%%
    %%%%%%%%%%%%%%%%%%%%%%%%%%%%%%%%%%%%%%%%%%%%%%%%%
%%%%%%%%%%%%%%%%%%%%%%%%%%%%%%%%%%%%%%%%%%%%%%%%%
%%%%%%%%%%%%%%%%%%%%%%%%%%%%%%%%%%%%%%%%%%%%%%%%%

\subsubsection{An example in the Lidov-Kozai limit}
\label{sec:Example_Kepler}

%The main phenomenological result we have found in this section is that
%when GR precession is switched on and a binary reaches very high eccentricity,
%short-timescale fluctuations may cause it to be `kicked' to a qualitatively new
%phase space trajectory. 
It is important to note that RPSD is not
exclusive to the non-Keplerian potentials that we have investigated so far, but
is present in the Lidov-Kozai problem as well, \textit{at the test-particle quadrupole level} (though to our knowledge, nobody
has mentioned it explicitly). 

\begin{figure*}\centering
          \includegraphics[width=0.95\linewidth]{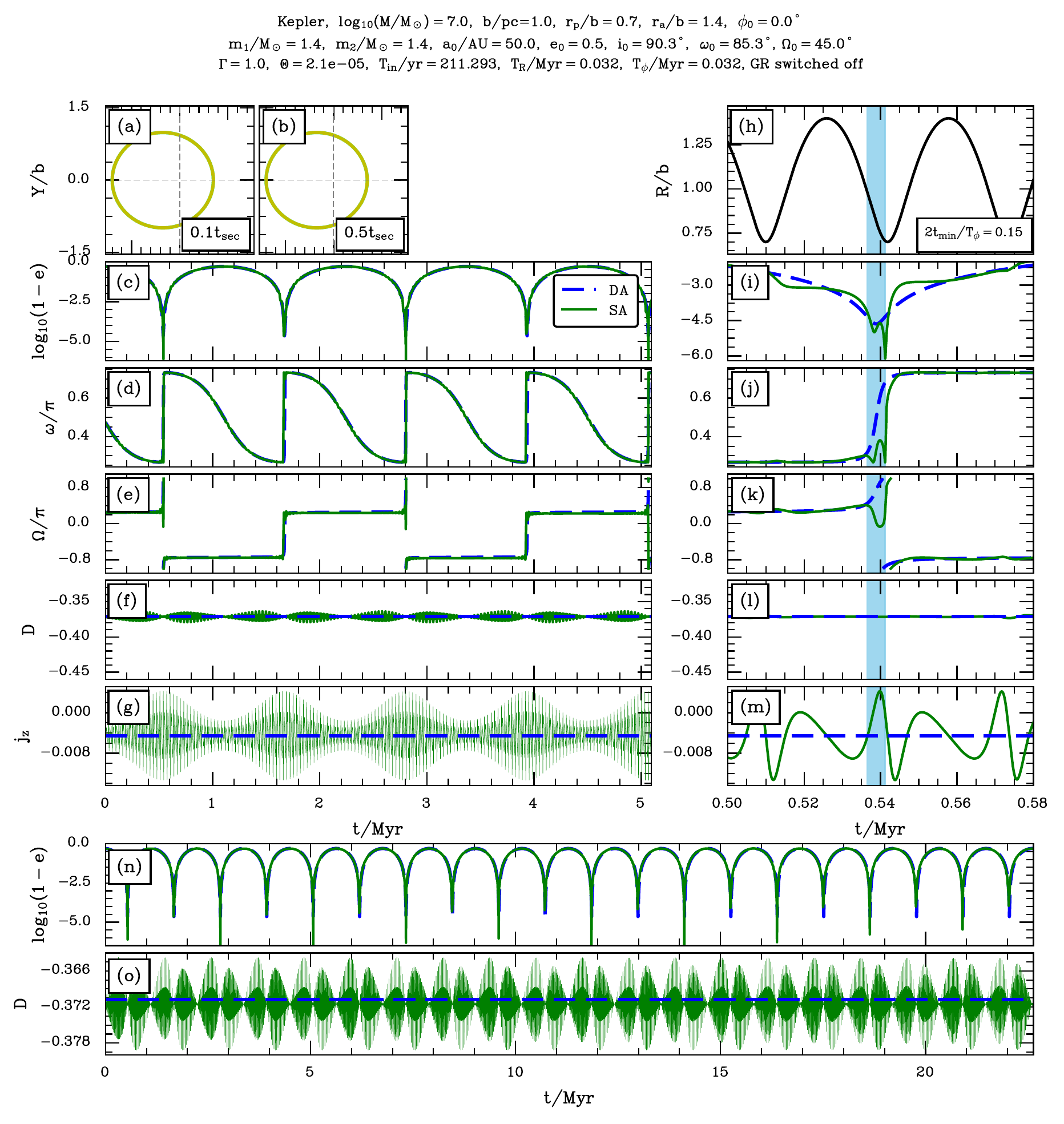}
      \caption{As in Figure
 \ref{fig:Example_Hernquist}, except we use the Kepler potential (so we are studying the classic test-particle quadrupole Lidov-Kozai mechanism).
 GR precession is switched off here.  }
      \label{fig:Example_Kepler}
   \end{figure*}%
   %\hfill <-- it is superfluous 

To demonstrate this, in Figure \ref{fig:Example_Kepler} we show a calculation with
   exactly the same initial condition as in Figure
   \ref{fig:Example_Hernquist}, except that we change the
   potential to the Keplerian one, $\Phi = -G\mathcal{M}/{R}$.  In other words,
   we are now investigating the classic test particle quadrupole Lidov-Kozai
   problem, relevant to e.g. a NS-NS binary orbiting a SMBH (e.g.
   \citealt{Antonini2012,Hamers2018,Bub2020}), except that we have GR precession switched off.
   In panels (a) and (b) we simply see the outer orbital ellipse, which has
   semimajor axis $a_\mathrm{g} = (\ra + \rp)/2 = 1.05b$ and eccentricity
   $e_\mathrm{g} = (\ra - \rp)/(\ra+\rp) = 0.33$.  %The outer orbit repeats this
   %elliptical trajectory indefinitely with period $T_R=T_\phi = 0.032
   %\mathrm{Myr}$.  
   Panels (c)-(m) {exhibit} behavior {which is qualitatively similar to that in Figure
   \ref{fig:Example_Hernquist}}.  Since $2t_\mathrm{min}/T_\phi = 0.15$, the
   highest eccentricity episode lasts for significantly less than one outer
   orbital period. Despite this, the SA result tracks the secular (DA) result
   indefinitely. {(We have also checked that the SA result shown here is indistinguishable from 
   that found by direct {orbit} integration).}

   \begin{figure*}\centering
\includegraphics[width=0.95\linewidth]{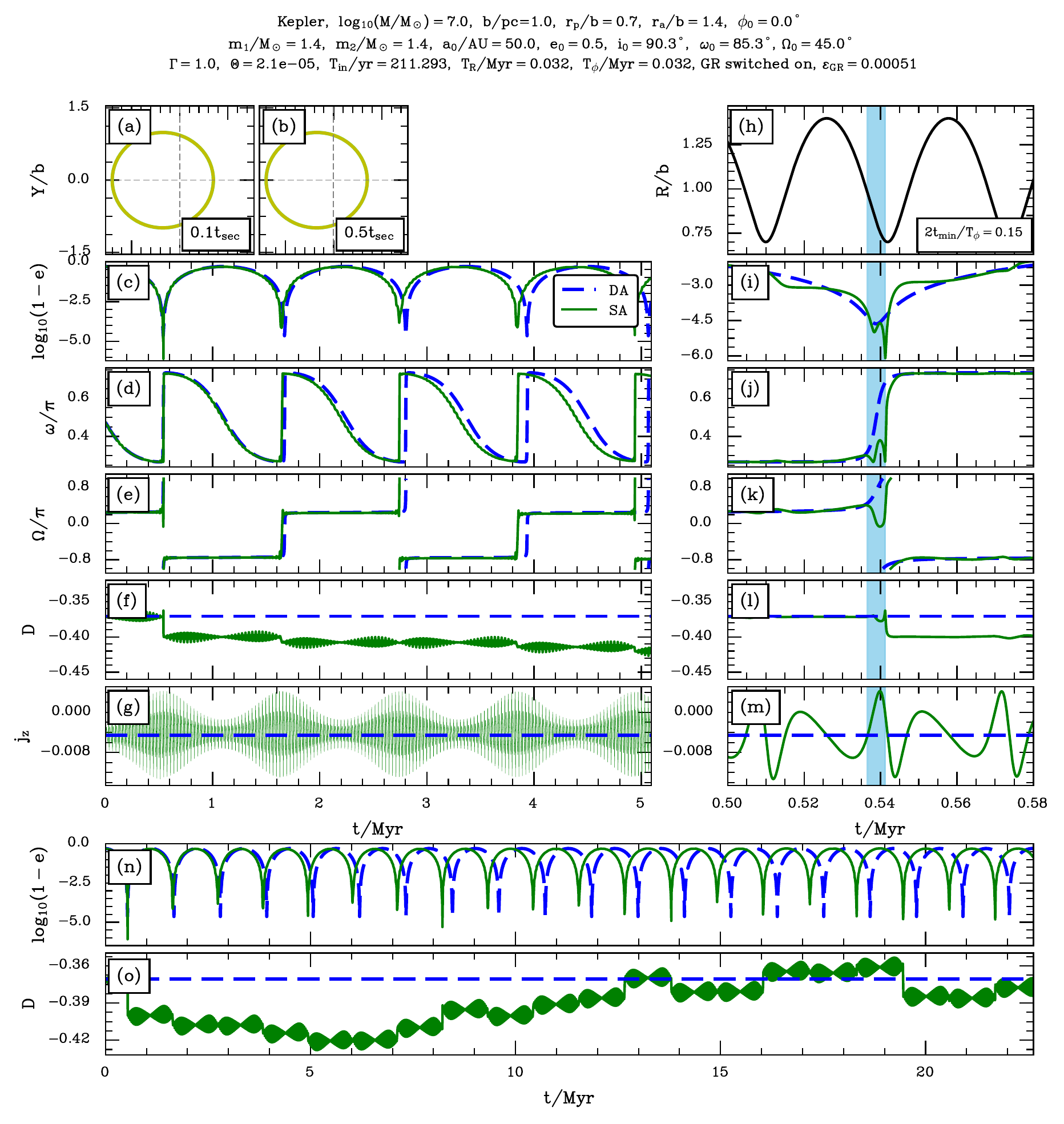}
      \caption{As in Figure \ref{fig:Example_Kepler} except with GR precession
 switched on.  This Figure shows that RPSD is present even in the test-particle quadrupole LK problem as long as GR is included.}
      \label{fig:Example_Kepler_GR}
   \end{figure*}

   Next, in Figure \ref{fig:Example_Kepler_GR} we perform the same
calculation as in Figure \ref{fig:Example_Kepler}, but now with GR precession switched on. This causes significant and
repeated diffusion of $D_\mathrm{SA}$ around the secular eccentricity peaks,
meaning the binary ultimately follows a completely different SA trajectory compared to the
one we would have predicted had we only used DA theory. This confirms that RPSD
is present as a phenomenon in LK theory {at the test-particle quadrupole
level} {as long as 1pN GR precession is included}.

\subsection{Physical interpretation of RPSD and quantitative analysis}
\label{sec:GR_physical_quantitative}

The aim of this section is
to provide some physical understanding {of RPSD} 
and to attempt
a rough quantitative analysis of this phenomenon.

\begin{figure*}
   \centering
   \includegraphics[width=0.99\linewidth]{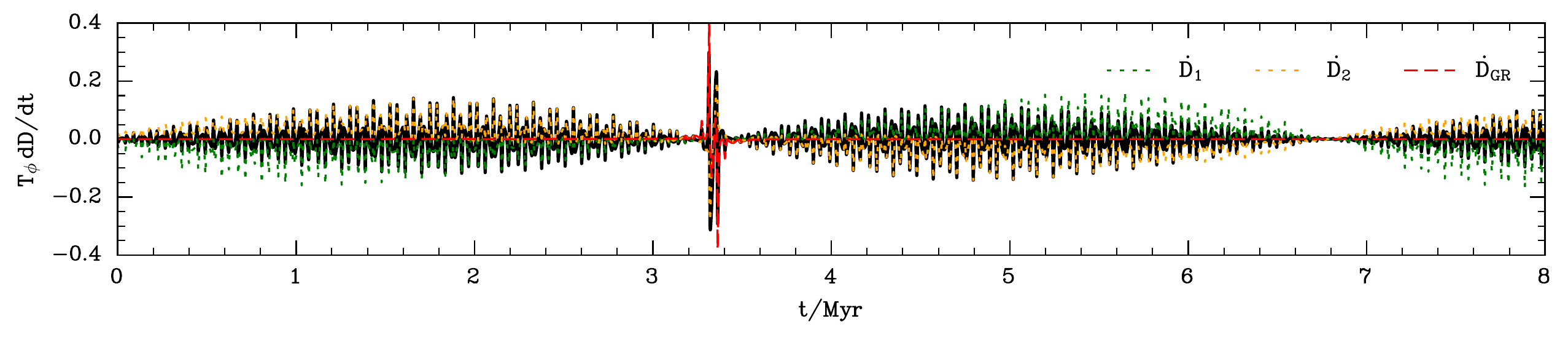}
   \caption{{Time series of $\md D/\md t$ (equation \eqref{eqn:dDdt_SA_withGR}) for the first $8$ Myr of evolution from Figure \ref{fig:Example_Hernquist_GR}; in other words, this is the counterpart of Figure \ref{fig:dDdtHernquist} with GR included. The green and orange dotted lines again show the contributions from {$\dot{D}_1$ and $\dot{D}_2$ (equations \eqref{eqn:Ddot_1}-\eqref{eqn:Ddot_2})}, 
   whereas the red dashed line shows \eqref{eqn:dDdtGR}; the black solid line is the total $\md D/\md t$.}}
   \label{fig:dDdtHernquist_GR}
   \end{figure*}
   
   \begin{figure}
   \centering
   \includegraphics[width=0.99\linewidth]{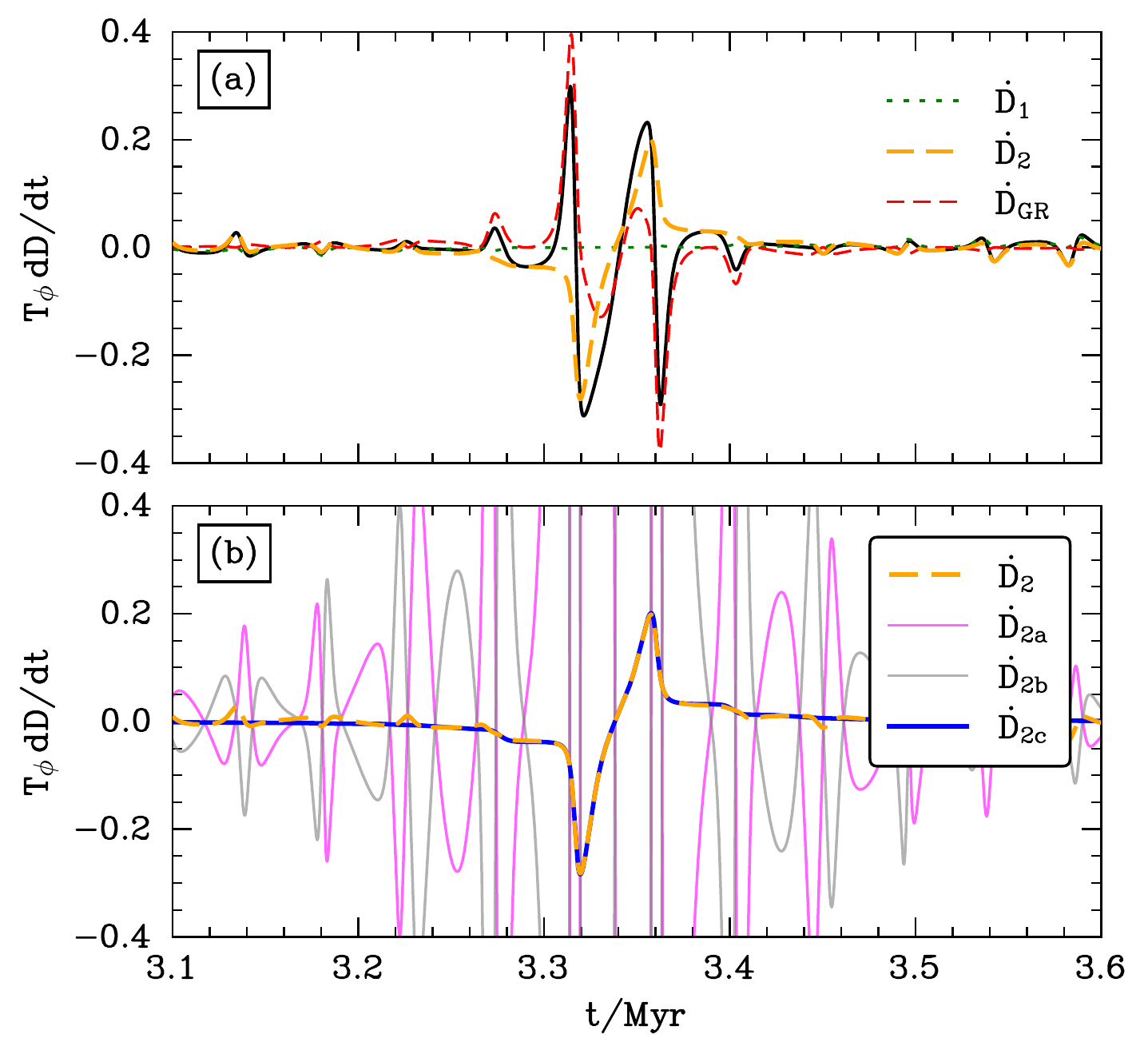}
   \caption{{Panel (a) is the same as in Figure \ref{fig:dDdtHernquist_GR}, but zoomed in around the first eccentricity peak. In panel (b) we ignore the red, green and black lines from panel (a), leaving only the contribution from the term $\dot{D}_2$ (orange), 
   which we decompose into its constituent parts \eqref{eqn:2nd_A}-\eqref{eqn:2nd_C}.
   The terms $\dot{D}_\mathrm{2a}$ and $\dot{D}_\mathrm{2b}$ have large amplitudes but cancel eachother almost perfectly at high eccentricity.}}
   \label{fig:dDdtZoom2abHernquist_GR}
   \end{figure}

   \begin{figure*}
   \centering
   \includegraphics[width=0.99\linewidth]{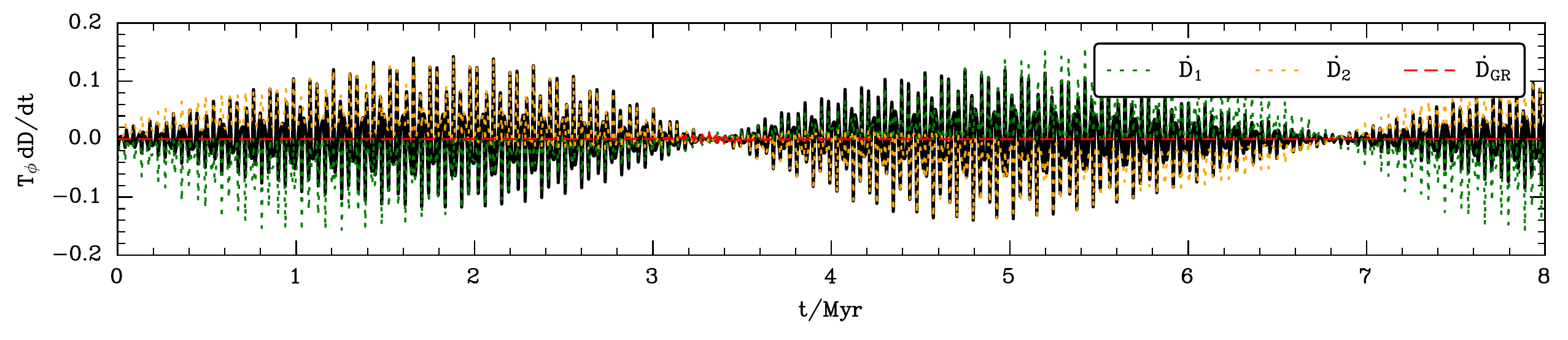}
   \caption{{As in Figure \ref{fig:dDdtHernquist_GR}, but using data from Figure 
   \ref{fig:Example_Hernquist_4_GR}, which does not exhibit RPSD.}}
   \label{fig:dDdtHernquist_4_GR}
   \end{figure*}
 
      \begin{figure}
   \centering
   \includegraphics[width=0.99\linewidth]{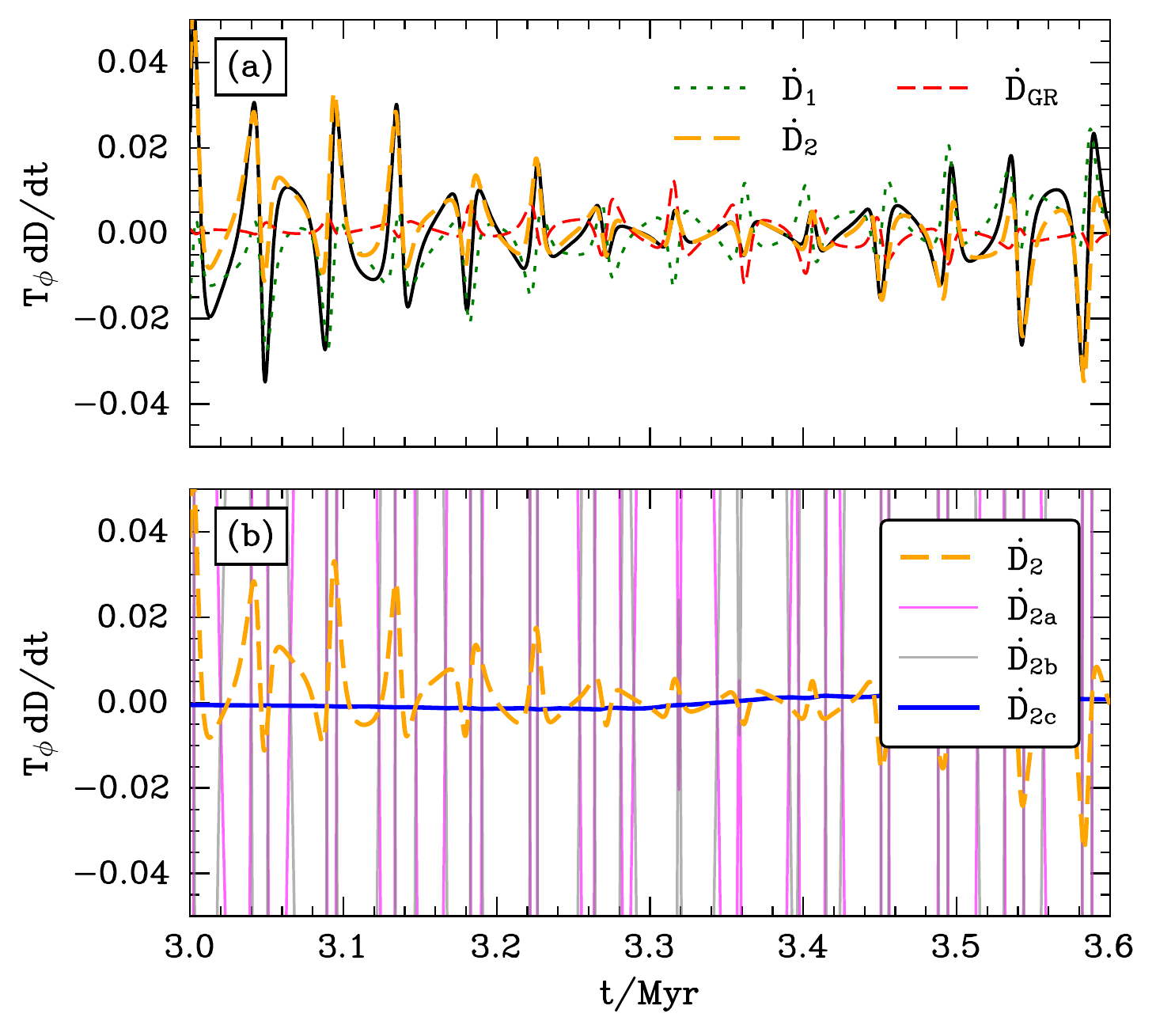}
   \caption{{Panel (a) is as in Figure \ref{fig:dDdtHernquist_4_GR}, but zoomed in around the first eccentricity peak. Panel (b) shows $\dot{D}_2$ (orange), and its constituent parts \eqref{eqn:2nd_A}-\eqref{eqn:2nd_C}.}}
   \label{fig:dDdtZoom2abHernquist_4_GR}
   \end{figure}

\subsubsection{Mechanism behind RPSD}
\label{sec:RPSD_Mechanism}

To understand why RPSD occurs, we take note of {two} pieces of {empirical} evidence
from the {above examples (which we confirmed in several additional
numerical experiments not shown here):
\begin{itemize}
    \item When GR precession is switched off, there is no RPSD.
    \item RPSD can occur when $2t_\mathrm{min}/T_\phi \lesssim 1$ but {we do not observe it to occur} when $2t_\mathrm{min}/T_\phi \gg 1$.
%\item There is no
%diffusion in the time-averaged value of $j_{z,\mathrm{SA}}$ to accompany the
%diffusion of $D_\mathrm{SA}$ (recall that $j_z$ and $D$ are both integrals of
%motion in the DA approximation).
\end{itemize}
Taking these strands of evidence together will allow us to identify the 
physical mechanism
behind RPSD, as we now explain.  }
 
 Whether we consider SA or DA dynamics, at eccentricities far from unity,
 significant changes in the orbital elements occur only on secular timescales,
 i.e. timescales much longer than $T_\phi$. Of course, $D$ (equation
 \eqref{eqn:def_D}) is a function of these orbital elements. Thus
 at eccentricities far from unity, $D_\mathrm{SA}(t)$ invariably exhibits
 relatively small and rapid (timescale $\sim T_\phi$) oscillations around the
 constant value $D_\mathrm{DA}$. However, at the highest eccentricities $e\to
 1$, significant changes in orbital elements can occur {on timescales much shorter than $\tsec$}.
 This is true even in DA theory: {for instance}, in
Appendix C of Paper III (which concerned high-$e$ behavior in the limit of weak, but nonzero, GR precession)
we saw several numerical examples of binaries
 whose $\omega_\mathrm{DA}$ value is turned through $\sim 90^\circ$ or more on
 the timescale $2t_\mathrm{min}$. Now, when $2t_\mathrm{min}/T_\phi \gg 1$ this {evolution}
 is still {slow from the point of view of the outer orbit}. But in the regime $2t_\mathrm{min}/T_\phi \lesssim 1$, the DA
 theory essentially predicts its own demise, since it tells us that $\mathcal{O}(1)$ relative changes
 in the orbital elements occur on the timescale of the outer orbit, contrary to
 the fundamental assumption of the {DA} approximation, {namely outer orbit-averaging} (Paper I). In that case, it is possible that the fluctuations of
 SA theory may no longer be considered small, and {would no longer} oscillate
 rapidly around the DA orbital elements while those DA elements change
 significantly.  

As an example, consider e.g. Figure
 \ref{fig:Example_Hernquist_GR}j, for which $2\tmin/T_\phi =0.42$.
 In this case $\omega_\mathrm{DA}$ undergoes a large `swing' on the timescale $\sim 2t_\mathrm{min}$
 near peak DA eccentricity.  This time
 range is so short that $\omega_\mathrm{SA}$ does not have a chance to {fluctuate} around $\omega_\mathrm{DA}$ during it. By contrast, consider Figure
 \ref{fig:Example_Hernquist_4_GR}j, for which $2\tmin/T_\phi
 =4.55$. In this instance $\omega_\mathrm{DA}$ undergoes a swing of very similar
 magnitude but on a much longer timescale, allowing $\omega_\mathrm{SA}$ to
 perform multiple small fluctuations { around $\omega_\mathrm{DA}$}  while the `swing' is in progress (i.e. within the blue
 range).
 
{
%This lack of timescale separation between $\tmin$ and $T_\phi$ affects
%$D_\mathrm{SA}$ as follows. 
Now let us incorporate GR precession into the discussion.
We know (see Figures \ref{fig:Example_Hernquist}, \ref{fig:Example_Kepler} and \ref{fig:Example_Plummer}) that when GR is switched off, the aforementioned
 short timescale fluctuations do not affect $D_\mathrm{SA}$
 very dramatically --- instead, $D_\mathrm{SA}$ just oscillates
 around the constant value $D_\mathrm{DA}$ indefinitely. Indeed, in the absence of GR,
$\md D_\mathrm{SA}/\md t$ is minimized at very high eccentricity, as we already showed for a particular example in Figure \ref{fig:dDdtHernquist}.
 When we \textit{do} include GR precession, $\md D_\mathrm{SA}/\md t$
 gets an additional contribution
 {arising from the last term in
 equation \eqref{eqn:def_D}: rather than \eqref{eqn:dDdt_SA} we now have
 %%%%%%%%%%%
\begin{align}
   \label{eqn:dDdt_SA_withGR}
   \frac{\md D}{\md t}  =  \dot{D}_\mathrm{1} +  \dot{D}_\mathrm{2} + 
   \dot{D}_\mathrm{GR},
\end{align}
%%%%%%%%%%%
with $\dot{D}_1$, $\dot{D}_2$ still given by \eqref{eqn:Ddot_1}-\eqref{eqn:Ddot_2}, and
%%%%%%%%%%%
\begin{align}
   \label{eqn:dDdtGR}
    \dot{D}_\mathrm{GR} \equiv \frac{\epsGR}{3(1-5\Gamma)}\frac{1}{j^2} \frac{\md j}{\md t}.
\end{align}
%%%%%%%%%%%
However, the extra term \eqref{eqn:dDdtGR}} cannot be directly responsible for RPSD, because when integrated across an eccentricity peak it {gives} zero.
(Equivalently, the third term in \eqref{eqn:def_D} is unchanged by passing through an 
eccentricity peak, since it only depends on the value of $j$).
Nevertheless, when coupled with the lack of timescale separation $2\tmin/T_\phi \lesssim 1$,
GR \textit{is} responsible for RPSD, as we will now demonstrate.

In Figure \ref{fig:dDdtHernquist_GR}a we plot $\md D/\md t$ (black line)
using data from the first $8$ Myr of evolution from Figure \ref{fig:Example_Hernquist_GR}
(c.f. Figure \ref{fig:dDdtHernquist}). The contributions from {$\dot{D}_1$ and $\dot{D}_2$ (equations \eqref{eqn:Ddot_1}-\eqref{eqn:Ddot_2})}
are overplotted with green and orange dotted lines respectively, while
the term $\dot{D}_\mathrm{GR}$ (equation \eqref{eqn:dDdtGR}) is shown with a red dashed line.
We see that contrary to Figure \ref{fig:dDdtHernquist}, there is a large spike in the signal at maximum eccentricity, around $t = 3.35$ Myr (recall that $t = 6.8$ Myr is an eccentricity \textit{minimum}).
In Figure \ref{fig:dDdtZoom2abHernquist_GR} we zoom in on this eccentricity peak, 
and find that $\dot{D}_1$ (green dotted line) 
contributes negligibly to $\md D/\md t$ at this time.
Since we just argued above that the GR term (red dashed line) cannot be responsible for RPSD, 
the total change in $D$ that occurs across the eccentricity peak 
must come from the integral over the orange curve, i.e. it must arise due to {$\dot{D}_2$},
which is proportional to $\md(\sin^2 i \sin^2 \omega)/\md t$.

Let us now dig even more deeply into this term.
{We write it as 
\begin{equation}
\dot{D}_2   = \dot{D}_{2\mathrm{a}} + \dot{D}_{2\mathrm{b}} + \dot{D}_{2\mathrm{c}},
\end{equation}
where
\begin{eqnarray}
   \dot{D}_{2\mathrm{a}} &\equiv& \frac{10\Gamma e^2}{1-5\Gamma}\sin 2i \, \frac{\md i}{\md t}\, \sin^2\omega \label{eqn:2nd_A} \\
    \dot{D}_{2\mathrm{b}} &\equiv& \frac{10\Gamma e^2}{1-5\Gamma}\sin^2 i \, \sin 2\omega \left[ \frac{\md \omega}{\md t} - \dot{\omega}_\mathrm{GR} \right], \\
    \dot{D}_{2\mathrm{c}} &\equiv& \frac{10\Gamma e^2}{1-5\Gamma}\sin^2 i \, \sin 2\omega \, \dot{\omega}_\mathrm{GR},
    \label{eqn:2nd_C}
\end{eqnarray}}
and
\begin{align}
    \dot{\omega}_\mathrm{GR} = \frac{AL^5 \epsGR}{8\mu^2 J^2} = \frac{A\epsGR T_\mathrm{in}}{16\pi j^2},
    \label{eqn:omegadot_GR}
\end{align}
is the direct contribution of GR precession 
(see equation \eqref{eqn:domegadt_SA}).
%Thus $\dot{D}_{2\mathrm{a}}$ depends on the rate of change of inclination $i$, while the other terms depend on the rate of change of $\omega$, and we have split them such that 
In Figure \ref{fig:dDdtZoom2abHernquist_GR}b we replot the orange dotted curve from Figure \ref{fig:dDdtZoom2abHernquist_4_GR}, but this time we also plot its 
component parts, $\dot{D}_{2\mathrm{a}}$, $\dot{D}_{2\mathrm{b}}$ and $\dot{D}_{2\mathrm{c}}$.
It is very striking 
that both contributions $\dot{D}_{2\mathrm{a}}$ and $\dot{D}_{2\mathrm{b}}$
have large amplitudes (they would individually give rise to fluctuations $\vert \Delta D\vert  \sim 5 $ on a timescale $T_\phi$), but that they almost perfectly cancel one another.  
{In fact, by 
using equations \eqref{eqn:domegadt_SA}, \eqref{eqn:dJdt_SA} and \eqref{eqn:dJzdt_SA}
it is possible to show that in the high-eccentricity limit $L \gg J \gtrsim J_z$ (the same limit we took in \S\ref{sec:scaling_highe}), the combination 
$\dot{D}_{2\mathrm{a}} + \dot{D}_{2\mathrm{b}} \to 0$.\footnote{{
More precisely, one finds
\begin{equation}
\dot{D}_{2\mathrm{a}} + \dot{D}_{2\mathrm{b}} = \sum_{\alpha\beta}\Phi_{\alpha\beta}(t) Q_{\alpha\beta}(\bw_\mathrm{SA}),
\end{equation}and each of the $Q_{\alpha\beta}(\bw_\mathrm{SA})$ individually vanish in the high eccentricity limit $j, j_z \to 0$, assuming nothing about $\omega, \Omega$ or the ratio $\cos i \equiv j_z/j$.
For instance, without any approximations we get
\begin{equation}
    Q_{zz} \propto (1-j^2) j \sin^2 2i \sin 2\omega \sin^2 \omega,
\end{equation}which obviously tends to zero at high eccentricity.
The other $Q_{\alpha\beta}$ are more complicated but all tend to zero in the limit $j, j_z\to 0$.}} 
As a result, {at high eccentricity} the 
 orange dotted line is comprised entirely of contributions from term $\dot{D}_{2\mathrm{c}}$ (equation \eqref{eqn:2nd_C}),
 i.e. the part driven directly by GR precession.} }

In Figures \ref{fig:dDdtHernquist_4_GR}-\ref{fig:dDdtZoom2abHernquist_4_GR} we repeat the exercise of 
Figures \ref{fig:dDdtHernquist_GR}-\ref{fig:dDdtZoom2abHernquist_GR}, except using
the data from Figure \ref{fig:Example_Hernquist_4_GR}, i.e. the example with $2t_\mathrm{min}/T_\phi = 4.55$
which does not exhibit RPSD.  This time there is no spike in any contribution to $\md D/\md t$ 
at high eccentricity. The contributions $\dot{D}_{2\mathrm{a}}$ and $\dot{D}_{2\mathrm{b}}$ 
cancel each other again as they must,
but this time there is no significant contribution from $\dot{D}_{2\mathrm{c}}$ either.

From these figures, the root cause of RPSD finally becomes clear.
In the absence of sufficiently rapid GR precession at high eccentricity, 
there is near perfect synchronicity between the 
evolution of the inclination $i_\mathrm{SA}$ and that of the argument of pericenter
$\omega_\mathrm{SA}$ built into the SA equations of motion.  
As a result, the factor $\sin^2i \sin^2\omega$ is essentially a constant during a high eccentricity episode.
But when GR gets switched on, it acts only on $\omega_\mathrm{SA}$ and not (directly) on $i_\mathrm{SA}$ --- instead $i_\mathrm{SA}$ can only `react' indirectly to the changes in {$\omega_\mathrm{SA}$}.  Thus there is a GR-driven lack of synchronization between {$\omega_\mathrm{SA}$} and $i_\mathrm{DA}$ on the timescale $T_\phi$, meaning
$\sin^2 i \, \sin^2 \omega$ is not precisely constant.
This `phase-lag' between $i$ and $\omega$ is instigated as the binary enters the eccentricity peak around $j\sim \jmin$, since this is where GR is most effective, and it is driven for a time $\sim 2\tmin$.
If $2\tmin/T_\phi \gg 1$ then the lack of synchronization will be negligible and there will be no RPSD.
But if $2\tmin/T_\phi \lesssim 1$, the GR-driven evolution of {$\omega$}
gives rise to RPSD before the value of {$i$} can catch up.
%where strictly we should evaluate all quantities at their SA values. 
%Moreover, we
%know that $\md j_\mathrm{SA}/\md t$  does not strongly depend on $j$ as the binary approaches
%very high eccentricity ({the doubly-averaged part of $j$ is stationary at maximum
%DA eccentricity by definition, while the fluctuating part $\delta j$ satisfies
%equation} \eqref{eqn:scalings}). As a result $\md
%D_\mathrm{SA}/\md t \propto \epsGR j^{-2}$ as $j\to 0$.  If $\epsGR \jmin^{-2}$
%is sufficiently large, this can easily lead to a fluctuation in $D$ around peak
%eccentricity that is far larger than the ones we saw when GR was switched off.
In terms of $D$, like we saw in Figure \ref{fig:dDdtHernquist_GR}l, the value of  
$D_\mathrm{SA}$ has no time to oscillate back and forth around its
`parent' $D_\mathrm{DA}$ value before the high eccentricity episode is over, so
that upon
emerging from the high-eccentricity blue stripe it `settles' on a new parent
$D_\mathrm{DA}$ value (compare this with Figure
\ref{fig:Example_Hernquist_4_GR}l). Equivalently, the binary
`jumps' to a new contour in the $(\omega,e)$ phase space while at high $e$ (still at
fixed $\Gamma$, $\Theta$ and $\epsGR$), aided by the fact that contours at high
$e$ are bunched so closely together ({Figure \ref{fig:Phase_Space_Hernquist_GR}}).

% In summary, the remarkable and non-intuitive truth is that in the weak GR regime,
% GR has essentially no effect on the DA dynamics, and yet modifies the SA
% dynamics --- and hence the real binary evolution --- fundamentally.
 
%Lastly, since GR does not directly drive any evolution in $j_z$, there is no RPSD-like phenomenon in the evolution of
%$j_{z,\mathrm{SA}}$.

 %Thus, cluster tides on their own do not drive sufficiently fast DA evolution
 %to provoke any $D_\mathrm{SA}$ diffusion by the mechanism described above.
 %Indeed, without GR, fluctuations in $D_\mathrm{SA}$ are minimised at maximum
 %eccentricity, as explained in \S\ref{sec:quantitative}. 

 \subsubsection{Criteria for producing significant RPSD}
 \label{sec:CriteriaRPSD}
 
Using {the above argument,
we can 
estimate the `jump' that $D_\mathrm{SA}$ sustains across the eccentricity peak, 
and in so doing find rough criteria for this jump to be significant.
We know that the jump in $D_\mathrm{SA}$ is equal to the time integral of \eqref{eqn:2nd_C} across the eccentricity peak, i.e. to the 
area under the black curve in Figure \ref{fig:dDdtZoom2abHernquist_GR}:
 \begin{align}
 D^\mathrm{jump} = \frac{10\Gamma}{1-5\Gamma} 
 \int \md t \, e^2 \sin^2 i\,\sin 2\omega \, \dot{\omega}_\mathrm{GR},
 \label{eqn:Djump_integral}
\end{align}
where $\dot{\omega}_\mathrm{GR}$ is given in equation \eqref{eqn:omegadot_GR}.
Here everything should have an `SA' suffix, and 
the width of the integration domain should be several $\tmin$,
centered on the eccentricity peak.
To get a characteristic value for the integral, we first approximate $e\approx 1$.
We also know that $\sin^2 i \, \sin 2\omega$ will be oscillating in a complicated way as the binary passes through the eccentricity peak, but as long as $2t_\mathrm{min}/T_\phi \lesssim 1$ these oscillations will not average out to zero.
To order of magnitude,
we therefore replace $\int \md t \sin^2 i \, \sin 2\omega\, j^{-2} \to 0.1 \tmin \jmin^{-2}$
for a very rough estimate.
This gives
%%%%%%%%%%%%%%%%%%%%%%%%%%%%%%%
\begin{align}
  \vert D^\mathrm{jump} \vert \sim  0.1 \times \Bigg\vert \frac{10\Gamma}{1-5\Gamma} \Bigg\vert \frac{\epsGR T_\mathrm{in} A \tmin}{16\pi \jmin^2},
  \label{eqn:Djump_integral_2}
\end{align}
%%%%%%%%%%%%%%%%%%%%%%%%%%%%%%%
{though the crudeness of our approxmations means that the prefactor $0.1$ is chosen somewhat arbitrarily.}
If we now recall that $A \sim 4\pi^2/T_\phi^2$ (Paper I),
and that $t_\mathrm{min} \sim
j_\mathrm{min} \tsec \sim \jmin T_\phi^2/T_\mathrm{in}$
(Paper II), 
we find
\begin{align}
 \vert  D^\mathrm{jump} \vert \sim 0.1 \times \Bigg\vert \frac{20\Gamma}{1-5\Gamma} \Bigg\vert  \epsGR \left( \frac{2\tmin}{T_\phi}\right)^{-1} \left(\frac{T_\phi}{T_\mathrm{in}}\right).
  \label{eqn:Djump_final}
\end{align}
For the examples of RPSD shown in Figures \ref{fig:Example_Hernquist_GR}, \ref{fig:Example_Plummer_GR}
and \ref{fig:Example_Kepler_GR}, equation \eqref{eqn:Djump_final} returns the values 
$\vert D^\mathrm{jump} \vert \sim 0.8 ,\,5$ and $0.3$ respectively\footnote{{Note that we use the values of $\tmin$ calculated at $t=0$,
before the binary has undergone any RPSD.  Once it has shifted to a new parent $D_\mathrm{DA}$ value,
this value will change (since both the 
minimum angular momentum $\jmin$ and the secular period $\tsec$ will have changed).}}.}

{Finally, we return to the 
assumption that $2t_\mathrm{min}/T_\phi \lesssim 1$.
Assuming weak GR so that $\jmin \sim
\Theta^{1/2}$, this  requirement 
may be recast as:}
%%%%%%%%%%%%
\begin{align}
        \Theta &\lesssim \left(\frac{T_\mathrm{in}}{T_\phi}\right)^2 
       \\
        &\lesssim 5 \times 10^{-5} \times
   \left( \frac{m_1+m_2}{2.8M_\odot}\right)^{-1}
  \left( \frac{a}{50 \mathrm{AU}}\right)^{3} \nn \\ &\times
  \left( \frac{\mathcal{M}}{10^7M_\odot}\right)^{1}
  \left( \frac{R}{1\mathrm{pc}}\right)^{-3},
   \label{eqn:Theta_condition_for_D_diff}
\end{align}
%%%%%%%%%%%
where we assumed a binary on a circular outer orbit of radius
 $R$ in a spherical cluster of mass $\mathcal{M}$, i.e. we put $T_\phi^2 \sim
 G\mathcal{M}/R^3$.
 For binaries that are not initially very eccentric,
 $\Theta \sim \cos^2 i_0$, meaning that
 that RPSD operates
 in the inclination window:
 %%%%%%%%%%%%%%%%%
 \begin{align}
    \vert \cos i_0 \vert  & \lesssim 0.007 \times
    \left( \frac{m_1+m_2}{2.8M_\odot}\right)^{-1/2}
   \left( \frac{a}{50 \mathrm{AU}}\right)^{3/2} \nn \\ &\times
   \left( \frac{\mathcal{M}}{10^7M_\odot}\right)^{1/2}
   \left( \frac{R}{1\mathrm{pc}}\right)^{-3/2}.
   \label{eqn:cosi0_RPSD}
 \end{align}
 %%%%%%%%%%%%%%%%%
 Throughout this paper we have considered clusters with mass $10^7M_\odot$,
 binaries with the mass of a NS-NS binary, $m_1=m_2=1.4M_\odot$, and outer
 orbits with semimajor axis $a_\mathrm{g} = 1$pc.  Putting $R \sim a_\mathrm{g}$
 with these numbers into equation \eqref{eqn:cosi0_RPSD} gives
 $\vert \cos i_0 \vert < 0.007$, which corresponds approximately to $i_0 \in
 (89.6^\circ, 90.4^\circ)$. This is concomitant with what we found numerically;
 all examples that exhibited RPSD had $i_0 = 90.3^\circ$, while the example for
 which there was no RPSD (Figure \ref{fig:Example_Hernquist_4_GR})
 had $i_0 = 93.3^\circ$. 

 {Moreover, the requirement that $2t_\mathrm{min}/T_\phi \lesssim 1$
 is essentially the same as the requirement that short timescale fluctuations are important at high eccentricity, $(\delta j)_\mathrm{max} \gtrsim \jmin$ --- see equation \eqref{eqn:SA_strength_estimate_2}.
 This leads us to the important conclusion that in situations with GR precession included, 
 \textit{if short-timescale fluctuations are dominating the high-eccentricity evolution then some level of RPSD is inevitable.}}
 We discuss the astrophysical implications of RPSD further in \S\ref{sec:AstroRPSD}.

{In conclusion, RPSD occurs if \eqref{eqn:cosi0_RPSD} is satisfied; if so, the resulting jumps in $D_\mathrm{SA}$ have a characteristic size \eqref{eqn:Djump_final}.
Obviously, if $\epsGR=0$ then there is no RPSD (i.e. $D^\mathrm{jump} = 0$ regardless of initial inclination).}

{We note here that although $j_z$ is a more physically transparent
quantity than $D$, and follows a simpler evolution equation, 
analyzing the evolution of $j_{z,\mathrm{SA}}$ did not give rise to any additional striking insights into RPSD.
Indeed, when  RPSD does occur,
sometimes the envelope of $j_{z,\mathrm{SA}}$
fluctuations undergoes abrupt shifts at eccentricity peaks (as in Figure \ref{fig:Example_Plummer_GR}g), but most of the time it does not (as in Figure \ref{fig:Example_Hernquist_GR}g). 
On the other hand, RPSD \textit{always} seems to coincide with a significant jump in $D$.  
We were not able to deduce any simple pattern relating jumps in $j_{z,\mathrm{SA}}$ to those in $D$ or any other quantity.}

\subsubsection{Statistical analysis of RPSD}
\label{sec:RPSD_Statistical}

We know from \S\ref{sec:phase_dependence} and \S\ref{sec:Phase_Angles_GR} that the details of the erratic high
eccentricity behavior in the presence of GR precession are highly 
dependent on the outer orbital phase, i.e. on the 
values of $R$ and $\phi-\Omega$ as the binary approaches the eccentricity peak.
This makes it very difficult to predict
the behavior of $D_\mathrm{SA}$ {precisely} for a given set of initial
conditions.  Thus, a natural next step is to investigate the
\textit{statistics} of jumps in $D_\mathrm{SA}$, by creating an ensemble of
$D^\mathrm{jump}$ values for the same binary on the same outer orbit, but set off with different initial values of $R$ and $\phi - \Omega$.

To do this, it is
necessary to first define more precisely what we mean by a `jump in $D$'.
This immediately raises some technical issues: (i) the value of $D_\mathrm{SA}$
is never actually fixed, and (ii) a steady-state approximation of
$D_\mathrm{SA}$ clearly breaks down as $e_\mathrm{SA}$ approaches unity. We
therefore choose to study time-averages of $D_\mathrm{SA}$ before and after the
eccentricity peak, taken over time intervals that 
do not include the peak itself, i.e. sufficiently far from the
peak for the averaged value to be meaningful\footnote{In practice, it is
sufficient to average $D$ over several outer orbital periods during a time
range corresponding to $e_\mathrm{SA} \notin (0.99,1)$. }.
In particular, before the first eccentricity peak this guarantees that the time average
$\langle
D_\mathrm{SA}\rangle$ coincides with the `parent' value
$D_\mathrm{DA}$. We then define the jump in $D$ across the peak to be
%%%%%%%%%%%%%%%%
\begin{align}
   \label{eqn:Djump_def}
    D^\mathrm{jump} \equiv \langle  D_\mathrm{SA}\rangle_\mathrm{after} - \langle  D_\mathrm{SA}\rangle_\mathrm{before}.
\end{align}
%%%%%%%%%%%%%%%%

\begin{figure*}
   \centering
\includegraphics[width=0.85\linewidth]{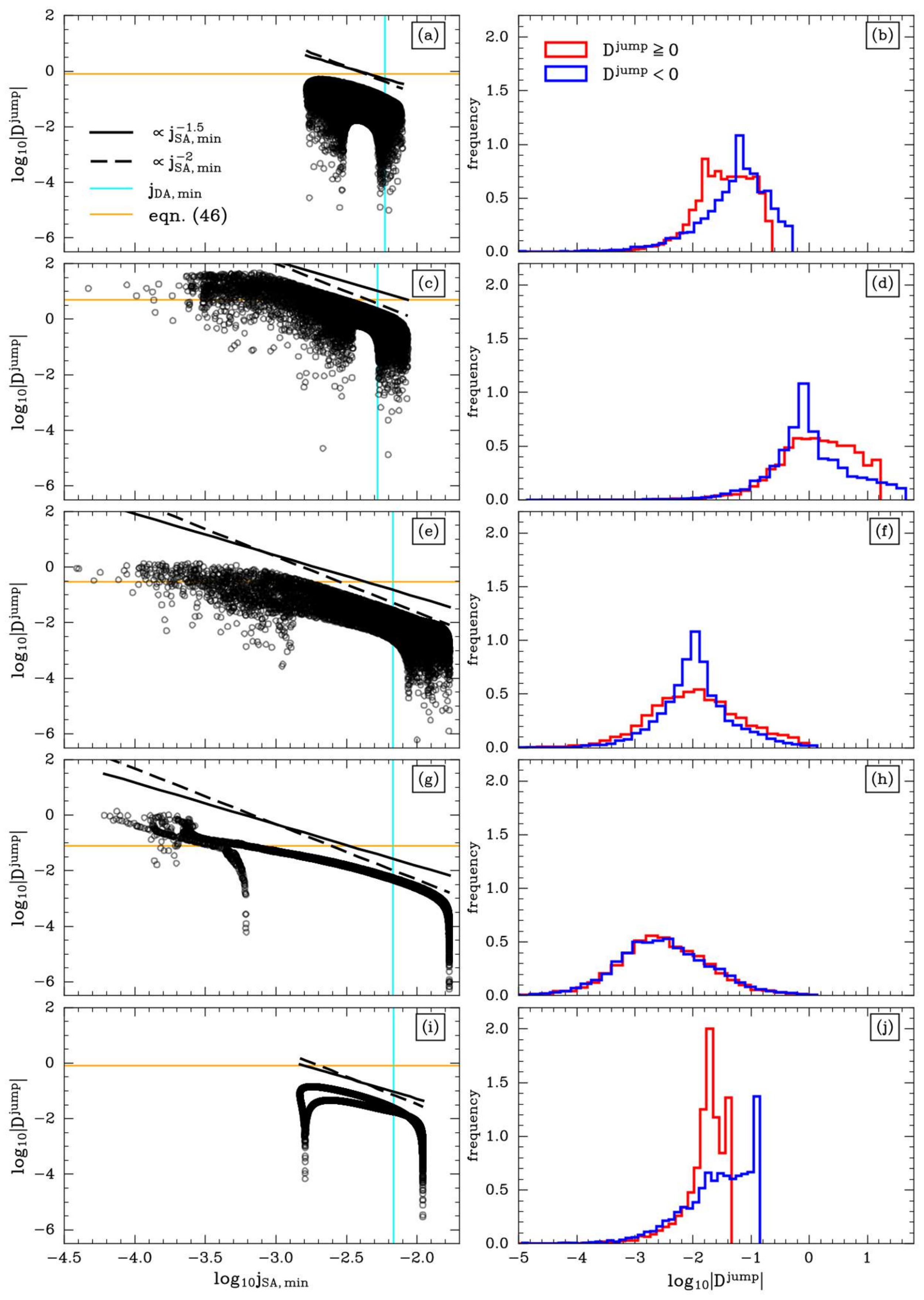}
   \caption{In panels (a)-(b) we rerun the first secular period from Figure
   \ref{fig:Example_Hernquist_GR}, except for $10^4$ randomly drawn outer orbital phases, i.e. values of $\phi - \Omega$ and $R$. Panel (a) shows {a scatter plot of}
   $\vert D^\mathrm{jump} \vert$ values from each run against the minimum $j$ value achieved
   in the SA calculation. Panel (b) shows a histogram of the $\vert
   D^\mathrm{jump} \vert$ values from panel (a). Note the initial DA value of $D$ here is $D_\mathrm{DA} = -1.036$.
   Panels (c)-(d) are the same except for the Plummer potential (c.f. Figure \ref{fig:Example_Plummer_GR}), with $D_\mathrm{DA} = 17.624$.
   Panels (e)-(f) are for the Kepler potential (c.f. Figure \ref{fig:Example_Kepler_GR}), with  $D_\mathrm{DA} = -0.371$. Panels (g)-(h) are again for the Kepler potential as in Figure \ref{fig:Example_Kepler_GR} except we change the outer orbit from $(\rp/b,\ra/b) = (0.7, 1.4) $ to $(\rp/b,\ra/b) = (0.7, 0.702) $.  Panels (i)-(j) are the same except this time 
   we change the outer orbit to $(\rp/b,\ra/b) = (1.4, 1.402) $.}
   \label{fig:Djump_Multipanel}
   \end{figure*}

In order to probe the statistics of $D^\mathrm{jump}$,
we {first} considered an ensemble of $10^4$ systems with exactly the same setup and
initial conditions as in Figure \ref{fig:Example_Hernquist_GR},
except in each case we drew a random initial value for $\phi - \Omega$ (uniformly in $(0, 2\pi)$) and a random initial {radial phase} (i.e. for fixed $\rp, \ra$ 
we chose at random the initial radius $R_0$,
correctly weighted by the time spent at each radius, i.e. $\md N
\propto \md R_0/v_R(\rp, \ra)$).  We stopped each integration at $t \approx t_\mathrm{sec}$, {i.e. after one full eccentricity oscillation, meaning  $D^\mathrm{jump}$
was well-established according to equation \eqref{eqn:Djump_def}.}

The results of this exercise are shown in Figure \ref{fig:Djump_Multipanel}a,b.
In panel (a) of this Figure we {present a scatter} {plot} of the values of $\vert D^\mathrm{jump} \vert$ against the minimum $j_\mathrm{SA}$
value that was achieved in the corresponding random realization. 
{With a vertical cyan line we show the DA prediction $j_\mathrm{DA,min}$,
and with a horizontal orange line we show the 
characteristic estimate \eqref{eqn:Djump_final}.}
%Red (blue) dots are for positive (negative) $D^\mathrm{jump}$.
%In Figure \ref{fig:Djumps_Hernquist}a we plot the resulting
%$D^\mathrm{jump}$ values against the minimum $j_\mathrm{SA}$ value achieved
%during the corresponding eccentricity peak. The dark red dots are for $\Omega_0
%= 0$, and they progress towards yellow and then green as $\Omega_0$ approaches
%$2\pi$. 
In panel (b) we marginalize over $j_\mathrm{SA,min}$ in order to produce a histogram of 
$D^\mathrm{jump}$ values, or rather two histograms, one for
positive $D^\mathrm{jump}$ (red) and one for negative $D^\mathrm{jump}$ (blue).
From these panels it is clear that 
there is 
no very simple dependence of $D^\mathrm{jump}$ on $j_\mathrm{SA,min}$,
nor is there necessarily any symmetry between the distribution of positive and negative $D^\mathrm{jump}$ values, nor does the distribution converge to any simple form. 
In fact, the value of $\vert D^\mathrm{jump} \vert$ varies over several orders of magnitude,
and equation \eqref{eqn:Djump_final} merely provides {an estimate (though often not a particularly accurate one)} of its maximum value.
On the other hand, the upper limit of the envelope of $D^\mathrm{jump}$ values
is fairly well-fit by a power law $\vert D^\mathrm{jump} \vert \propto j_{\mathrm{SA,min}}^{-1.5}$ {(see the black solid line in panel (a)).}

In the remaining panels of Figure \ref{fig:Djump_Multipanel} we repeat this exercise for several other examples known to exhibit RPSD.
In panels (c)-(d) and (e)-(f) respectively we change the potential to the Plummer potential (i.e. taking the initial conditions from Figure \ref{fig:Example_Plummer_GR}) and the Kepler potential (taking initial conditions from Figure \ref{fig:Example_Kepler_GR}).
Again we see that the $\vert D^\mathrm{jump} \vert$ upper envelopes are fairly well-fit by power laws
$\vert D^\mathrm{jump} \vert \propto j_{\mathrm{SA,min}}^{-1.5}$,
but aside from this no robust trend that can be pulled out.
Although the histograms of $D^\mathrm{jump}$ values in the Kepler case 
may seem like they have a promising symmetry (panel (f)), this also proves unreliable.
To show this, we ran two further examples in the Kepler potential, this time for nearly circular outer orbits.  Precisely, in panels (g)-(h) and (i)-(j) we use the same initial conditions as in Figure \ref{fig:Example_Kepler_GR} except we 
change 
the outer orbit's peri/apocenter from $(\rp/b,\ra/b) = (0.7, 1.4) $ to $(0.7, 0.702) $ and $(1.4, 1.402) $ respectively.  
In these cases, the tracks through $(j_\mathrm{SA,min}$, $D^\mathrm{jump})$ space approximately follow one-dimensional {curves} (panels (g) and (i)) 
rather than a broad two-dimensional {distribution}, due to the fact that we have removed a degree of freedom (the initial radial phase) by making the outer orbit near-circular.
Despite this, there seems to be no simple trend in the distribution of $D^\mathrm{jump}$ values (panels (h) and (j)).

In fact, we
plotted several such figures for several different sets of initial conditions,
cluster potentials, and so on, 
tried increasing the number of random realizations substantially, and
experimented with different ways of binning the values of $D^\mathrm{jump}$, but we were unable to uncover any striking insights.
{We also tried taking the Fourier transform of $\delta j(t)$
near the eccentricity peak and 
isolating the important frequencies that contribute to the signal, hoping that way to gain insight into RPSD ({the same exercise that we performed in panel (r) of Figures \ref{fig:Example_Hernquist}, \eqref{fig:Example_Hernquist_4} and \ref{fig:Example_Plummer}}).
Naturally, we found that these Fourier spectra are concentrated at frequencies $n_1\Omega_R + n_2\Omega_\phi$ for pairs of integers $n_1$, $n_2$ (here $\Omega_i \equiv 2\pi/T_i$ is the outer orbital frequency). 
Unfortunately, there were typically several important frequencies at play simultaneously, especially for 
outer orbits that were far from circular, and we did not extract anything useful from this effort.}

%%%%%%%%%%%%%%%%%%%%%%%%%%%%%%%%%%%%%%%%%%%%%%%%%%%%%%%%%%%%%%%%%%%%%%%%%%%%%%%%%%%%%%%%%%%%%%%%%%%%%%%%%%%%%%%%%%%%%%%%%%%%%%%%%%%%%%%%%%%%%%%%%%%%%%%%%%%%%%%%%%%%%%%%%%%%%%%%%%%%%%%%%%%%%%%%%%%%%%%%%%%%%%%%%%%%%%%%%%%%%%%%%%%%%%%%%%%%%%%%%%%%%%%%%%%%%%%%%%%%%%%%%%%%%%%%%%%%%%%%%%%%%%%%%%%%%%%%%%%%%%%%%%%%%%%%

%%%%%%%%%%%%%%%%%%%%%%%%%%%%%%%%%%%%%%%%%%%%%%%%%
\section{Discussion}
\label{sec:discussion}

%%%%%%%%%%%%%%%%%%%%%%%%%%%%%%%%%%%%%%%%%%%%%%%%%%%

\subsection{Astrophysical relevance of RPSD}
\label{sec:AstroRPSD}

Since we have uncovered a new effect in this paper the obvious question is:
how relevant is it to astrophysical systems? 
Let us assume that our compact object binary satisfies
$\epsGR \ll \epsweak$ and hence reaches very high eccentricity $\jmin \sim
\Theta^{1/2}$ (Paper III). Assume also that its initial eccentricity is not so large. Then
the important necessary condition for significant RPSD to occur is equation
\eqref{eqn:cosi0_RPSD}.
In 
\cite{Hamilton2019c} we considered compact object mergers driven by spherical cluster tides, so
 we will
use that as a test case.  There, $10^7M_\odot$ was the upper limit
on sensible cluster masses; $m_1=m_2=1.4M_\odot$ was the lower limit on
compact object masses; and $50\mathrm{AU}$ was the upper limit on any
sensible distribution of (still rather soft) binaries. Since $\cos i_0$ is
distributed uniformly $\in (0,1)$ for isotropically oriented binaries, we
conclude that RPSD would have affected much less than $1\%$ of our sample. Moreover,
this fraction will be even smaller when we consider e.g. BH-BH binaries with
$m_1=m_2=30M_\odot$.
%Moreover, those binaries that are most affected by it (i.e. those with $i_0
%\approx 90^\circ$) are those that were most likely to have merged anyway. 
Thus, we do not expect RPSD to be important for the bulk populations of binary
mergers that we considered in \cite{Hamilton2019c}.  

Nevertheless, it is easy to find numerical examples of compact object 
binaries orbiting in clusters for which RPSD {is a contributing effect}
\citep{Rasskazov2023}. 
In these cases the analytic description of secularly-driven inspirals developed in 
\cite{hamilton2022anatomy} breaks down completely, and the merger timescale can be wildly different (either longer or shorter) from that used in \cite{Hamilton2019c}.

%{One should not conclude from this that }

We speculate that RPSD may be important for certain exotic phenomena that
 involve even more extreme eccentricities, such as head-on collisions of white
 dwarfs in triple systems \citep{Katz2012}. Note that RPSD occurs even for
 circular outer orbits and for equal mass binaries, i.e. in triples where the
 octupole contributions to the potential are very small, which are not usually
 considered promising for producing high merger rates. 
 {However, further analysis of this possibility will require incorporating the effects 
 GW emission, which we have neglected throughout this paper.
This is left as an avenue for future work.}

\subsection{Breakdown of the SA approximation}
\label{sec:SA_breakdown}

\begin{figure*}
   \centering
   \includegraphics[width=0.99\linewidth]{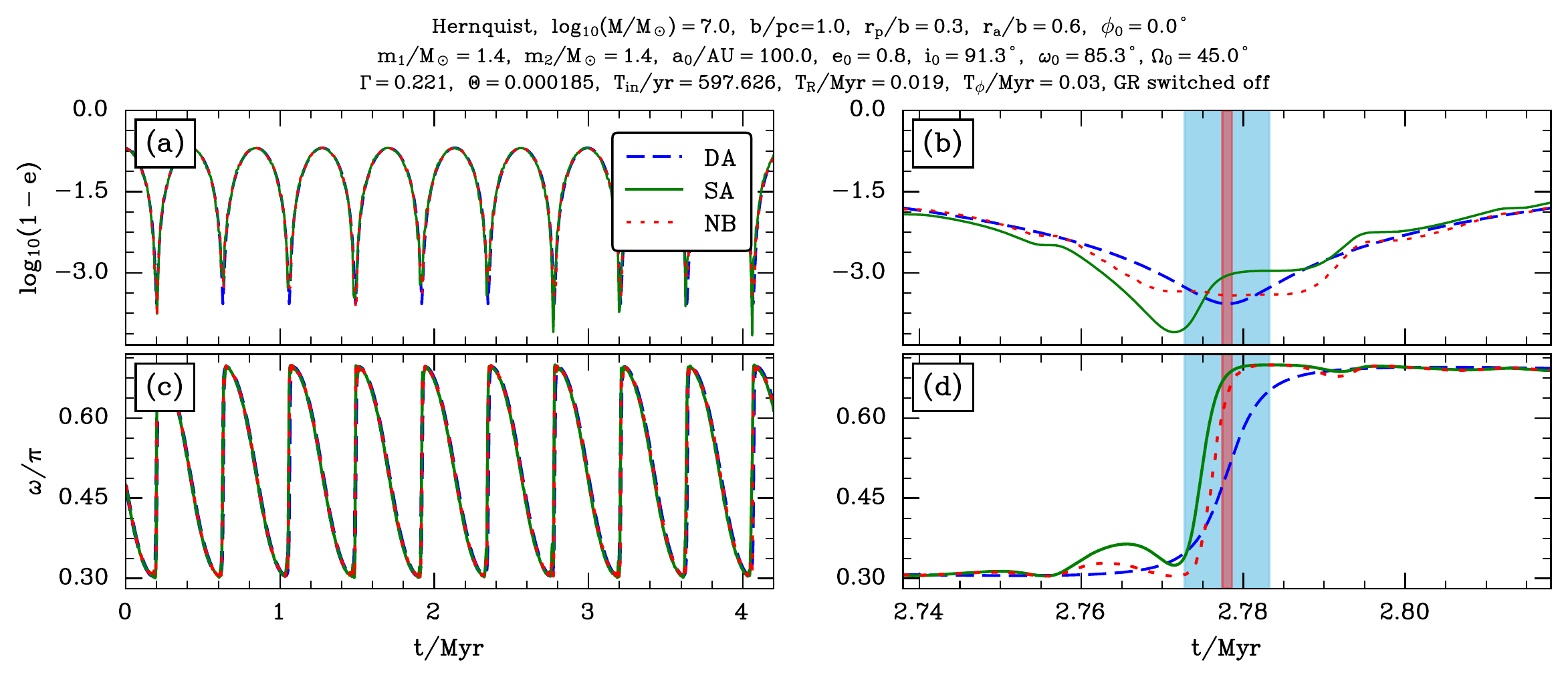}
   \caption{Example
   of SA and {direct (`N-body')} integrations disagreeing at high eccentricity.  In the right column
   we zoom in to the seventh eccentricity peak, where the two approaches for
   calculating $1-e$ disagree by as much as $\sim 0.5$ dex. 
   {The blue band in the right hand panels spans $2\tmin$ while the 
   red band spans $2T_\mathrm{in}$.}
   %Here we took $T_\phi/\Delta t
   %= 5000$ to be sure the results had converged. 
   {Note that we started the outer orbit at apocenter, rather than pericenter, in this example.}}
   \label{fig:SA_Validity_B_5000}
   \end{figure*}

In this paper we have considered the SA dynamics of binaries driven by cluster
tides, particularly at very high eccentricity. We have implicitly assumed that
these equations are accurate. It is worth noting, however, that even the SA
approximation can itself break down, in situations where any of $(\omega, J, \Omega,
J_z)$ evolve significantly on the timescale of the inner orbital period
\citep{Antonini2014}.  Such a situation is shown in Figure
\ref{fig:SA_Validity_B_5000}. In this figure we consider a soft NS-NS binary with $T_\mathrm{in} \sim 600$ yr,
that reaches $1 - e \sim 10^{-4}$ in the Hernquist potential.  (We do not switch on GR
precession in this example, to make it clear that the breakdown of the SA approximation has nothing to do with GR effects). 
%We ran this integration with very high resolution, in particular
%employing a very large number of timesteps for the outer orbit
%integration, $T_\phi/\Delta t =5000$. 
{The blue band in the right hand panels spans $2\tmin$ while the 
   red band spans $2T_\mathrm{in}$.}
{We see that}, by the time of the 7th
eccentricity peak, the SA {result deviates significantly from that found with direct orbit integration}, with $1-e$ differing by up to $\sim 0.5$ dex.

Note
that the binary component masses are equal in this example, $m_1=m_2$, so any
octupole terms in the tidal force ought to be zero, meaning any corrections are
of hexadecapole or higher order. Nevertheless, we checked that the disagreement
is really due to a breakdown in the orbit-averaging approximation, rather than
these higher order terms, by running a {direct orbit integration using the N-body code \texttt{REBOUND}, as in earlier examples, except with the perturbing potential $\Phi(\mathbf{R}_\mathrm{g}(t))$ truncated at 
quadrupole order}. The agreement with the full, untruncated {direct orbit integration} was
excellent.

The SA approximation is based on the assumption that $T_\mathrm{in}$ is small compared to any other timescale of interest. 
Thus, the breakdown of the SA approximation occurs if the binary orbital elements, particularly $\omega$ and $\Omega$, change sufficiently rapidly near peak eccentricity, which occurs if $\tmin$
{is not sufficiently large compared to} $ T_\mathrm{in}$.
For instance, in Figure \ref{fig:SA_Validity_B_5000}, $\omega$ changes by $\sim \pi$ on
a timescale of $\sim 5000$ yr.  Given $T_\mathrm{in} \approx 600$ yr, this
corresponds to a rate $\dot{\omega} \sim 20^\circ$ per inner orbital period.
{Such a large value is really an indication that the
orbit is not truly Keplerian, and thus that the Keplerian orbital elements themselves are not particularly well-defined;
it is perhaps unsurprising that in this regime even the SA approximation breaks
down.}
%Comparing this 
%{Let us roughly stipulate that the SA approximation breaks down if
%$\tmin < 10 T_\mathrm{in}$.
%The using the same estimate 
%$\tmin \sim \cos i_0 T_\phi^2/T_\mathrm{in}$ as led to equation \eqref{eqn:SA_strength_estimate_2}}, we %find that the SA approximation breaks down if $\vert \cos i_0 \vert \lesssim %(T_\mathrm{\phi}/T_\mathrm{in})^2$, that is if (c.f. equation \eqref{eqn:cosi0_RPSD}):%
% \begin{align}
%    \vert \cos i_0 \vert  & \lesssim 10^{-4} \times
%    \left( \frac{m_1+m_2}{2.8M_\odot}\right)^{-1}
%   \left( \frac{a}{50 \mathrm{AU}}\right)^{3} \nn \\ &\times
%   \left( \frac{\mathcal{M}}{10^7M_\odot}\right)^{1}
%   \left( \frac{R}{1\mathrm{pc}}\right)^{-3}.
%   \label{eqn:cosi0_SA_breakdown}
% \end{align}

\subsection{Numerical accuracy}

{We mention here that in the presence of RPSD, 
the SA numerical integrations become very expensive, because a very tiny timestep is required in order to get results that are robust over many secular periods. Moreover, any lack of convergence often does not become apparent until late in the integration, when RPSD happens to kick a binary to extremely high eccentricity. For instance, when running the calculation shown in Figure \ref{fig:Example_Plummer_GR}, using a slightly larger timestep for the SA integration gave indistinguishable results for the first $\sim 85$ Myr, but during the subsequent high-eccentricity spike there was a noticeable difference in the behavior, after which the two calculations diverged completely.
The reason for this divergence is that even tiny errors in the numerical scheme will always accumulate over very long timescales, and will then tend to be amplified during a very high eccentricity episode.}

{Essentially the same conclusions were recently reached by \cite{Rasskazov2023}, who found that because of RPSD the cluster-tide-driven evolution of highly eccentric binaries in the SA approximation (including both GR precession and GW emission) is extremely sensitive to the way in which the cluster tide is computed along the outer orbit.}

{The difficulty of performing accurate numerical integrations 
means that for practical purposes, even test-particle, quadrupolar SA evolution may be best thought of as a stochastic process.}

\subsection{Relation to LK literature}
\label{sec:literature}
    
The question of how short-timescale fluctuations affect high eccentricity
evolution in LK theory was first addressed by \citet{Ivanov2005}, who estimated the amplitude of the
angular momentum fluctuations experienced near maximum eccentricity by a binary
undergoing LK oscillations. 
\citet{Ivanov2005}'s result and other results very similar to it
have since been used extensively for modelling hierarchical triples (e.g.
\citealt{Katz2012,Bode2014,Antognini2014,Silsbee2017,Grishin2018}). 
Their high-$e$ fluctuating torque formula is a special case of our equation \eqref{SAFluctuations_dJdtfinal}.
To evaluate this formula, \cite{Ivanov2005} took the outer orbit to be circular; {hence} our analysis in \S\ref{sec:deltaj_circular_approx} encompasses theirs as a special case.  On the other hand, in a Keplerian potential the circular outer orbit approximation is not necessary; \citet{Haim2018} generalized the result of \cite{Ivanov2005} to eccentric outer orbits.  No such generalization is possible for arbitrary spherical cluster potentials.

More recently, as we mentioned in \S\ref{sec:phase_dependence}, \citet{Luo2016} took a perturbative approach to the SA LK problem for
arbitrary inner and outer eccentricities. 
They showed that short-timescale fluctuations captured by the
SA equations of motion can accumulate over many secular periods, resulting in
secular evolution that does not resemble the original DA prediction.  In other
words, the time-averaged SA solution does not agree with the DA solution, but
instead diverges from it gradually {(see \citealt{tremaine2023hamiltonian}
for a more rigorous mathematical treatment of this phenomenon)}.
This divergence occurs in a predictable way, and \cite{Luo2016} were able to derive a `corrected DA' Hamiltonian (essentially a ponderomotive potential, see \citealt{tremaine2023hamiltonian}) that accurately reproduces the time-averaged SA dynamics. 
%(It is worth noting that in their calculations \cite{Luo2016} implicitly worked in the limit 
%$2\tmin/T_\phi \gg 1$, because they needed to `freeze' the DA solution on the outer orbital timescale).
Although \cite{Luo2016} only considered the LK problem up to quadrupolar order in the tidal expansion, their results were generalized to arbitrary (octupole, hexadecapole, ...) order by \citet{Lei2018}, and then further extended to include fluctuations on the inner orbital timescale $\sim
T_\mathrm{in}$ by \citet{Lei2019}.
%To rectify this they calculated the effect of the accumulated SA fluctuations
%and derived a corrected DA Hamiltonian that accounted for them on average.
%However, they also assumed that the doubly-averaged quantities could be frozen
%on an outer orbital timescale, although their approach was significantly more
%advanced than that of \citet{Ivanov2005}.  
Moreover, \citet{Grishin2018} applied the formalism of \citet{Luo2016} to high eccentricity
behavior {(see also \citealt{Mangipudi2022})}.  Assuming a circular outer orbit they calculated the new maximum
eccentricity arising from \citet{Luo2016}'s `corrected' secular theory, as well
as the magnitude of angular momentum fluctuations at highest eccentricity.
Though we have not done so here, in the special case of circular outer orbits
the results of \citet{Luo2016} and \citet{Grishin2018} could be trivially extended
to arbitrary axisymmetric cluster potentials of the sort considered in this
paper. %Unfortunately, for non-circular outer orbits we lack an analytic
%prescription for $\Rg(t)$ (Paper I), which renders the calculation
%intractable.  

However, in the key papers mentioned above
\citep{Ivanov2005,Luo2016,Grishin2018}, GR precession was not directly
included when calculating the fluctuating behavior at high eccentricity. Those
authors also all implicitly assumed the timescale separation $2t_\mathrm{min}/T_\phi \gg 1$,
allowing them to freeze the time-averaged values of $(\omega,J,\Omega,J_z)$ on the timescale $T_\phi$ while they calculated the fluctuations. Our
work is different in that we have included GR precession and, crucially, 
investigated systems
with $2t_\mathrm{min} / T_\phi \lesssim 1$.  In this case, the RPSD effect that
we have uncovered means that SA dynamics do not converge to the original DA
prediction on average, just as was found by \citet{Luo2016} in the non-GR LK
theory.  However, unlike \citet{Luo2016}'s discovery, RPSD depends critically on
the strength of GR precession and also happens very rapidly (on a
timescale $t_\mathrm{min} \lesssim 2T_\phi$) rather than accumulating over many secular
periods.

It is also worth contrasting the behavior found in this paper with that from \citet{hamilton2022anatomy}. In that case, we found that the secular behavior of 
binary orbital elements 
changed over time due to bursts of GW emission at eccentricity peaks, in a purely DA framework --- 
we did not include short-timescale fluctuations.
By contrast, in the present paper we have found secular behavior that changes over time due to bursts of RPSD at high eccentricity, which is driven by short-timescale fluctuations and GR precession, 
but we have not included GW emission.  Thus, although we motivated our study by using parameters typical of compact object binaries, in our case the binaries will never merge since there is no dissipation of energy.
{Since GW emission occurs at a higher (2.5pN) post-Newtonian order than the (1pN) GR precession considered here, we do not expect that it will strongly affect the dynamics of \textit{individual} RPSD episodes, unless one of those episodes happens to kick the binary into what we called the `strong GR regime' in \cite{hamilton2022anatomy}. However, GW emission can certainly make a large difference
cumulatively over many secular cycles}, {as has been recently confirmed by \cite{Rasskazov2023}.}

Finally, we have considered only quadrupole terms in the tidal potential,
i.e. we have considered an SA problem whose DA counterpart is completely integrable {(though we note that our {direct orbit} integrations required no tidal approximation and still provide an excellent match to the SA results)}.
The octupole
terms (see Appendix E of Paper I) are very small in most cases we consider, since $a/R$ is very small in applications to binaries in clusters (e.g. \citealt{Hamilton2019a}). In
fact, in all the numerical examples shown in this paper the octupole terms are identically zero,
since we always took the binary components to have equal masses $m_1=m_2$.  In LK theory,
octupole and higher-order terms are expected to become important when the outer
orbit is significantly eccentric and the component masses are not equal, and can
lead to {non-integrability and hence chaos, both in the SA and DA approximations} \citep{Li2015}.  We have shown
that {in the SA approximation}, 
quasi-chaotic phase space behavior can arise even at the pure test particle quadrupole level via RPSD,
provided 1PN GR precession is included.
Indeed, perhaps the reason that quadrupole-level RPSD has not been mentioned before in the LK literature is that 
the majority of numerical integrations of the LK equations with GR \textit{do} include octupolar or higher order terms and/or relax the test-particle approximation,
and any {complicated} behavior that results is then attributed to these effects.
%On the contrary, we have demonstrated here that they are present even at the quadrupolar level.

%%%%%%%%%%%%%%%%%%%%%%%%%%%%%%%%%%%%%%%%%%%%%%%%%
\section{Summary}
\label{sec:summary}

In this paper we have investigated the role of short-timescale fluctuations upon (test particle quadrupole) tide-driven
evolution of binary systems, particularly with
regard to their high eccentricity behavior. 
Our main results can be summarized as
follows.

\begin{itemize}

\item We analyzed the behavior of fluctuations of binary orbital elements in the singly-averaged (SA) approximation, in the absence of 1pN GR apsidal precession.  In particular, we derived an expression for
    the magnitude of angular momentum fluctuations
    at high eccentricity for binaries orbiting in arbitrary spherically symmetric cluster potentials.
    Roughly, these fluctuations are comparable in magnitude 
    to the minimum angular momentum predicted by doubly-averaged (DA) theory whenever the cosine of the initial inclination is comparable to or smaller than the ratio of inner and outer orbital periods (equation \eqref{eqn:SA_strength_estimate}).

\item We then investigated the high eccentricity SA behavior including 1pN GR precession, and found that the evolution can be dramatically different {from the case without GR.  This can be true even in the weak GR regime (Paper III), where GR makes negligible difference to the DA dynamics}.  In particular, relativistic phase space diffusion
   (RPSD) may kick the binary to a new phase space contour on the timescale of the outer orbit, potentially
   leading to quasi-chaotic evolution and extreme eccentricities, and a full breakdown of the naive DA theory. The rough criterion for RPSD to occur is essentially the same as for angular momentum fluctuations to be comparable to the DA minimum angular momentum (equation \eqref{eqn:cosi0_RPSD}), with the additional requirement that GR precession be switched on.
   The size of the typical RPSD kick
   is proportional to the dimensionless strength of GR precession $\epsGR$.
   
\item   RPSD likely affects only a very small fraction of binaries in population synthesis studies of LIGO/Virgo gravitational wave sources, but may be a crucial ingredient for e.g. head-on collisions of white dwarfs.

\end{itemize}

%%%%%%%%%%%%%%%%%
%%%%%%%%%%%%%%%%%%%%%%%%%%%%%%%%%%%%%%%%%%%

\appendix

%\section{Doubly-averaged equations of motion}
%\label{sec:DA_equations}

%The doubly-averaged tidal Hamiltonian $H_\mathrm{1,DA}$ is given by
%\eqref{eqn:H1_Doubly_Averaged}. The Hamiltonian $H_\mathrm{GR}$ encoding GR
%precession is given in \eqref{eqn:HGR}. The DA equations of
%motion that arise from the total DA Hamiltonian $H_\mathrm{DA} = H_\mathrm{1,DA}
%+ H_\mathrm{GR}$ are
%\begin{align} 
%\label{eqn:domegadt_DA}
%    \frac{\md \omega}{\md t} &= \frac{\partial H_\mathrm{DA}}{\partial J} =  \frac{6C}{L^2}\frac{[5\Gamma L^2  J_z^2 - J^4 +5\Gamma(J^4-L^2J_z^2)\cos2\omega]}{J^3} + \frac{CL \epsGR}{J^2}, \\
    %%%%%%%%%%%%%%
%    \label{eqn:dJdt_DA}
%    \frac{\md J}{\md t} &= -\frac{\partial H_\mathrm{DA}}{\partial \omega} = -\frac{30 C\Gamma}{L^2} \frac{(J^2-J_z^2)(L^2-J^2)}{J^2}\sin 2\omega, \\
%%%%%%%%%%%%%%
%      \frac{\md \Omega}{\md t} &= \frac{\partial H_\mathrm{DA}}{\partial J_z} = \frac{-6C\Gamma}{L^2} \frac{[5 L^2-3J^2-5\cos 2\omega (L^2-J^2)]J_z}{J^2}, 
%      \label{eqn:dOmegadt_DA}
%      \\
 %%%%%%%%%%%%%%%
%     \frac{\md J_z}{\md t} &= -\frac{\partial H_\mathrm{DA}}{\partial \Omega} = 0.
%     \label{eqn:dJzdt_DA}
%\end{align}
%%%%%%%
%%%%%%%%%%%%%%%%%%%%%%%%%%%%%%%%%%%%%%%%%%%%%%%%%%%%%%%%%%%%%%%%%%%%%%%%%%%%%%%%%%%%%%%%%%%%%%%%%%%%%%%%%%%%%%%%%%%%%%%%%%%%%%%%%%%%%%%%%%%%%%%%%%%%%%%%%%%%%%%%%%%%%%%%

\section{Singly-averaged equations of motion in the test-particle, quadrupole limit for arbitrary cluster potentials}
\label{sec:SA_equations}

The full SA Hamiltonian is $H_\mathrm{SA}=H_\mathrm{1,SA} + H_\mathrm{GR}$ where $H_\mathrm{GR}$ is given by 
\eqref{eqn:HGR} and $H_{1,\mathrm{SA}}$ is given by \eqref{eqn:H1SA}.
The corresponding SA equations of motion are
%%%%%%%%%%%%%%%%%%%%%%%%%%%%
%%%%%%% domega/dt %%%%%%%%%%%%%%
%%%%%%%%%%%%%%%%%%%%%%%%%%%%
%\begin{footnotesize}
\begin{align}
\frac{\md \omega}{\md t} \nn   = &\frac{\partial H_\mathrm{1,SA}}{\partial J} + \frac{\partial H_\mathrm{GR}}{\partial J} \\
  \nn =&[L^2/(8J^4 \mu^2  )] \times
%%%%%%%%%%%%%%%%%%%%
\\
 & \nn  
 \Bigg\{ (\Phi_{xx}+\Phi_{yy}) \Big[- J^5 (3+5\cos 2\omega) - 5JJ_z^2L^2(1-\cos 2\omega) \Big]
     %%%%%%%%%%%%%%%%%%%%%%%%%%%%
     \\ 
 & \nn  
 + (\Phi_{xx}-\Phi_{yy}) \Big[- J^5 (3+5\cos 2\omega) \cos 2\Omega 
 + 5JJ_z^2L^2(1-\cos 2\omega)\cos 2\Omega 
+ 5J^2J_z(J^2+L^2)\sin2\Omega \sin2\omega \Big] \nn
%%%%%%%%%%%%%%%%%%%%
 \\
& \nn +
\Phi_{zz} \Big[-6 J^5  + 
    10 J J_z^2  L^2 + 
    10 (J^5 - J J_z^2  L^2)  \cos 2 \omega \Big] 
    %%%%%%%%%%%%%%%%%%%%%
 \\
 & \nn + 
 \Phi_{xy} \Big[-10 J^2 J_z( J^2 + L^2) 
      \cos 2 \Omega \sin 2 \omega
-J^5 \sin 2\Omega(6+10\cos2\omega) 
+ 10JJ_z^2L^2\sin 2\Omega (1-\cos 2\omega)
\Big] 
        %%%%%%%%%%%%%%
 \\
 & \nn  + 
 \Phi_{xz} (1 - J_z^2/J^2)^{-1/2} \Big[J_z (6 J^4 + 
    10 J^2 L^2 
 - 20 J_z^2 L^2 )\sin\Omega \\ \nn
 &+ J (-20 J^4 + 10 J^2 J_z^2 + 10 J_z^2 L^2)  \sin 2\omega \cos\Omega
       +  J_z (- 
    10 J^4  - 
    10 J^2 L^2   + 
    20 J_z^2 L^2 )\cos 2 \omega \sin\Omega\Big] 
      %%%%%%%%%%%%%%%%%%%%%%%%%%%%
\\
 & \nn    + 
 \Phi_{yz} (1 - J_z^2/J^2)^{-1/2} \Big[- J_z (6 J^4 + 10 J^2 L^2 - 20 J_z^2 L^2) \cos \Omega 
  \\ &+  J(- 
    20 J^4  + 
    10 J^2 J_z^2  + 
    10 J_z^2 L^2)\sin 2\omega \sin\Omega
 - 
    J_z (-10 J^4 - 10 J^2 L^2 + 20 J_z^2 L^2) \cos 2 \omega \cos\Omega 
\Big] 
    \Bigg\} + \frac{AL^5 \epsGR}{8\mu^2 J^2},
       %%%%%%%%%%%%%%%%%%%%
       \label{eqn:domegadt_SA}
\end{align}
%%%%%%%%%%%%%%%%%%%%   
%\end{footnotesize}

%%%%%%%%%%%%%%%%%%%%%%%%%%%%
%%%%%%% dJ/dt %%%%%%%%%%%%%%
%%%%%%%%%%%%%%%%%%%%%%%%%%%%
\begin{align}
\label{eqn:dJdt_SA}
  \nn  \frac{\md J}{\md t} =& -\frac{\partial H_\mathrm{1,SA}}{\partial \omega} \\ =&
    \nn [5L^2/(2 J^2\mu^2)](J^2 - L^2) \times
   \\
        & \nn \Bigg\{- 0.25 (\Phi_{xx}+\Phi_{yy})\Big[(J^2 - J_z^2) \sin2\omega \Big] - 0.25(\Phi_{xx}-\Phi_{yy})\Big[ (J^2 + J_z^2)\cos2\Omega\sin2\omega + 
      2JJ_z\cos2\omega\sin2\Omega\Big] \nn
   \\
  & \nn + \Phi_{zz} \Big[0.5(J^2 - J_z^2)\sin2\omega \Big]
+ 
    \Phi_{xy}\Big[JJ_z\cos2\omega\cos2\Omega - 0.5(J^2 + J_z^2)\sin2\omega\sin2\Omega\Big]
   \\
     & +  \Phi_{xz}\Big[J\sqrt{1 - J_z^2/J^2}(J\cos2\omega\cos\Omega - J_z\sin2\omega\sin\Omega)\Big]
+ \Phi_{yz}\Big[J\sqrt{1 - J_z^2/J^2}(J \cos 2\omega\sin\Omega + J_z\sin2\omega\cos\Omega)\Big]  \Bigg\},
\end{align}
%%%%%%%%%%%%%%%%%%%%%%%%%%%%%%%
%%%%%%% dOmega/dt %%%%%%%%%%%%%
%%%%%%%%%%%%%%%%%%%%%%%%%%%%%%%
\begin{align}
\label{eqn:dOmegadt_SA}
     \nn  \frac{\md\Omega}{\md t} 
     =& \frac{\partial H_\mathrm{1,SA}}{\partial J_z} \\
  \nn =& [L^2/(4 J^3\mu^2)] \times
%%%%%%%%%%%%%%%%%%%%
\\
 & \nn  \Bigg\{ (\Phi_{xx}+\Phi_{yy}) \Big[ 0.5JJ_z(-3J^2+5L^2+5(J^2-L^2)\cos2\omega)\Big]\\
& \nn + (\Phi_{xx}-\Phi_{yy}) \Big[- 0.5JJ_z\cos 2\Omega(-3J^2+5L^2+5(J^2-L^2)\cos2\omega) + 5J^2(J^2-L^2)\sin 2\omega\sin2\Omega]\Big]
        %%%%%%%%%%%%%%%%%%%%%%%%%%%%
 \\
 & \nn - \Phi_{zz} \Big[ JJ_z(-3J^2+5L^2+5(J^2-L^2)\cos 2\omega) \Big]
        %%%%%%%%%%%%%%%%%%%%%%%%%%%%
\\
 & \nn - \Phi_{xy} \Big[ J[5J(J^2-L^2)\cos2\Omega\sin2\omega + J_z(-3J^2+5L^2+5(J^2-L^2)\cos2\omega)\sin 2\Omega] \Big]
        %%%%%%%%%%%%%%%%%%%%%%%%%%%%
 \\
 & \nn + \Phi_{xz} (1 - J_z^2/J^2)^{-1/2} \Big[ 5JJ_z(J^2-L^2)\sin 2\omega\cos\Omega 
    \\
 & \nn + [3J^4 + 10 J_z^2L^2 + J^2(-6J_z^2-5L^2)+(-5J^4 -10J_z^2L^2 + J^2(10J_z^2+5L^2))\cos2\omega]\sin\Omega \Big]
        %%%%%%%%%%%%%%%%%%%%%%%%%%%%
  \\
 & \nn + \Phi_{yz} (1 - J_z^2/J^2)^{-1/2} \Big[ 5JJ_z(J^2-L^2)\sin 2\omega\sin\Omega 
    \\
 &  - [3J^4 + 10 J_z^2L^2 + J^2(-6J_z^2-5L^2)
+(-5J^4 -10J_z^2L^2 + J^2(10J_z^2+5L^2))\cos2\omega]\cos\Omega \Big] \Bigg\} ,
\end{align}
%%%%%%%%%%%%%%%%%%%%%%%%%%%%%%%%%%%%%%%%%%%
%%%%%%%%%%%%%%%%%%%%%%%%%%%%
%%%%%%% dJz/dt %%%%%%%%%%%%%%
%%%%%%%%%%%%%%%%%%%%%%%%%%%%
\begin{align}
 \nn \frac{\md J_z}{\md t} =&  -\frac{\partial H_\mathrm{1,SA}}{\partial \Omega}
 \\=&
  \nn  -[L^2/(4J^2\mu^2)] \times
 \\ \nn
& \Bigg\{(\Phi_{xx}-\Phi_{yy})\Big[5JJ_z(J^2 - L^2)\cos2\Omega\sin2\omega + 0.5((J^2 - J_z^2)(3J^2 - 5L^2) 
+ 5(J^2 + J_z^2)(J^2 - L^2)\cos2\omega)\sin2\Omega\Big]
%%%%%%%%%%%%%%%%%%%%%%%%%
  \\    
  & \nn     + \Phi_{xy}\Big[((-J^2 + J_z^2) (3 J^2 - 5 L^2) - 
     5 (J^2 + J_z^2) (J^2 - L^2) \cos 2 \omega ) \cos 2 \Omega 
+ 
  10 J J_z (J^2 - L^2) \sin 2\omega \sin 2 \Omega\Big]
%%%%%%%%%%%%%%%%%%%%%%%%%  
  \\ 
& \nn - \Phi_{xz}\Big[ J\sqrt{1 - J_z^2/J^2}(-5J(J^2 - L^2)\sin\Omega\sin2\omega 
+ 
       J_z(-3J^2 + 5L^2 + 5(J^2 - L^2)\cos2\omega)\cos\Omega)\Big]
%%%%%%%%%%%%%%%%%%%%%%%%%       
  \\  
& - \Phi_{yz}\Big[ J\sqrt{1 - J_z^2/J^2}(5J(J^2 - L^2)\cos\Omega\sin2\omega 
+ 
       J_z(-3J^2 + 5L^2 + 5(J^2 - L^2)\cos2\omega)\sin\Omega)\Big]
          \Bigg\}.
          \label{eqn:dJzdt_SA}
\end{align}
%%%%%%%%%%%%%%%%%%%%%%%%%       
{In the case of axisymmetric cluster potentials,} it is straightforward to recover the DA equations from the SA equations
\eqref{eqn:dJdt_SA}-\eqref{eqn:dOmegadt_SA} by
replacing the time-dependent quantities $\Phi_{\alpha \beta}$ with their
time-averages $ \overline{\Phi}_{\alpha\beta}$, and using the identities
$\overline{\Phi}_{xx} = \overline{\Phi}_{yy}$ and  $\overline{\Phi}_{xy} =
\overline{\Phi}_{xz} = \overline{\Phi}_{yz} = 0$ (see equations (33)-(36) of Paper I).

\section{A note on phase space morphology}
\label{sec:Note_Morphology}

The phase space trajectories described by the DA Hamiltonian \eqref{eqn:H1_Doubly_Averaged}
fall into two categories --- librating trajectories, which loop around fixed points at $\omega = \pm \pi/2$ in the $(\omega, j)$ phase plane, and circulating trajectories, which traverse all $\omega \in (-\pi, \pi)$.  Whether a binary's trajectory will librate or circulate depends on its initial orbital elements as well as the value of $\Gamma$.  In Paper II we showed how to determine the family to which a binary's phase space trajectory belongs,
explained how this affects the resulting minimum/maximum eccentricity achievable, etc.  We also explained there how altering $\Gamma$ changes the phase space morphology, and hence the relative importance of librating versus circulating trajectories.  In particular, $\Gamma = 0, \pm 1/5$ turn out to be critical values, such that e.g. systems with $\Gamma > 1/5$ have a qualitatively different phase space morphology to those with $0 < \Gamma < 1/5$, a result which has strong implications for the types of cluster which are able to easily excite high eccentricity behavior \citep{Hamilton2019c,hamilton2022anatomy}.
Moreover,  In Paper III we extended these results to include the effect of GR precession, which also alters the phase space morphology.

However, {for the purposes of studying short-timescale fluctuations at extremely high eccentricity, it is largely irrelevant whether a binary is on a librating or circulating trajectory, or which $\Gamma$ regime it happens to be in.  This is because 
the details of the high eccentricity fluctuations depend predominantly on very short
short-timescale ({of order the outer orbital period} $\sim T_\phi$) torquing that the binary experiences as $e\to 1$, and not on the averaged behavior over the rest of the secular cycle.}
For instance, the numerical examples we present in this paper all happen to be for librating trajectories, but we have found qualitatively indistinguishable results for circulating examples.  Similarly, we do show examples with both $\Gamma > 1/5$ and $0 < \Gamma < 1/5$ in this paper, but the distinction between these two $\Gamma$ regimes is not central to our discussion, so we mostly neglect to mention it {in the main text}.

\section{{Additional numerical examples of high eccentricity behavior without GR precession}}
    \label{sec:App_Further_Examples}
    
    \subsection{Changing the initial inclination to $i_0 = 93.3^\circ$}

          \begin{figure*}\centering
         \includegraphics[width=0.95\linewidth]{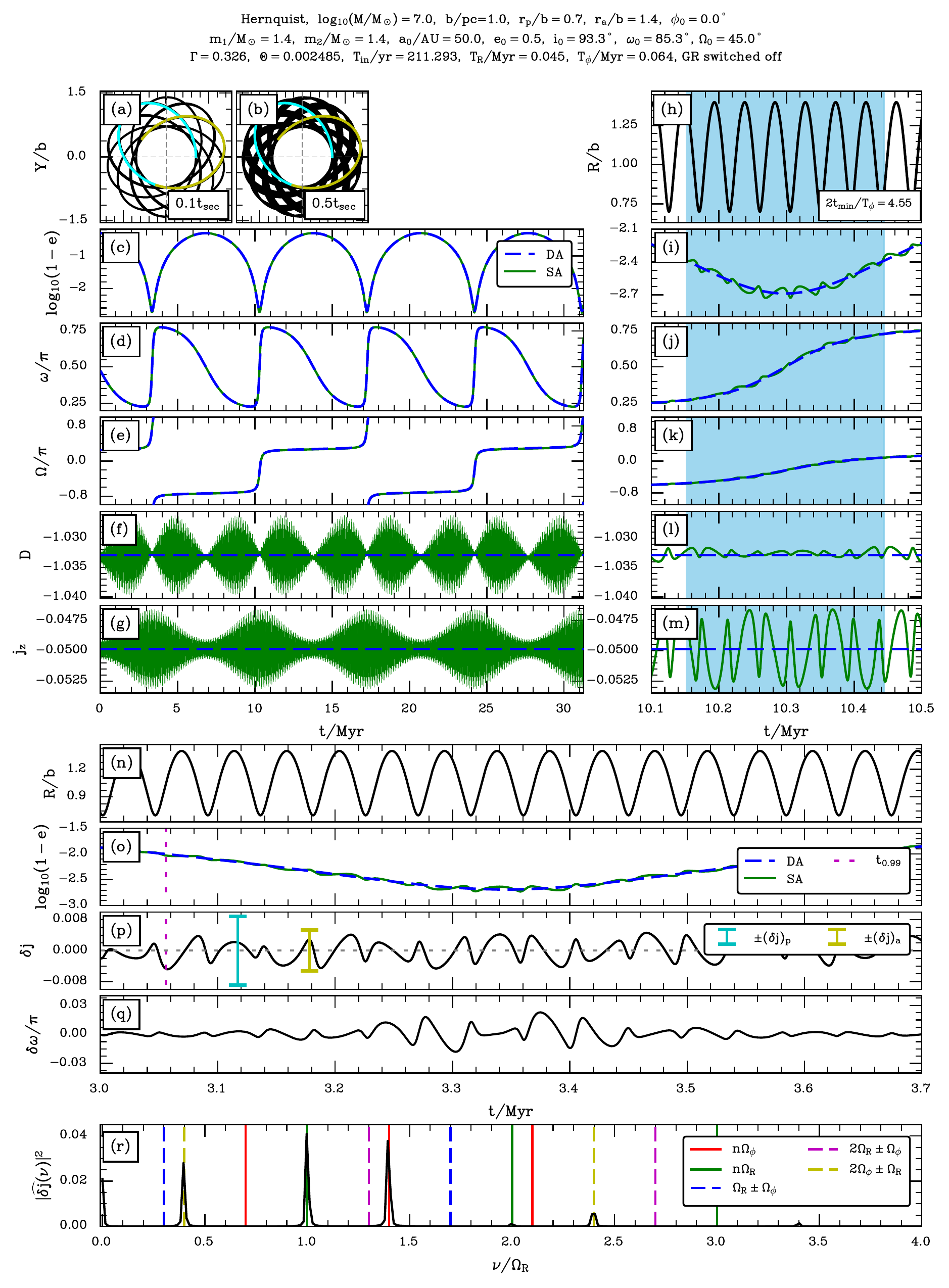}
         \caption{As in Figure \ref{fig:Example_Hernquist} except we change $i_0$ to $93.3^\circ$. As a result $e_\mathrm{max}$ is decreased and $2t_\mathrm{min}$ becomes significantly larger than $T_\phi$.
         {The DA secular period is again $t_\mathrm{sec} \approx 6.9 $ Myr.}}
         \label{fig:Example_Hernquist_4}
      \end{figure*}
    
    In Figure \ref{fig:Example_Hernquist_4} we run the
    same calculation as in Figure \ref{fig:Example_Hernquist},
    except we take $i_0=93.3^\circ$ rather than $90.3^\circ$.  \footnote{To keep the plots clean, we do not show {the results of any direct orbit integration (`N-body') runs} for this example, or {for} the example in \S\ref{sec:Example_Plummer}.  However, we have checked that {in this case, direct orbit integration gives results that are indistinguishable from the SA results.}}
    The main effect
    of this choice is to reduce the maximum eccentricity significantly, so that
    $1-e_\mathrm{max} \approx 10^{-2.7}$. As a result, DA evolution near maximum
    eccentricity is slower than in Figure
    \ref{fig:Example_Hernquist}, while the outer orbit is
    unchanged; hence we find $2t_\mathrm{min}/T_\phi = 4.55$ in this case. The
    qualitative fluctuating behavior of $D$, $j_z$ and $\delta j$ is quite similar between the two
    figures, although in Figure \ref{fig:Example_Hernquist_4} many more
    fluctuations fit into the `blue stripe' surrounding maximum eccentricity.  The
    fluctuations $\delta \omega$ are very different: the brief `pulse' that
    lasted for only $\sim 2T_R$ in Figure
    \ref{fig:Example_Hernquist}q has been replaced with a much
    broader signal with a {slightly} smaller amplitude.
    
    %The fact that the outer orbit is circular and that we are in the limit
    %$2t_\mathrm{min}/T_\phi \gg 1$ means that we can derive an approximate
    %analytic expression for $\delta j(t)$, provided we know the value of the
    %phase angle $\phi-\Omega$ at the instant $t(j_\mathrm{min})$ --- see
    %\S\ref{sec:non_stationary}. 
    %This solution is plotted as a blue dashed line in panel (p).
    
    %%%%%%%%%%%%%%%%%%%%%%%%%%%%%%%%%%%%%%%%%%%%%%%%%%%%%%%%%%%%%%%%%%%%%%%%%%%%%%%%%%%%%%%%%%%%%%%%%%%%%%%%%%%%%%%%%%%%%%%%%%%%%%%%%%%%%%%%%%%%%%%%%%%%%%%%%%%%%%%%%%%%%%%%%%%%%%%%%%%%%%%%%%%%%%%%%%%%%%%%
    %%%%%%%%%%%%%%%%%%%%%%%%%%%%%%%%%%%%%%%%%%%%%%%%%
    \subsection{An example in the Plummer potential}
    \label{sec:Example_Plummer}

    The phenomenology reported above is rather characteristic of binaries
    orbiting in cusped potentials, and will be analysed more quantitatively in
    \S\ref{sec:quantitative}.  First, however, we
    perform the same calculation except this time with a cored potential,
    namely the Plummer sphere.
    
      %\hfill <-- it is superfluous 
  
          \begin{figure*}\centering
                   \includegraphics[width=0.95\linewidth]{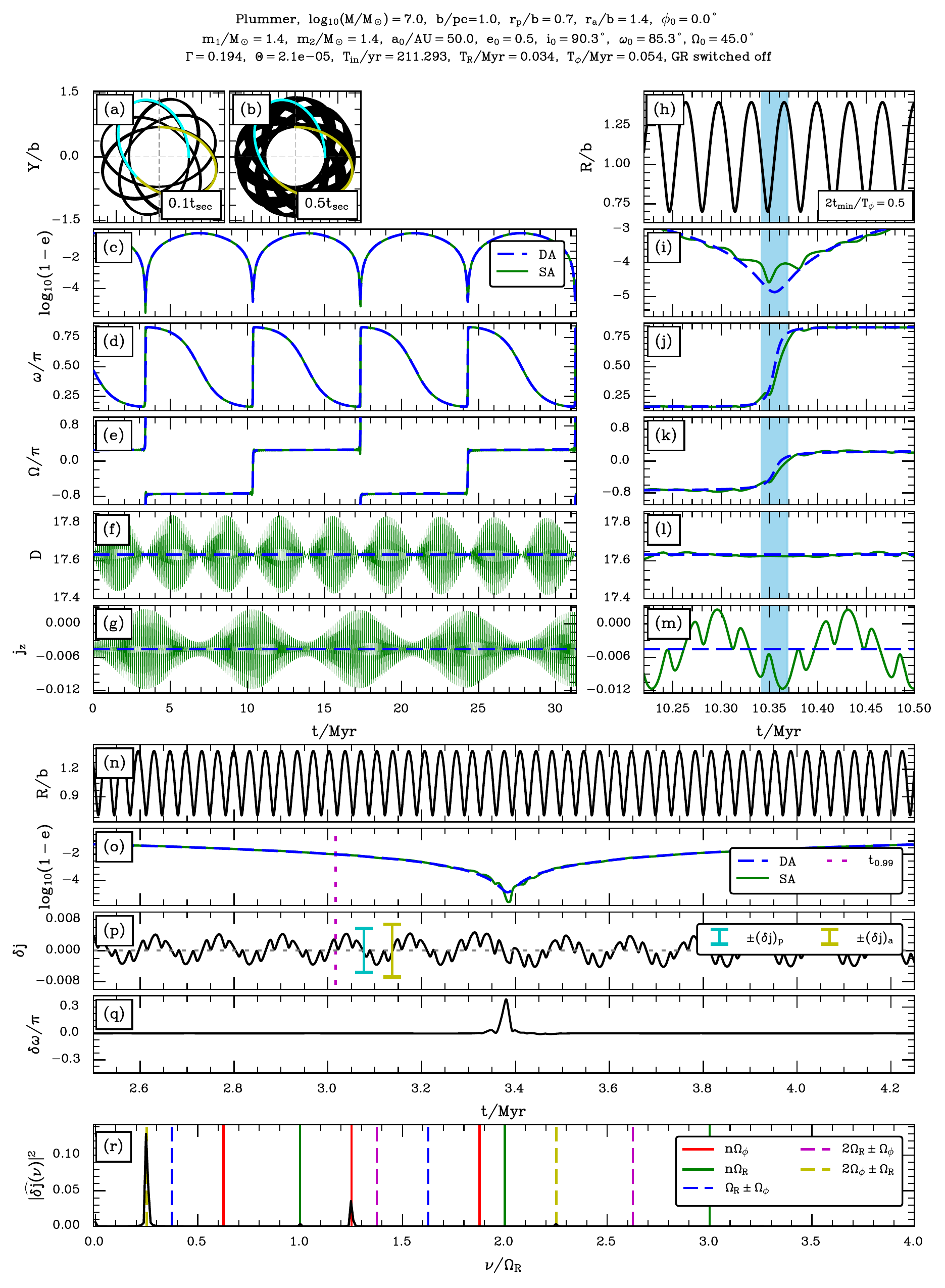}
         \caption{As in Figure \ref{fig:Example_Hernquist}, except this time we use the Plummer potential. Here {$\tsec \approx 7.0$ Myr}.}
         \label{fig:Example_Plummer}
      \end{figure*}%

          \begin{figure}\centering
                   \includegraphics[width=0.49\linewidth]{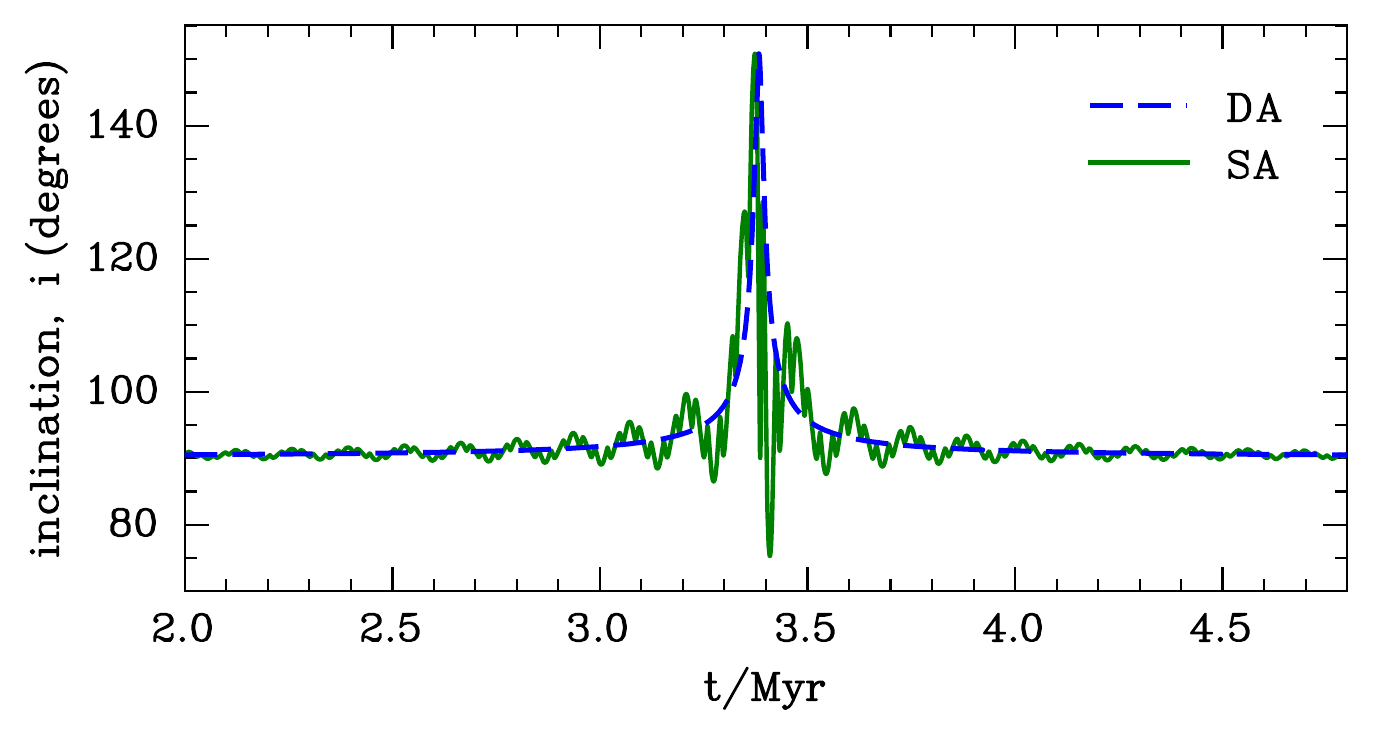}
         \caption{{Plotting the inclination $i = \arccos(j_z/j)$
         around the first eccentricity peak from Figure \ref{fig:Example_Plummer}.
         The SA result reveals that near peak eccentricity, 
         the binary experiences orbital `flips' whereby its inclination crosses $90^\circ$.}}
         \label{fig:Inclination_Plummer}
      \end{figure}%

    In Figure \ref{fig:Example_Plummer} we provide an example of
    a binary exhibiting short-timescale fluctuations in a cored potential. All
    input parameters are the same as in Figure
    \ref{fig:Example_Hernquist} except we change the potential
    from Hernquist to Plummer, so $\Phi = -G\mathcal{M}/\sqrt{b^2+r^2}$ {(we again use $\mathcal{M}=10^7M_\odot$ and $b=1$pc)}.
    The change of potential
    means that we now have $\Gamma = 0.194 <1/5$, which leads to a rather large
    $D_\mathrm{DA}$ value of around $17.6$ (equation
    \eqref{eqn:def_D}). 
    %Once again, SA and N-body results agree very nicely even at extreme eccentricities.
    
    Noteworthy in this case is the morphology of the fluctuations of
    $j_{z,\mathrm{SA}}$ near highest eccentricity, around the DA value
    $j_{z,\mathrm{DA}} = -\sqrt{\Theta} = -0.0045$ ({panel (m)}).  In this case,
    small oscillations on the timescale $\sim T_R$ are superimposed
    upon a larger `carrier signal' oscillation, which has amplitude $\sim 0.006$
    and its own period $\sim 4T_R$. 
    In this case these fluctuations can actually change the sign of $j_z$, which corresponds to the SA binary inclination $i$ temporarily passing through $90^\circ$, {a so-called `orbital flip' --- see \citet{Naoz2016,Grishin2018}.
    In Figure \ref{fig:Inclination_Plummer}
    we show the inclination $i(t)$ explicltly for SA and DA integrations; we see that near peak eccentricity the DA approximation fails entirely to capture the flip behavior.}
    
    A similar morphology is exhibited by the
    $\delta j$ time series (panel (p)).
    % spectrum (panel (r)) demonstrates that the majority of the power lies in
    % the $2\Omega_\phi-\Omega_R$ mode, with a small amount at $2\Omega_\phi$.
    % There is also power in the zero frequency mode reflecting the
    % non-centering of $\delta j$ fluctuations around zero.  
    Again the
    fluctuations $\delta \omega$ (panel (q)) are negligible until the very
    highest eccentricities are reached, where there is a sharp, negative pulse
    of maximum amplitude $\sim 0.2\pi$, that lasts for $\sim 2T_R$ in total
     before decaying back to zero.

\subsection{Dependence on phase angles}
\label{sec:Phase_Angles_no_GR}

    It is also worth emphasising here the dependence of these results on the choice of various \textit{phase angles}, namely the initial radial phase of the outer orbit, the initial azimuthal angle of the outer orbit $\phi$, and the initial choice of longitude of ascending node $\Omega$ of the inner orbit\footnote{In fact, {the SA equations for binaries perturbed by spherically symmetric clusters} only depend on the difference $\phi - \Omega$, {rather than on} $\phi$ and $\Omega$ individually --- see equation \eqref{eqn:DeltaH_spherical}.}. 
    These 
    choices feed into the solutions of the SA equations of motion
    \eqref{eqn:domegadt_SA}-\eqref{eqn:dJzdt_SA}, though of course they do not affect the DA solutions.
  {By} inspecting numerical examples without GR precession (such as that in Figure \ref{fig:Example_Hernquist_Divergence}, as well as several others not shown here),
    we found that the choice of these phase angles can significantly affect e.g. the maximum value of 
    $e_\mathrm{SA}$ that is reached by a binary, but that the qualitative behavior is very similar from one realization to the next, and the averaged values of $D$ and $j_z$ are conserved\footnote{{Apart from the gradual drift away from the DA solution that occurs over many secular timescales \citep{Luo2016}, which is distinct from RPSD, as discussed in \S\S\ref{sec:phase_dependence} and \ref{sec:literature}.}}.
    However, this ceases to be true when GR precession is
    included, as we saw in Figure \ref{fig:Example_Hernquist_GR_Divergence} (see also \S\ref{sec:Effect_of_GR}).

\section{Fluctuating Hamiltonian}
\label{sec:fluctuating_Hamiltonian}

By subtracting the DA Hamiltonian from the SA Hamiltonian, assuming them to be
functions of the same variables $(J,\omega,...)$, and using $\Phi_{xx} =
\overline{\Phi}_{yy}$ and $\overline{\Phi}_{xy} = \overline{\Phi}_{xz} =
\overline{\Phi}_{yz} =0$, we get an expression for the `fluctuating Hamiltonian'
(equation \eqref{eqn:DeltaH_HSA_minus_HDA}):
%%%%%%%%%%%%%%%%%%%%%%%%%%%%
%%%%%%%%%%%%%%%%%%%%%%%%%%%%
\begin{align}
    \Delta H \equiv &H_\mathrm{SA}(J,\omega,...) - H_\mathrm{DA}(J,\omega,...) 
    = \frac{1}{2}\sum_{\alpha\beta}(\Phi_{\alpha\beta}(t)-\overline{\Phi}_{\alpha\beta})\langle r_\alpha r_\beta \rangle_M  
    \nn \\ =& \frac{1}{2}\Bigg\{
    (\Phi_{xx}-\overline{\Phi}_{xx}) \langle x^2  \rangle_M
    + (\Phi_{yy}-\overline{\Phi}_{xx}) \langle y^2  \rangle_M
    + (\Phi_{zz}-\overline{\Phi}_{zz}) \langle z^2  \rangle_M 
  + \Phi_{xy} \langle xy \rangle_M
    + \Phi_{xz} \langle xz \rangle_M
    + \Phi_{yz} \langle yz \rangle_M
    \Bigg \}.
    \label{eqn:deltaH_general}
\end{align}
%%%%%%%%%%%%%%%%%%%%%%%%%%%%%%%%%%%%%%%%%%%%%%%%%%%%%%%%
%%%%%%%%%%%%%%%%%%%%%%%%%%%%v
(Note that the term involving $\epsGR$ has disappeared, since it is the same in
SA and DA theory). Equation \eqref{eqn:deltaH_general} holds for
binaries in arbitrary axisymmetric potentials.

%%%%%%%%%%%%%%%%%%%%%%%%%%%%%%%%%%%%%%%%%%%%%%%%%%%%%%%%
%%%%%%%%%%%%%%%%%%%%%%%%%%%%%%%%%%%%%%%%%%%%%%%%%%%%%%%%
%%%%%%%%%%%%%%%%%%%%%%%%%%%%%%%%%%%%%%%%%%%%%%%%%%%%%%%%

We can simplify matters significantly if we restrict ourselves to spherical
potentials $\Phi(r) = \Phi(\sqrt{R^2+Z^2})$. Let us define
\begin{align}
    f_\pm(\Rg(t)) \equiv \frac{1}{2} \left[ \left( \frac{\partial^2\Phi}{\partial R^2}\right)_{\Rg}\pm \left( \frac{1}{R}\frac{\partial\Phi}{\partial R}\right)_{\Rg}\right],
    \label{eqn:f_pm_definition}
    \end{align}
   and assume without loss of generality that $\Rg$ is confined to $Z=0$. Then
   it is easy to show (see equations
   (33)-(36) of Paper I) that:
%%%%%%%%%%%%%%%%%%%%%%%%%%%%
%%%%%%%%%%%%%%%%%%%%%%%%%%%%
%%%%%%%%%%%%%%%%%%%%%%%%%%%%
\begin{align} 
\label{SAFluctuations_phixxnon} 
&\Phi_{xx} = f_+ + f_-\cos 2\phi , \,\,\,\,\,\,\,\,\,
\Phi_{yy} = f_+ - f_-\cos 2\phi, \,\,\,\,\,\,\,\,\,
\Phi_{zz} =f_+-f_-,  \,\,\,\,\,\,\,\,\,\,\,
\Phi_{xy} =  f_-\sin 2\phi, \,\,\,\,\,\,\,\,\,\,
\Phi_{xz}=  
\Phi_{yz}=0.
\end{align}
%%%%%%%%%%%%%%%%%%%%%%%%%%%%
%%%%%%%%%%%%%%%%%%%%%%%%%%%%
(Note that we have dropped the `g' subscript for ease of notation).
If we also define $\Delta f_\pm \equiv f_\pm - \overline{f}_\pm$ where $\overline{f}_\pm$ is the annulus-averaged value of $f_\pm$ then the fluctuating Hamiltonian can be written concisely as 
\begin{align}
    \Delta H =& \frac{1}{2}\left[ \Delta f_+ \langle x^2+y^2+z^2 \rangle_M - \Delta f_- \langle z^2 \rangle_M + f_-\left( \langle x^2-y^2 \rangle_M \cos 2\phi + 2\langle xy \rangle_M \sin 2\phi \right) \right]. \nn
    \\
    =& \frac{L^2}{8J^2\mu^2}\Bigg\{ (3 J^2 - 5 L^2) \Big[(\Delta f_- - 2 \Delta f_+) J^2 - \Delta f_- J_z^2 + 
     f_- (-J^2 + J_z^2) \cos[2 (\Omega - \phi)]\Big] \nn \\ &- 
  5 (J^2 - L^2) \cos
    2 \omega \Big[\Delta f_- (J^2 - J_z^2) + 
     f_- (J^2 + J_z^2) \cos[2 (\Omega - \phi)]\Big] + 
  10 f_- J J_z (J^2 - L^2) \sin 2 \omega \sin[2 (\Omega - \phi)]
  \Bigg\}.
  \label{eqn:DeltaH_spherical}
\end{align}
%%%%%%%%%%%%%%%%%%%%%%%%%%%%%%%%%%%%%%%%%%%
%From now on the outer orbit will be denoted $(R,\phi)$, with the understanding that these quantities should be evaluated at the position $\Rg(t)$.
%%%%%%%%%%%%%%%%%%%%%%%%%%%%%%%%%%%%%%%%%%%%%%%
As explained in \S\ref{sec:quantitative}, equations for the evolution of fluctuations in orbital elements
be derived from the fluctuating Hamiltonian $\Delta H(\omega, J, ..)$ by taking
its partial derivatives (equation \eqref{eqn:diag}). In
particular, we have
%%%%%%%%%%%%%%%%%%%%%%%%%%%%
%%%%%%% SA - DA %%%%%%%%%%%%%%
%%%%%%%%%%%%%%%%%%%%%%%%%%%%
\begin{align}
\frac{\partial \Delta H}{\partial \omega} &= \frac{5}{4}\frac{L^2}{J^2\mu^2}(J^2-L^2)\Bigg\{ \sin 2\omega (J^2-J_z^2)\Delta f_- +f_-\Big((J^2+J_z^2)\sin2\omega\cos[2(\phi-\Omega)] 
 -2JJ_z\cos2\omega\sin[2(\phi-\Omega)]\Big)\Bigg\}.
\label{eqn:minusddeltaHdomega}
\end{align}
Assuming this to be a good approximation {to $-\md \delta j/\md t$ at high
$e$}, as we do in \S\ref{sec:characteristic_delta_j}, it
constitutes a generalisation of equation (B4) of \citet{Ivanov2005}. The result
of \citet{Ivanov2005} is recovered if one assumes the outer orbit to be circular
(so $\Delta f_\pm =0$), the perturbing potential to be Keplerian, and evaluates
\eqref{eqn:minusddeltaHdomega} at $\omega = \pm\pi/2$ and $e\to
1$.  We also have
\begin{align} \nn
   \frac{\partial \Delta H}{\partial \Omega}
     \nn =& \frac{L^2}{4J^2\mu^2} f_- \Bigg\{10JJ_z(J^2 - L^2)\sin2\omega \cos[2(\phi-\Omega)] \nn \\ & - ((J^2 - J_z^2)(3J^2 - 5L^2) + 5(J^2 + J_z^2)(J^2 - L^2)\cos2\omega)\sin[2(\phi-\Omega)]\Bigg\},
   %%%%%%%%%%%%%%%%%%%%%%%%%
   \label{eqn:minusddeltaHdOmega}
   \end{align}
   which coincides precisely with (minus) the right hand side of \eqref{eqn:dJzdt_SA} 
   if one assumes $\Phi$ to be spherical.  This is as it must be, since in DA dynamics $J_z$ is perfectly conserved, and so $\md J_z / \md t = \md \delta J_z / \md t$, as we argued below equation \eqref{eqn:fluctuating_quantities_not_equal}.
%%%%%%%%%%%%%%%%%%%%%%%%%%%%
%%%%%%% dJz/dt %%%%%%%%%%%%%%
%%%%%%%%%%%%%%%%%%%%%%%%%%%%

%%%%%%%%%%%%%%%%%%%%%%%%%%%%
%%%%%%% domega/dt %%%%%%%%%%%%%%
%%%%%%%%%%%%%%%%%%%%%%%%%%%%
%\begin{footnotesize}
%\begin{align}
%\frac{\partial \Delta H}{\partial J}  \nn =&\frac{L^2}{4J^3 \mu^2}
% \nn  
% \Bigg\{ \Delta f_- \Big[-5J_z^2L^2(1-\cos2\omega)-5J^4\cos2\omega \Big] +
%[-f_-+2f_+] (-3J^4) 
%     %%%%%%%%%%%%%%%%%%%%%%%%%%%%
%     \\ 
% & 
% + f_- \Big[-5J J_z(J^2+L^2)\sin2\omega\sin[2(\phi-\Omega)] + (5J_z^2L^2(1-\cos2\omega)-J^4(3+5\cos2\omega))\cos[2(\phi-\Omega)]  \Big]
%    \Bigg\},
%\end{align}

%%%%%%%%%%%%%%%%%%%%%%%%%%%%%%
%%%%%%% dOmega/dt %%%%%%%%%%%%%
%%%%%%%%%%%%%%%%%%%%%%%%%%%%%%%
%\begin{align}
%\frac{\partial \Delta H}{\partial J_z}
%     =& \frac{L^2}{4 J^2\mu^2}
%\nn  \Bigg\{ \Delta f_- \Big[ J_z(-3J^2+5L^2+5(J^2-L^2)\cos2\omega)\Big]\\
% & + f_- \Big[ -J_z(-3J^2+5L^2+5(J^2-L^2)\cos 2\omega)\cos [2(\phi-\Omega)]
% -10J(J^2-L^2)\sin2\omega \sin [2(\phi-\Omega)]\Big]
% \Bigg\}.
% \label{eqn:ddeltaHdJz}
%\end{align}

%%%%%%%%%%%%%%%%%%%%%%%%%%%%%%%%%%%%%%%%%%%%%%%%%%%%%%%%%%%%
%%%%%%%%%%%%%%%%%%%%%%%%%%%%%%%%%%%%%%%%%%%%%%%%%%%%%%%%%%%%
%%%%%%%%%%%%%%%%%%%%%%%%%%%%%%%%%%%%%%%%%%%%%%%%%%%%%%%%%%%%

\section{{Justifying the approximation $\delta \bw \approx \Delta \bw(\bw_\mathrm{SA})
\approx \Delta \bw(\bw_\mathrm{DA})$}}
\label{sec:Deltaw_Explanation}
\begin{figure*}
   \centering
   \includegraphics[width=0.99\linewidth]{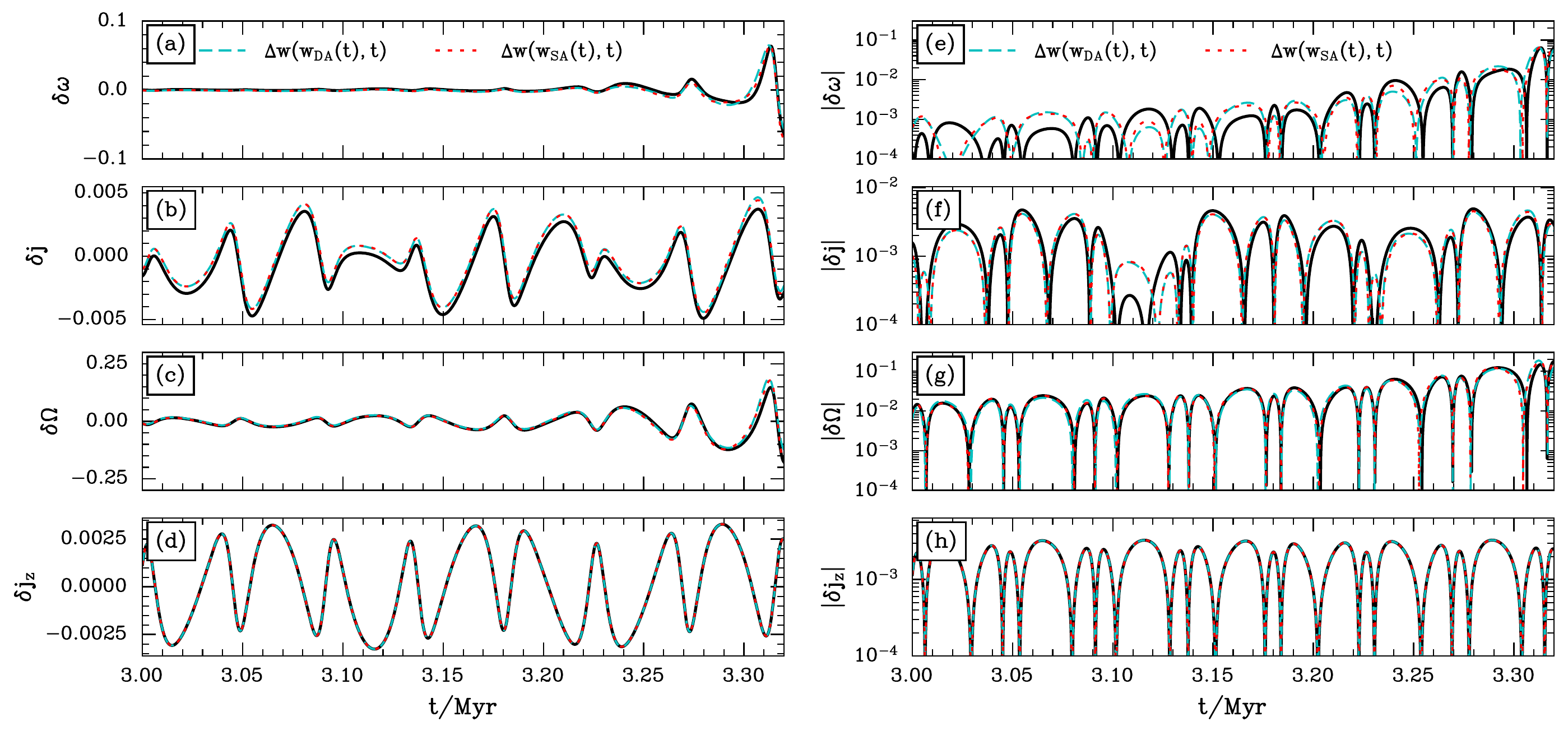}
   \caption{Justifying the approximation $\delta \mathbf{w}(t) \approx \Delta \mathbf{w}(\bw_\mathrm{SA},t)
   \approx \Delta \mathbf{w}(\bw_\mathrm{DA},t)$.  {In panels (a)-(d),
   black solid lines 
   show the evolution of $\delta\bm{\bw} \equiv \bw_\mathrm{SA} - \bw_\mathrm{DA}$ using the data from 
   Figure \ref{fig:Example_Hernquist} during the time interval $t \in (3, 3.32)$Myr (the first peak in DA eccentricity occurs at $t\approx 3.34$Myr).  Cyan dashed lines and red dotted lines show the approximations $\Delta \mathbf{w}(\bw_\mathrm{SA},t)$ and $\Delta \mathbf{w}(\bw_\mathrm{SA},t)$ respectively.
   In panels {(e)-(h)} we show the same data using a logarithmic scale.}}
   \label{fig:Difference_Equations_Hernquist}
   \end{figure*}
{If we} feed a numerical result $\mathbf{w}(t) =
\bw_\mathrm{SA}(t)$ into equation
\eqref{eqn:diag} {and integrate forwards in time, we do not in general}
reproduce the `SA minus DA' solution
\eqref{eqn:deltaw_SA_minus_DA};
for example:
%%%%%%%%%%%%%%%%%%%%%%%%%%%%%%%%%%%%%%%%%%%%%%%%%%%%%%%%%%%%%%%%%%
\begin{align}
   \delta \omega(t) &\equiv  \omega_\mathrm{SA}(t)  - \omega_\mathrm{DA}(t) \nn \\ &\neq \Delta \omega(\bw_\mathrm{SA}(t), t) \equiv \int_0^t \md t' \frac{\partial \Delta H(\bw_\mathrm{SA}(t'),t')}{\partial J_{\,\mathrm{SA}}}.
   \label{eqn:fluctuating_quantities_not_equal}
\end{align}
%%%%%%%%%%%%%%%%%%%%%%%%%%%%%%%%%%%%%%%%%%%%%%%%%%%%%%%%%%%%%%%%%%
Similarly, $\delta J \neq \Delta J$ and $\delta \Omega \neq \Delta \Omega$.  The
exception is $J_z$, for which there is no DA evolution, so that $\delta J_z(t) =
 \Delta J_z(\bw_\mathrm{SA}(t),t)$.
We feel it is important to make this distinction since it affects perturbative calculations
--- for example, \cite{Luo2016}
implicitly used $\delta \bw(t) = \Delta \bw(\bw_\mathrm{DA}(t), t)$
when calculating the cumulative impact of short-timescale fluctuations over many secular cycles,
by `freezing' the DA elements on the timescale of the outer orbit.
See \S\ref{sec:literature} for more.

{Nevertheless, these quantities are \textit{approximately} equal, as we wrote in equation
\eqref{eqn:deltaw_DeltawSA_DeltawDA}.
There are two basic reasons why this approximation holds.  The first is that the torques felt by an extremely eccentric binary are not particularly sensitive to its exact orbital elements.
For instance, the torque on a binary at $e_\mathrm{DA}=0.999$ is not particularly different from that on a binary with $e_\mathrm{SA}=0.9995$, since in each case the inner-orbit-averaged version 
of that binary essentially looks like a one-dimensional line. Second, there \textit{is} a sensitive dependence of the precession of $\omega$ and $\Omega$ upon the exact value of eccentricity as $e\to 1$ (see equation \eqref{eqn:scalings}), but the high-$e$ episode is over so quickly that this does not lead to a significant breakdown of equation \eqref{eqn:deltaw_DeltawSA_DeltawDA}.}

{To demonstrate the accuracy of the approximation \eqref{eqn:deltaw_DeltawSA_DeltawDA}, we 
took the quantities from Figure \ref{fig:dDeltawdtHernquist}
and integrated them forwards in time {using \eqref{eqn:diag}}.
We plot the result in
Figure \ref{fig:Difference_Equations_Hernquist} on top of  
the `true' fluctuation $\delta \bw(t)$, shown with black lines. {We see that the approximation
\eqref{eqn:deltaw_DeltawSA_DeltawDA} works very well in this case;
we also confirmed {its accuracy}
in several other examples not shown here, including in cases with GR precession switched on (relevant for \S\ref{sec:Effect_of_GR}).}
%This is useful because we have
%explicit expressions for $\md \Delta \bw/\md t$, via equations
%\eqref{eqn:diag} and the Hamiltonian
%\eqref{eqn:deltaH_general}.}

%In blue we show $\Delta
%\bw(\bw_\mathrm{DA}(t),t)$, i.e. the result of feeding the DA solution to
%equations \eqref{eqn:diag} and integrating forwards in time.
%Analagously, in green we plot $\Delta \bw(\bw_\mathrm{SA}(t),t)$.  Finally,
%overplotted in black we show $\delta \bw(t) = \bw_\mathrm{SA}(t) -
%\bw_\mathrm{DA}(t)$. Panels (a)-(d) correspond to fluctuations in $\omega, j,
%\Omega, j_z$ respectively. We see in each panel that the blue and green curves
%agree nicely, so that we may approximate $\Delta \bw(\bw_\mathrm{DA}(t),t)
%\approx \Delta \bw(\bw_\mathrm{SA}(t),t)$. Furthermore, panel (d) shows that for
%$j_z$ fluctuations, all three curves agree very precisely, as expected. On the
%other hand, in panels (a)-(c) there is a visible offset between the blue/green
%curves and the black curves. This is particularly true for fluctuations in
%$\omega$ (panel (a)). However, the differences are small enough --- particularly
%for fluctuations in $j$ (panel (b)), which are what we care about most --- 

 %The second can in principle useful if
%one wishes to feed in the explicit analytic solution $\bw_\mathrm{DA}(t)$ to
%the DA problem at high eccentricity (\S\ref{PaperIII_sec:analytic}). However,
%in practice such an approach turns out to have little predictive power, so we
%will not pursue it here.

\section{{Characteristic amplitude of angular momentum fluctuations}}
\label{sec:App_deltaj}

For simplicity we will assume that $\Phi$ is spherically symmetric. 
%Then the formula for $(\delta j)_\mathrm{max}$ will be a function of the same variables
%that allow us to specify the minimum DA anglular momentum $j_\mathrm{min}$,
%namely $(\Phi,M,b)$ and the vector
%\begin{align}
%\label{eqn:x}
%    \mathbf{x} = \left(\frac{\rp}{b},\frac{\ra}{b},a_0,e_0,i_0,\omega_0,m_1,m_2\right),
%\end{align}
%where the subscript `$0$' denotes initial values. 
Then to evaluate the torque at high eccentricity we can use
\eqref{eqn:minusddeltaHdomega}, which by
\eqref{eqn:diag} and
\eqref{eqn:deltaw_DeltawSA_DeltawDA} is a good approximation to
$-\md \delta j/\md t$ if we evaluate it using DA quantities. The maximum
eccentricity as predicted by the DA theory is
$e_\mathrm{DA}=e_\mathrm{max}\approx 1$, and it always occurs either at
$\omega_\mathrm{DA} = \pm \pi/2$ or at $\omega_\mathrm{DA}=0$. Let the
corresponding minimum inclination be $i_\mathrm{min}$.  Evaluating
\eqref{eqn:minusddeltaHdomega} at these (assumed fixed) DA
values, we find
%%%%%%%%%%%%%%%%%%%%%%%%%%%%%%%%%%%%%%
  %%%%%%%%%%%%%%%%%%%%%%%%%%%%%%%%%%%%%%%%%
  \begin{align} 
   \frac{\md \delta J}{\md t} \bigg\vert_{\omega=\pm\pi/2}  
   =& \frac{5}{4}a^2 \cos i_\mathrm{min} \times  2f_-(R) \sin [2(\phi-\Omega)], 
   \label{SAFluctuations_dJdtfinal}
\end{align} 
   %%%%%%%%%%%%%%%%%%%%%%%%%%%%%%%%%%%%%%%%%
   or the same thing with an additional minus sign if evaluating at $\omega=0$.
   Note that the function $f_-(R)$, defined in equation
   \eqref{eqn:f_pm_definition}, depends on the instantaneous
   value of the outer orbital radius $R(t)$. Finally, one can check that for a
   Keplerian potential $\Phi = -G\mathcal{M}/R$ we recover equation (B4) of \citet{Ivanov2005}.   
   
  %From now on we will drop the $\Rg$ notation and just refer to the outer orbit trajectory as $R(t),\phi(t)$.
   %, who were the first to calculate the short-timescale fluctuations in torque at high eccentricity in the Lidov-Kozai problem.
   {Next we use the fact that in DA theory  $i_\mathrm{min}$ does not vary from one eccentricity peak to the next,  and we
 assume that $\Omega_\mathrm{DA}$ is stationary on the timescale $T_\phi$.}
 (This assumption, along with that of stationary $\omega_\mathrm{DA}$ and $e_\mathrm{DA}$ on the timescale $T_\phi$, breaks down whenever $\tmin \lesssim T_\phi$ --- {see Figures \ref{fig:Example_Hernquist}, \eqref{fig:Example_Hernquist_4} and \ref{fig:Example_Plummer}, as well as Appendix C of Paper III}.  Nevertheless, these assumptions are good enough in order to get for a simple estimate of $(\delta j)_\mathrm{max}$ which is all we need here).
 Placing the maximum DA eccentricity at $t=0$ without loss of generality, 
 we set $\Omega=\Omega(0)$. Then the only time dependence in equation
   \eqref{SAFluctuations_dJdtfinal} comes from $R(t)$ and $\phi(t)$.  
   Furthermore, $f_-(R) < 0$ for all $R$ in sensible cluster
   potentials\footnote{To see this, suppose the cluster has density profile
   $\rho(r)$.  From Poisson's equation $\nabla^2\Phi = 4\pi G\rho$ it is
   straightforward to show that
%%%%
\begin{align}
\frac{\partial^2 \Phi}{\partial R^2} - \frac{1}{R}\frac{\partial \Phi}{\partial R} = R\frac{\partial}{\partial R}\left(\frac{G\mathcal{M}(R)}{R^3}\right),
\label{SAFluctuations_eqn_negative}
\end{align}
%%%%
where $\mathcal{M}(R) = \int_0^R 4\pi r^2 \md r \rho(r)$ is the mass enclosed inside a
     sphere of radius $R$.  {In particular,} for any model in which $\rho$ is a
     monotonically decreasing function of radius, the expression
     \eqref{SAFluctuations_eqn_negative} is negative for all $R$.}. As a result,
     the sign of the torque at highest eccentricity (equation
     \eqref{SAFluctuations_dJdtfinal}) is dictated entirely by the instantaneous
     value of the phase angle $2(\phi-\Omega)$.  The fluctuation $(\delta
     j)_\mathrm{max}$ is therefore accumulated over a quarter period in azimuth,
     say from $\phi(t_1)-\Omega = 0$ to $\phi(t_2)-\Omega = \pi/2$, after which
     the torque changes sign. Integrating \eqref{SAFluctuations_dJdtfinal} over
     time 
     %and recalling that $j \equiv J/L =  [G(m_1+m_2)a]^{-1/2}J$ 
     we find
%%%%%%%%%
\begin{align}
(\delta j)_\mathrm{max} &= \frac{5}{4}\left[ \frac{a^{3}}{G(m_1+m_2)}\right]^{1/2} \cos i_\mathrm{min} \, F(\rp,\ra)
\label{SAFluctuations_eqn_deltaj} \\
=& 10^{-4} \times \left( \frac{\sqrt{\Theta}/j_\mathrm{min}}{1}\right)\left(\frac{m_1+m_2}{M_\odot}\right)^{-1/2} \left(\frac{a}{10\mathrm{AU}}\right)^{3/2} 
\left(\frac{F^*}{0.8}\right) \left(\frac{M}{10^5M_\odot}\right)^{1/2}\left(\frac{b}{\mathrm{pc}}\right)^{-3/2}, \label{SAFluctuations_eqn_deltajnum}
\end{align}
%%%%%%%%%
where all the details of the potential and outer orbit have been absorbed by the
function
%%%%%%%%%
\begin{align}
F(\rp,\ra) = \int_{t_1}^{t_2} \md t \, \left|\frac{\partial^2 \Phi}{\partial R^2} - \frac{1}{R}\frac{\partial\Phi}{\partial R}\right|\, \sin [2(\phi-\Omega)],
\label{SAFluctuations_eqn_Frpra}
\end{align}
%%%%%%%%%
and in the numerical estimate \eqref{SAFluctuations_eqn_deltajnum} we defined
the dimensionless number 
\begin{align}
   F^* \equiv \left(\frac{G\mathcal{M}}{b^3}\right)^{-1/2}F.
\end{align}
%%%%%%%%%%%%%%%%%%%%%%%%%%%%%%%%%%%%%%%%
% Note that for $\sqrt{\Theta}/j_\mathrm{min} \sim 1$ and $F^* \sim 1$,  equation \eqref{SAFluctuations_eqn_deltaj} gives $(\delta j)_\mathrm{max} \sim Ma^3/([m_1+m_2]b^3) \sim T_\phi/T_\mathrm{in}$, roughly the ratio of outer to inner binary orbital periods.

\subsection{Circular outer orbits}
\label{sec:deltaj_circular_approx}
%%%%%%%%%%%%%%%%%%%%%%%%%%%%%%%%%%%%%
%%%%%%%%%%%%%%%%%%%%%%
The simplest (and practically speaking, only) way to proceed more quantitatively than this 
is to estimate $F$ by imagining that the binary is
on a circular outer orbit with radius $R$. Then $F=F_\mathrm{circ}(R)$ where
%%%%%%%%%%%%%%%%%%%%%%
\begin{align}F_\mathrm{circ}(R) &= 2\Omega_\mathrm{circ} \Bigg\vert \frac{\partial \ln \Omega_\mathrm{circ}}{\partial \ln R} \Bigg\vert,
\label{SAFluctuations_eqn_FofR}
\end{align}
%%%%%%%%%%%%%%%%%%%%%%%%%%%%%%%%%%%%%
%%%%%%%%%%%%%%%%%%%%%%%%%%%%%%%%%%%%%%%%%%%%%%%%%%%
{where $\Omega_\mathrm{circ}(R) = [R^{-1}\partial \Phi/\partial R]^{1/2}$ is the angular frequency of a circular orbit of radius $R$.
In the LK case of Keplerian potentials, the result
arising from \eqref{SAFluctuations_eqn_deltaj} with the circular approximation
\eqref{SAFluctuations_eqn_FofR} was already derived by \citet{Ivanov2005}.}

In Figure \ref{SAFluctuations_DimensionlessF} we plot the dimensionless number
$F_\mathrm{circ}^* \equiv (G\mathcal{M}/b^3)^{-1/2}F_\mathrm{circ}$ as a function of $R/b$
for circular outer orbits in various spherically symmetric cluster potentials with scale radius $b$.
For reference we also plot $F_\mathrm{circ}^*$ for the Kepler potential $\Phi =
-G\mathcal{M}/R$.
%%%%%%%%%%%%%%%%%%%%%%%%%%%%%%%%%%%%%%%%%%%%%%%%%%%
   %%%%%%%%%%%%%%%%%%%%%%%%%%%%%%
\begin{figure}
   \centering
   \includegraphics[width=0.4\linewidth]{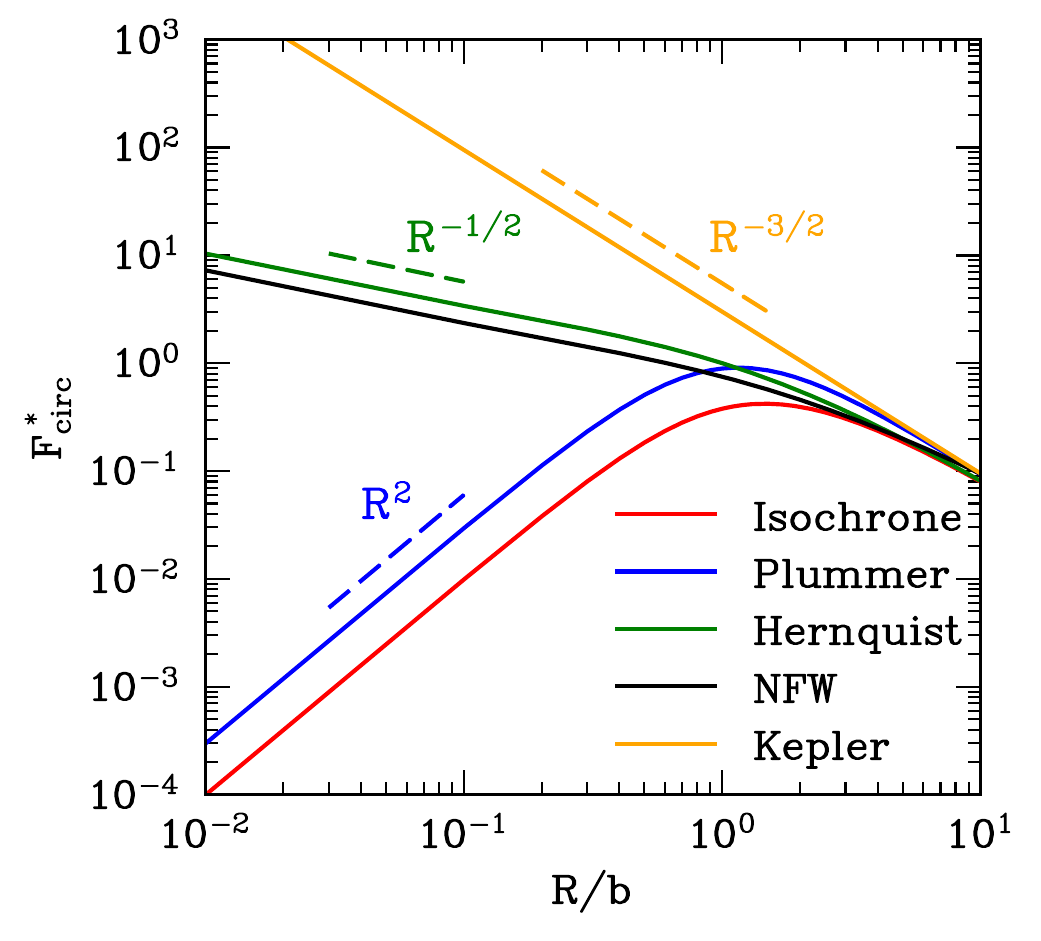}
   \caption{Plot of the dimensionless function $F_\mathrm{circ}^* \equiv
   F_\mathrm{circ}(R)/(G\mathcal{M}/b^3)^{1/2}$, where $F$ is defined in equation
   \eqref{SAFluctuations_eqn_FofR}, for four spherical potentials with scale radius $b$, 
   as well as the Kepler potential
   $\Phi = -G\mathcal{M}/r$.}
   \label{SAFluctuations_DimensionlessF}
   \end{figure}
   %%%%%%%%%%%%%%%%%%%%%%
We see that in the cored (Plummer and isochrone) models $F_\mathrm{circ}^*$ has
a maximum value of order unity which is realised when $R\sim b$, and that it
falls sharply to zero towards the centre of the cluster.  For centrally cusped
potentials (Hernquist and NFW) we again have $F_\mathrm{circ}^* \sim 1$ at
intermediate radii $R\sim b$, but $F_\mathrm{circ}^*$ diverges towards the
centre as $\sim R^{-1/2}$, typically reaching $F_\mathrm{circ}^* \sim 10$ at the
smallest sensible radii. At very large radii $R\gg b$, the isochrone, Plummer
and Hernquist potentials tend toward Keplerian behavior, $F_\mathrm{circ}^*
\sim R^{-3/2}$. (The logarithm in the NFW potential means it never quite becomes Keplerian at these
radii). 

From Figure \ref{SAFluctuations_DimensionlessF} we learn that (i) the magnitude
of short-timescale angular momentum fluctuations is roughly independent of
potential type for $R \gtrsim b$, (ii) short-timescale fluctuations are
significantly larger in cusped potentials than in cored potentials for $R < b$,
and (iii) very large values of $F^*$ can be reached at small radii in the Kepler
potential.  
%The effect of each point (i)-(iii) on the merger fractions of
%compact object binaries in spherical globular and nuclear star clusters is
%examined in Hamilton \& Rafikov (2020, in prep.). 

\subsection{Further discussion and examples}

Although it is strictly valid for circular outer orbits only, Figure \ref{SAFluctuations_DimensionlessF}
can also teach us something about non-circular outer orbits.
For instance, in cusped potentials, the scaling of $F_\mathrm{circ}^*$ with $R$ suggests that as long as the outer orbit is not too eccentric, a decent approximation to the dominant short-timescale
fluctuation can be found by employing the circular approximation with
$F_\mathrm{circ}$
(equation \eqref{SAFluctuations_eqn_FofR}) evaluated at $R=\rp$. In cored potentials this
is no longer true because of the turnover in $F_\mathrm{circ}^*$ at $R \sim b$.
Then, for example, for orbits with $r_\mathrm{a} \lesssim b$ the dominant $j$
fluctuations clearly arise around apocenter passage, since this is where $F_\mathrm{circ}^*$ is largest and this is where the outer orbit spends the most time.  However, outer orbits in cored
potentials with $\ra \lesssim b$ tend to have $\Gamma < 1/5$ (Paper I), so they tend not to reach such high eccentricities
anyway\footnote{Of course, the $\Gamma < 1/5$ numerical examples shown in this
paper \textit{do} reach very high $e$ (Figures \ref{fig:Example_Plummer_GR} and \ref{fig:Example_Plummer}), but that is because we have purposely
chosen a rather special set of initial conditions in order to make this happen.}, and
besides, the values of $F_\mathrm{circ}^*$ never exceeds $\sim 1$ regardless of
$R$ for these potentials. Hence, for an order of magnitude estimate we may
choose simply to evaluate $(\delta j)_\mathrm{max}$ using equation
\eqref{SAFluctuations_eqn_FofR} with $R=\rp$, regardless of the type of potential or
outer orbit.

In panel (p) of Figures
\ref{fig:Example_Hernquist} and \ref{fig:Example_Plummer}
we show as `error bars' the values of $\pm(\delta j)_\mathrm{p}$ (cyan) and
$\pm(\delta j)_\mathrm{a}$ (yellow), which are calculated by evaluating $\pm
(\delta j)_\mathrm{max}$ (equation \eqref{SAFluctuations_eqn_deltaj}) using the
circular approximation \eqref{SAFluctuations_eqn_FofR} at $R=\rp$ and $R=\ra$
respectively. We see that the circular approximation gives a {reasonable}
estimate of the amplitude of fluctuations $\delta j$. 

%However a simple improvement suggests itself, by considering that a maximal
%$\delta j$ fluctuation is not accumulated from $\delta j = 0$, but rather from
%the previous peak/trough in the time series.  This is perhaps most clearly shown
%%in Figure \ref{fig:Example_Hernquist_2}p, in which the
%sine-wave-like $\delta j(t)$ is centered around zero, so the maxiumum $\vert
%\delta j \vert $ is only ever around half the size of a maximal fluctuation
%$(\delta j)_\mathrm{max}$. This suggests that we include an extra factor $1/2$
%in our evaluations. With this in mind, we show the values of $ \pm 0.5\times
%(\delta j)_\mathrm{p/a}$ in panel (p) of Figures
%\ref{fig:Example_Hernquist_AB}-\ref{fig:Example_Plummer_AB}
%using red error bars, centered at the same positions as the cyan and yellow ones.
%We see that $0.5(\delta j)_\mathrm{p}$ is consistently a rather good
%approximation to the amplitude $\vert \delta j\vert$ as maximum eccentricity is
%approached.

%Two caveats are worth mentioning here.  The first is that whenever $\delta j(t)$
%has a `carrier wave' type signal --- i.e. when there ia significant power in the
%$2\Omega_\phi - \Omega_R$ mode (\S\ref{sec:Fourier_analysis}) ---
%then the fluctuations can accumulate over several $T_\phi$, as in Figures
%\ref{fig:Example_Plummer}-\ref{fig:Example_Plummer_2}.
%However this is not often an important effect, particularly because
%near-harmonic orbits, for which the effect is most pronounced, often have
%$\Gamma < 1/5$ so rarely reach extremely high eccentricities (Chapter
%\ref{PaperII_ch:PaperII}; \citet{Hamilton2019c}). 
One important caveat here is that while the $\delta j(t)$ behavior is often
rather regular up to $e_\mathrm{DA} \lesssim 0.99$, it often becomes rather
\textit{irregular} in the immediate vicinity of the eccentricity peak, as can be
seen in Figures
\ref{fig:Example_Hernquist} and \ref{fig:Example_Plummer}.
This is because of the rapid evolution, $\omega$, $\Omega$ {and $i$} when $j \approx
j_\mathrm{min}$ (see the light blue shaded bands in panels (j) and (k) of each
of those Figures, {as well as the inclination plot in Figure \ref{fig:Inclination_Plummer}}) which introduces a significant phase dependence into the
detailed fluctuation behavior. Of course, since
\eqref{SAFluctuations_eqn_deltaj} was derived by assuming stationary $\omega,
\Omega, i, j$, it necessarily fails to capture this irregular behavior.

{Finally, we attempted to capture the
complicated behavior of $\delta j(t)$ near the eccentricity peak for non-circular outer orbits,
and thereby move beyond the circular approximation, by
isolating the contribution of individual Fourier modes $\widehat{\delta j}(\nu)$ (panel (r) of Figures \ref{fig:Example_Hernquist}, \ref{fig:Example_Hernquist_4} and \ref{fig:Example_Plummer}).
However, even in the cases where a single dominant Fourier mode can be extracted (such as the $\nu = 2\Omega_\phi - \Omega_R$ mode in Figure \ref{fig:Example_Plummer}r), our
lack of knowledge of the outer orbital phase as high eccentricity was approached made it
near-impossible to predict in detail the SA behavior e.g. in Figure \ref{fig:Example_Plummer}o.
}

%otherwise.}.

\acknowledgements 

We thank Ulrich Sperhake and Bence Kocsis for their careful scrutiny of an earlier version of this work,
and Martin Pessah for helpful comments on \S\ref{sec:SA_breakdown}.
This work was supported by a grant from the Simons Foundation (816048, CH) and the John N. Bahcall Fellowship Fund,
{as well as STFC grant ST/T00049X/1 and the Ambrose Monell Foundation.}

%%%%%%%%%%%%%%%%%%%%%%%%%%%%%%%%%%%%%%%%%%%%%%%%%%



\bibliographystyle{apj}
\bibliography{Bibliography} % if your bibtex file is called example.bib

\begin{thebibliography}{}
\expandafter\ifx\csname natexlab\endcsname\relax\def\natexlab#1{#1}\fi

\bibitem[{Abbott {et~al.}(2021)Abbott, Abbott, Abraham, Acernese, Ackley,
  Adams, Adams, Adhikari, Adya, Affeldt, {et~al.}}]{abbott2021population}
Abbott, R., Abbott, T., Abraham, S., {et~al.} 2021, The Astrophysical journal
  letters, 913, L7

\bibitem[{Antognini {et~al.}(2014)Antognini, Shappee, Thompson, \&
  Amaro-Seoane}]{Antognini2014}
Antognini, J.~M., Shappee, B.~J., Thompson, T.~A., \& Amaro-Seoane, P. 2014,
  Monthly Notices of the Royal Astronomical Society, 439, 1079

\bibitem[{Antonini {et~al.}(2014)Antonini, Murray, \& Mikkola}]{Antonini2014}
Antonini, F., Murray, N., \& Mikkola, S. 2014, The Astrophysical Journal, 781,
  45

\bibitem[{Antonini \& Perets(2012)}]{Antonini2012}
Antonini, F., \& Perets, H.~B. 2012, The Astrophysical Journal, 757, 27

\bibitem[{{Bode} \& {Wegg}(2014)}]{Bode2014}
{Bode}, J.~N., \& {Wegg}, C. 2014, Monthly Notices of the Royal Astronomical
  Society, 438, 573

\bibitem[{{Bovy}(2015)}]{Bovy2015}
{Bovy}, J. 2015, The Astrophysical Journals, 216, 29

\bibitem[{Brown(1936)}]{brown1936stellar}
Brown, E.~W. 1936, Monthly Notices of the Royal Astronomical Society, 97, 56

\bibitem[{Bub \& Petrovich(2020)}]{Bub2020}
Bub, M.~W., \& Petrovich, C. 2020, The Astrophysical Journal, 894, 15

\bibitem[{Fabrycky \& Tremaine(2007)}]{Fabrycky2007}
Fabrycky, D., \& Tremaine, S. 2007, The Astrophysical Journal, 669, 1298

\bibitem[{Grishin {et~al.}(2018)Grishin, Perets, \& Fragione}]{Grishin2018}
Grishin, E., Perets, H.~B., \& Fragione, G. 2018, Monthly Notices of the Royal
  Astronomical Society, 481, 4907

\bibitem[{{Haim} \& {Katz}(2018)}]{Haim2018}
{Haim}, N., \& {Katz}, B. 2018, Monthly Notices of the Royal Astronomical
  Society, 479, 3155

\bibitem[{{Hamers}(2018)}]{Hamers2018}
{Hamers}, A.~S. 2018, Monthly Notices of the Royal Astronomical Society, 476,
  4139

\bibitem[{{Hamilton} \& {Rafikov}(2019{\natexlab{a}})}]{Hamilton2019c}
{Hamilton}, C., \& {Rafikov}, R.~R. 2019{\natexlab{a}}, The Astrophysical
  Journal Letters, 881, L13

\bibitem[{{Hamilton} \& {Rafikov}(2019{\natexlab{b}})}]{Hamilton2019a}
---. 2019{\natexlab{b}}, Monthly Notices of the Royal Astronomical Society,
  488, 5489

\bibitem[{{Hamilton} \& {Rafikov}(2019{\natexlab{c}})}]{Hamilton2019b}
---. 2019{\natexlab{c}}, Monthly Notices of the Royal Astronomical Society,
  488, 5512

\bibitem[{Hamilton \& Rafikov(2021)}]{Hamilton2021}
Hamilton, C., \& Rafikov, R.~R. 2021, Monthly Notices of the Royal Astronomical
  Society,
  https://academic.oup.com/mnras/advance-article-pdf/doi/10.1093/mnras/stab1284/37881081/stab1284.pdf,
  stab1284

\bibitem[{Hamilton \& Rafikov(2022)}]{hamilton2022anatomy}
---. 2022, The Astrophysical Journal, 939, 48

\bibitem[{{Ivanov} {et~al.}(2005){Ivanov}, {Polnarev}, \& {Saha}}]{Ivanov2005}
{Ivanov}, P.~B., {Polnarev}, A.~G., \& {Saha}, P. 2005, Monthly Notices of the
  Royal Astronomical Society, 358, 1361

\bibitem[{Katz \& Dong(2012)}]{Katz2012}
Katz, B., \& Dong, S. 2012, arXiv preprint arXiv:1211.4584

\bibitem[{{Kozai}(1962)}]{Kozai1962}
{Kozai}, Y. 1962, The Astronomical Journal, 67, 591

\bibitem[{Lei(2019)}]{Lei2019}
Lei, H. 2019, Monthly Notices of the Royal Astronomical Society, 490, 4756

\bibitem[{Lei {et~al.}(2018)Lei, Circi, \& Ortore}]{Lei2018}
Lei, H., Circi, C., \& Ortore, E. 2018, Monthly Notices of the Royal
  Astronomical Society, 481, 4602

\bibitem[{Leigh {et~al.}(2018)Leigh, Geller, McKernan, Ford, {Mac Low},
  Bellovary, Haiman, Lyra, Samsing, O'Dowd, Kocsis, \& Endlich}]{Leigh2018}
Leigh, N. W.~C., Geller, A.~M., McKernan, B., {et~al.} 2018, Monthly Notices of
  the Royal Astronomical Society, 474, 5672

\bibitem[{Li {et~al.}(2015)Li, Naoz, Kocsis, \& Loeb}]{Li2015}
Li, G., Naoz, S., Kocsis, B., \& Loeb, A. 2015, Monthly Notices of the Royal
  Astronomical Society, 451, 1341

\bibitem[{{Lidov}(1962)}]{Lidov1962}
{Lidov}, M.~L. 1962, Planetary Space Science, 9, 719

\bibitem[{{Luo} {et~al.}(2016){Luo}, {Katz}, \& {Dong}}]{Luo2016}
{Luo}, L., {Katz}, B., \& {Dong}, S. 2016, Monthly Notices of the Royal
  Astronomical Society, 458, 3060

\bibitem[{{Mangipudi} {et~al.}(2022){Mangipudi}, {Grishin}, {Trani}, \&
  {Mandel}}]{Mangipudi2022}
{Mangipudi}, A., {Grishin}, E., {Trani}, A.~A., \& {Mandel}, I. 2022, \apj,
  934, 44

\bibitem[{Miller \& Hamilton(2002)}]{Miller2002}
Miller, M.~C., \& Hamilton, D.~P. 2002, The Astrophysical Journal, 576, 894

\bibitem[{Naoz(2016)}]{Naoz2016}
Naoz, S. 2016, Annual Review of Astronomy and Astrophysics, 54, 441

\bibitem[{{Rasskazov} \& {Rafikov}(2023)}]{Rasskazov2023}
{Rasskazov}, A., \& {Rafikov}, R.~R. 2023, arXiv e-prints, arXiv:2310.15374

\bibitem[{{Rein} \& {Liu}(2012)}]{Rein2012}
{Rein}, H., \& {Liu}, S.~F. 2012, Astronomy and Astrophysics, 537, A128

\bibitem[{{Silsbee} \& {Tremaine}(2017)}]{Silsbee2017}
{Silsbee}, K., \& {Tremaine}, S. 2017, The Astrophysical Journal, 836, 39

\bibitem[{Tremaine(2023)}]{tremaine2023hamiltonian}
Tremaine, S. 2023, Monthly Notices of the Royal Astronomical Society, 522, 937

\bibitem[{Wen(2003)}]{Wen2003}
Wen, L. 2003, The Astrophysical Journal, 598, 419

\end{thebibliography}

%%%%%%%%%%%%%%%%% APPENDICES %%%%%%%%%%%%%%%%%%%%%

\end{document}